\documentclass[12pt,preprint]{emulateapj}
\shorttitle{Vibrationally excited HCN/HCO$^{+}$/HNC J=3--2 emission
  lines in IRAS 20551-4250} 
\shortauthors{Imanishi et al.}

\begin{document}


%\begin{CJK*}{UTF8}{gbsn}
\title{ALMA Investigation of Vibrationally Excited HCN/HCO$^{+}$/HNC
Emission Lines in the AGN-Hosting Ultraluminous Infrared Galaxy IRAS
20551$-$4250}


\author{Masatoshi Imanishi \altaffilmark{1,2}}
\affil{Subaru Telescope, 650 North A'ohoku Place, Hilo, Hawaii, 96720,
U.S.A.} 
\email{masa.imanishi@nao.ac.jp}

\author{Kouichiro Nakanishi \altaffilmark{2}}
\affil{National Astronomical Observatory of Japan, 2-21-1
Osawa, Mitaka, Tokyo 181-8588, Japan}

\and

\author{Takuma Izumi}
\affil{Institute of Astronomy, School of Science, The University of Tokyo,
2-21-1 Osawa, Mitaka, Tokyo 181-0015, Japan}

\altaffiltext{1}{National Astronomical Observatory of Japan, 2-21-1
Osawa, Mitaka, Tokyo 181-8588, Japan}
\altaffiltext{2}{Department of Astronomical Science,
The Graduate University for Advanced Studies (SOKENDAI), 
Mitaka, Tokyo 181-8588, Japan}
%\end{CJK*}

\begin{abstract}
We present the results of ALMA Cycle 2 observations of the ultraluminous  
infrared galaxy, IRAS 20551$-$4250, at HCN/HCO$^{+}$/HNC J=3--2 lines
at both vibrational-ground (v=0) and vibrationally excited (v$_{2}$=1)
levels. 
This galaxy contains a luminous buried active galactic nucleus (AGN), 
in addition to starburst activity, and our ALMA Cycle 0 data revealed a
tentatively detected vibrationally excited HCN v$_{2}$=1f J=4--3
emission line. 
In our ALMA Cycle 2 data, the HCN/HCO$^{+}$/HNC J=3--2 emission
lines at v=0 are clearly detected.
The HCN and HNC v$_{2}$=1f J=3--2 emission lines are also detected, but
the HCO$^{+}$ v$_{2}$=1f J=3--2 emission line is not. 
Given the high-energy level of v$_{2}$=1 and the resulting difficulty of
collisional excitation, we compared these results with those of the
calculation of infrared radiative pumping, using the available 
infrared 5--35 $\mu$m spectrum. 
We found that all of the observational results were reproduced, if the
HCN abundance was significantly higher than that of HCO$^{+}$ and HNC.
The flux ratio and excitation temperature between v$_{2}$=1f and v=0,
after correction for possible line opacity, suggests that infrared
radiative pumping affects rotational (J-level) excitation 
at v=0 at least for HCN and HNC. 
The HCN-to-HCO$^{+}$ v=0 flux ratio is higher than those of 
starburst-dominated regions, and will increase even more when the
derived high HCN opacity is corrected. 
The enhanced HCN-to-HCO$^{+}$ flux ratio in this AGN-hosting 
galaxy can be explained by the high HCN-to-HCO$^{+}$ abundance ratio and
sufficient HCN excitation at up to J=4, rather than the significantly
higher efficiency of infrared radiative pumping for HCN than HCO$^{+}$.
\end{abstract}


\section{Introduction}

The apparent ubiquity of supermassive blackholes 
(SMBHs) in the spheroidal components of present-day galaxies, and the
correlation between the mass of SMBHs and spheroidal stellar
components \citep{mag98,fer00,gul09,mcc13} indicate that 
the active galactic nucleus (AGN; SMBH-driven activity) and the
starburst (= active star-formation; plausibly the progenitors of
spheroids) are 
physically connected and have co-evolved. 
Observational constraints on how and when SMBHs grow in mass, in
relation to starbursts, is vital for fully understanding the galaxy
formation process in the universe. 

The widely accepted cold dark matter-based galaxy formation scenarios
postulate that mergers of gas-rich galaxies with SMBHs at their centers 
are common throughout the history of the universe \citep{hop08}.
Numerical simulations of such gas-rich galaxy mergers not only suggest 
that many stars form rapidly in dust/gas-obscured regions, but
also suggest that SMBHs have high mass accretion rates and become
luminous AGNs, deeply buried in the nuclear dust/gas, making such
gas-rich galaxy mergers luminous in the infrared \citep{hop06}.  
It is essential to evaluate the importance of {\it buried} AGNs
in dust/gas-rich infrared luminous galaxy mergers, by separating them
from the surrounding starbursts, if we are to observationally understand
the physical processes that occur during this key epoch in the
co-evolution of stars and SMBHs \citep{hop05,deb11}.  
However, unlike optically selected AGNs, where a central
mass-accreting SMBH is surrounded by toroidally shaped dust/gas (the
so-called dusty molecular torus) and the AGN's radiation can 
photo-ionize gas clouds along the torus axis above a torus scale height 
\citep{ant85}, deeply buried AGNs are optically elusive and their
detection requires observations at wavelengths of low dust
extinction. 

A buried AGN (mass accretion onto a SMBH) has two different properties
from a starburst (nuclear fusion inside stars). 
First, the efficiency of the AGN's radiative energy generation (6--42\%
of Mc$^{2}$, where M is the mass of the accreting material and c is the
speed of light) \citep{bar70,tho74} is much higher than that of nuclear 
fusion inside stars ($\sim$0.7\% of Mc$^{2}$),
so that an AGN has very high surface brightness emission, and can heat
a large amount of surrounding dust to high temperatures (several 100K), 
producing a strong mid-infrared 3--20 $\mu$m continuum emission.
Second, the X-ray to UV luminosity ratio is much higher in an AGN than
a starburst \citep{sha11,ran03}. 
These differences between an AGN and a starburst could create different
effects/feedback to the surrounding material.  
If we discover AGN-sensitive (sub)millimeter molecular lines, we can
disentangle the AGN from the starburst using the 
flux ratios of these lines, in a manner similar to the optical
AGN-starburst spectroscopic classification method, based on the optical
emission line flux ratios, which is extensively 
applied to less dusty galaxies \citep{vei87,kew01,kau03}. 
Once established, this (sub)millimeter molecular line energy diagnostic
method will create a large potential to scrutinize deeply 
buried AGNs in dust/gas-rich galaxies out to the distant universe,
because of (a) negligible dust extinction in the (sub)millimeter
wavelength range, and (b) the availability of highly sensitive ALMA
observing facility.  

Based on pre-ALMA (sub)millimeter interferometric 1--5'' resolution 
observations of the nuclear regions of nearby bright starburst and
Seyfert galaxies (= modest luminosity AGNs), the enhanced emission of
the dense gas tracer, HCN, at J=1--0 (rotational transition), relative
to HCO$^{+}$, has been found in AGNs and been proposed as a possible AGN
indicator \citep{koh05,kri08}.
This was mainly thanks to the spatially-resolved investigation of
AGN-affected molecular gas, by reducing the contamination from 
spatially-extended star-formation-affected molecular gas emission in
Seyfert host galaxies, 
because the contamination could hamper the detection of the signatures
of AGN-affected molecular gas emission in large-aperture single-dish
telescope observations \citep{gra06,baa08,sne11,cos11}. 
Subsequent interferometric observations have also confirmed enhanced HCN
J=1--0 emission in the nuclei of dust/gas-rich merging luminous infrared
galaxies (LIRGs; L$_{\rm IR}$ $>$ 10$^{11}$L$_{\odot}$), which are
diagnosed to be buried-AGN-important based on infrared spectroscopy, 
compared to starburst-dominated LIRGs
\citep{ima04,ima06,in06,ima07a,ima09a}. 
Based on recent single-dish telescope's observations, a similar
trend of HCN J=1--0 flux excess, relative to HCO$^{+}$ J=1--0, has been
confirmed for infrared-classified AGN-important LIRGs, compared to
starburst-dominated LIRGs \citep{pri15}.
ALMA Cycle 0--1 interferometric observations further support HCN
emission enhancement, relative to HCO$^{+}$, in the nuclei of
AGN-important LIRGs, compared to starburst regions at J=4--3
\citep{ima13a,ima13b,ion13,ima14,gar14,izu15}.  
It has now become clear that luminous AGNs tend to show enhanced
HCN emission.
However, the physical interpretation of enhanced HCN emission in AGNs is
still unclear.  
Although HCN abundance enhancement in an AGN is proposed in some
chemical calculations, because of high dust temperature and/or X-ray
irradiation \citep{mei05,lin06,har10}, it is also calculated that 
the HCN-to-HCO$^{+}$ abundance ratio in molecular gas around an AGN 
can vary dramatically, depending on assumed physical parameters
\citep{mei05,har13}. Additionally, even if HCN abundance is indeed
enhanced in an AGN, the widely assumed collisional excitation alone
seems not to be enough to reproduce the strong HCN rotational
J-transition emission lines observed in some AGNs \citep{yam07}. 

An infrared radiative pumping mechanism has been proposed to be a
plausible scenario for the enhancement of HCN rotational J-transition
emission in AGNs \citep{aal95,gar06,wei07}.
Specifically, HCN can be vibrationally excited to v$_{2}$=1 
by absorbing infrared 14 $\mu$m photons \citep{tow75}.
Through the decay back to the vibrational-ground level (v=0), 
the HCN rotational J-transition flux at v=0 can be
stronger than collisional excitation alone \citep{ran11}. 
Since an AGN can emit infrared 14 $\mu$m continuum emission much more
strongly than a starburst, due to AGN-heated hot (several 100 K) dust, 
such an infrared radiative pumping mechanism for HCN should work more
effectively in an AGN than in a starburst. 
In fact, the HCN v$_{2}$=1--0 {\it absorption} features were
seen at infrared 14 $\mu$m in several obscured AGNs \citep{lah07},
and the vibrationally excited HCN v$_{2}$=1, l=1f (hereafter v$_{2}$=1f)
J=4--3 and/or J=3--2 {\it emission} lines in the (sub)millimeter
wavelength have recently been detected in several AGN-hosting LIRGs 
\citep{sak10,ima13b,aal15a,aal15b,cos15}, supporting the scenario that
infrared radiative pumping actually works in some AGNs to vibrationally
excite HCN and could contribute to an increase in HCN J-transition flux
at v=0.    

However, the other dense gas tracers, HCO$^{+}$ and HNC, also have
absorption features at infrared $\sim$12 $\mu$m and $\sim$21.5 $\mu$m,
respectively \citep{dav84,kaw85,bur87}, and could therefore be
vibrationally excited to v$_{2}$=1 by infrared radiative pumping.  
If the infrared radiative pumping mechanism is responsible for the
enhanced HCN J-transition emission at v=0 in AGNs, its efficiency must be
particularly high for HCN only. 
A comparison of the strengths of the v$_{2}$=1 emission lines for HCN,
HCO$^{+}$, and HNC is the most straightforward way to test the scenario 
of whether infrared radiative pumping works most efficiently for HCN
and selectively boosts HCN J-transition flux at v=0.
No meaningful observational constraints are placed in external galaxies
for the strengths of HCO$^{+}$ and HNC v$_{2}$=1f J-transition lines, 
except for the report concerning the detection of HNC v$_{2}$=1f J=3--2
emission line in the AGN-hosting LIRG, NGC 4418 \citep{cos13}.  

We conducted ALMA Cycle 2 observations of the luminous
infrared galaxy, IRAS 20551$-$4250 (z=0.043), to investigate the
relative strength of vibrationally excited (v$_{2}$=1) HCN, HCO$^{+}$,
and HNC J-transition lines. 
IRAS 20551$-$4250 is categorized as an ultraluminous
infrared galaxy (ULIRG) with infrared (8--1000 $\mu$m) luminosity 
L$_{\rm IR}$ = 10$^{12.0}$L$_{\odot}$, and has far-infrared (40--500
$\mu$m) luminosity L$_{\rm FIR}$ = 10$^{11.9}$L$_{\odot}$ (see Table 1
of Imanishi \& Nakanishi 2013b).
It consists of main single nucleus with the signature of a long 
merging tail in the southern direction \citep{duc97,rot04}.
It is classified optically as a LINER/HII-region; namely, it has no
obvious optical AGN (Seyfert) signature \citep{duc97,yua10}.  
Based on several energy diagnostic methods using infrared and X-ray
spectroscopy, the signature of a luminous, energetically significant
(20--60\% of the bolometric luminosity) buried AGN was identified
in IRAS 20551$-$4250, in addition to starburst activity
\citep{fra03,ris06,san08,nar08,nar09,nar10,ima10,ima11,vei13}.  
The vibrationally excited HCN v$_{2}$=1f J=4--3 emission line was 
tentatively detected with our ALMA Cycle 0 data because of small
molecular line widths, which enable us to clearly separate the HCN
v$_{2}$=1f J=4--3 emission line from the nearby, much brighter, HCO$^{+}$
v=0 J=4--3 emission line \citep{ima13b}.
Thus, IRAS 20551$-$4250 is a very good target for investigating the
relative strength of vibrationally excited HCN/HCO$^{+}$/HNC
J-transition emission lines, and how the infrared radiative pumping
mechanism affects the rotational excitation of HCN/HCO$^{+}$/HNC at v=0. 

In the present paper, we report the results of ALMA Cycle 2 observations of
the AGN-hosting ULIRG, IRAS 20551$-$4250, at HCN, HCO$^{+}$, and HNC
lines at both the vibrational-ground (v=0) and vibrationally excited
(v$_{2}$=1) levels. 
Throughout this paper, we adopt H$_{0}$ $=$ 71 km s$^{-1}$ Mpc$^{-1}$, 
$\Omega_{\rm M}$ = 0.27, and $\Omega_{\rm \Lambda}$ = 0.73
\citep{kom09}, in which case the luminosity distance is 188 Mpc
and 1 arcsec corresponds to 0.84 kpc at the distance of IRAS
20551$-$4250.    
Molecular parameters are derived from the Cologne Database of Molecular
Spectroscopy (CDMS) \citep{mul05} via Splatalogue
(http://www.splatalogue.net). 
For molecular transitions, where there is no indication of v (the
vibrational-level), we mean v=0 (i.e., the vibrational-ground level). 
When we write HCN, HCO$^{+}$, and HNC in this manuscript, we mean 
H$^{12}$C$^{14}$N, H$^{12}$C$^{16}$O$^{+}$, and H$^{14}$N$^{12}$C,
respectively.

\section{Observations and Data Analysis}

Our ALMA Cycle 2 observations in band 6 (211--275 GHz) were
undertaken within the program 2013.1.00033.S (PI = M. Imanishi). 
The details of our observations are given in Table 1.
We adopted the widest 1.875 GHz band mode and 3840 total channel number
for our observations.
We requested $<$0$\farcs$6 spatial resolution in our proposal.

We employed three spectral windows in USB in band 6 to cover (a) HCN 
J=3--2 (rest-frame frequency is $\nu_{\rm rest}$ = 265.89 GHz), 
(b) HCO$^{+}$ J=3--2 ($\nu_{\rm rest}$ = 267.56 GHz) and HCN v$_{2}$=1f
J=3--2 ($\nu_{\rm rest}$ = 267.20 GHz), and (c) HCO$^{+}$ v$_{2}$=1f
J=3--2 ($\nu_{\rm rest}$ = 268.69 GHz) in each of the three spectral
windows. HCN and HCO$^{+}$ J=3--2 emission lines can therefore be
obtained at both the vibrational-ground (v=0) and vibrationally excited
(v$_{2}$=1f) levels concurrently.

We also conducted additional independent observations for 
HNC J=3--2 ($\nu_{\rm rest}$ = 271.98 GHz) and 
HNC v$_{2}$=1f J=3--2 ($\nu_{\rm rest}$ = 273.87 GHz).
We employed four spectral windows, of which two in USB were used to 
cover HNC J=3--2 and HNC v$_{2}$=1f J=3--2 lines.
The remaining two spectral windows in LSB were used to probe continuum
emission and serendipitously detected emission lines. 

We used CASA \citep{pet12} for our data analysis. 
We first checked the visibility plots in calibrated data provided by the
Joint ALMA Observatory and found that the presence of HCN, HCO$^{+}$,
and HNC J=3--2 emission lines was clearly recognizable.  
Signatures of other weak emission lines were also seen in some spectral
windows. 
We selected channels that were not clearly affected by detected emission
line features, and estimated the continuum flux levels, by using data in 
(a) three spectral windows for HCN/HCO$^{+}$ observations and 
(b) four spectral windows for HNC observations. 
After subtracting these estimated continuum levels, we performed the task 
``clean'' for detected emission line data.
We applied the ``clean'' procedure also for the continuum data. 
We employed 40 channel spectral binning ($\sim$20 km s$^{-1}$), and
adopted a pixel scale of 0$\farcs$1 pixel$^{-1}$ in this clean procedure. 
The maximum recoverable scale in our data is $\sim$10''.
In ALMA Cycle 2, the absolute calibration uncertainty is estimated to be
$<$10\% in band 6, which will not significantly affect our main
discussion.  

\section{Results}

From our observations, two continuum data were obtained.
Figure 1 displays continuum-a (taken with the HCN and HCO$^{+}$
J=3--2 observations) and continuum-b (taken with HNC J=3--2) maps.
Table 2 presents the continuum emission properties.
The achieved spatial resolution is $\sim$0$\farcs$5 ($\sim$420 pc)
(Table 2, column 5) and meets our request of $<$0$\farcs$6.  
Using the CASA task ``imfit'', we estimate the deconvolved,
intrinsic emission size to be 0$\farcs$28 $\times$ 0$\farcs$23 (240 pc
$\times$ 190 pc) for continuum-a and 0$\farcs$28 $\times$ 0$\farcs$24  
(240 pc $\times$ 205 pc) for continuum-b, with the uncertainties of
$<$20 pc, where the values along the major and minor axes are shown.
Continuum emission is estimated to have a spatial extent with
about half of the ALMA synthesized beam size.
These independently taken continuum data show consistent
results, in terms of peak position, peak flux, and morphology.
To understand the physical origin of the detected continuum emission,  
we assume that half of the far-infrared (40--500 $\mu$m) luminosity of
IRAS 20551$-$4250 comes from starburst activity, because (a) 
the AGN bolometric contribution in IRAS 20551$-$4250 is
estimated to be 20--60\% \citep{nar08,nar09,nar10,vei13} and (b) the
bolometric luminosity of IRAS 20551$-$4250 is dominated by 
far-infrared (40--500 $\mu$m) emission (L$_{\rm FIR}$ =
10$^{11.9}$L$_{\odot}$) ($\S$1).
The contribution from thermal free-free emission in starburst
HII-regions at $\sim$255 GHz is estimated to be $\sim$1.6 mJy, using
equation (1) of \citet{nak05}. 
This is a significant contribution to the observed continuum flux 
at $\sim$255 GHz with 3.6--3.8 mJy. 
If the starburst contribution to the observed far-infrared luminosity is 
higher than 50\%, the free-free emission flux at $\sim$255 GHz will be
even higher. 
Therefore, we infer that the detected continuum emission ($\sim$3.6--3.8
mJy) at $\sim$255 GHz is a combination of thermal free-free emission
from starburst HII-regions and dust emission heated by dust-obscured
energy sources (AGN and/or starburst).
Since the detected continuum emission is spatially compact with $<$250
pc (Fig. 1), the thermal free-free emission should be of nuclear 
starburst origin.  

A full spectrum taken with our ALMA observations at the continuum peak
position within the beam size is shown in Figure 2. 
In the spectrum taken with HCN/HCO$^{+}$ J=3--2 lines, 
the HCN v$_{2}$=1f J=3--2 emission line is visible, but the HCO$^{+}$
v$_{2}$=1f J=3--2 line is not (Fig. 2c). 
In the spectrum taken with the HNC J=3--2 line, 
the HNC v$_{2}$=1f J=3--2 emission line is detected (Fig. 2e). 
We also serendipitously detected several other emission lines.
In the spectrum taken with HCN/HCO$^{+}$ J=3--2 lines, 
an emission line is recognizable at the observed frequency 
$\nu_{\rm obs}$ $\sim$ 257.40 GHz, or rest-frame frequency 
$\nu_{\rm rest}$ $\sim$ 268.47 GHz. 
We ascribe this emission line to HOC$^{+}$ J=3--2 (v=0) at 
$\nu_{\rm rest}$ = 268.45 GHz, which was strongly detected in the
Galactic star-forming regions \citep{app97,app99,sav04}
and LIRGs \citep{aal15a,aal15b}.

In the spectrum taken with the HNC J=3--2 line, emission lines are 
identifiable at $\nu_{\rm obs}$ $\sim$ 245.75 GHz, 247.60 GHz, 
248.35 GHz, 261.65 GHz, and 262.15 GHz.
The third and fourth lines at $\nu_{\rm obs}$ $\sim$ 248.35 GHz and 
261.65 GHz correspond to the rest-frame frequencies of 
$\nu_{\rm rest}$ $\sim$ 259.0 GHz and 272.9 GHz, which 
we ascribe to H$^{13}$CN J=3--2 
($\nu_{\rm rest}$ $\sim$ 259.01 GHz) and HC$_{3}$N J=30--29 
($\nu_{\rm rest}$ $\sim$ 272.88 GHz) \citep{cos10,cos13}, respectively.
The first and second lines at $\nu_{\rm obs}$ $\sim$ 245.75 GHz and 
247.60 GHz correspond to $\nu_{\rm rest}$ $\sim$ 256.3 GHz and 258.2
GHz, which we tentatively attribute to SO$_{2}$ 5(3,3)--5(2,4)
($\nu_{\rm rest}$ $\sim$ 256.25 GHz) and SO 6(6)--5(5) 
($\nu_{\rm rest}$ $\sim$ 258.26 GHz), respectively, because these lines
are the strongest in these frequency ranges in the Galactic star-forming
region, Orion A \citep{bla86,gre91}. 
For the second line, HC$^{15}$N J=3--2 emission line 
($\nu_{\rm rest}$ $\sim$ 258.16 GHz) might make some contribution.
For the fifth line at $\nu_{\rm obs}$ $\sim$ 262.15 GHz, or 
$\nu_{\rm rest}$ $\sim$ 273.4 GHz, we tentatively assign it as 
CH$_{3}$CCH 16(2)--15(2) ($\nu_{\rm rest}$ $\sim$ 273.40
GHz), based on the extra-galactic detection in the Splatalogue database
(http://www.splatalogue.net). 

We note that the rest-frequencies of v$_{2}$=1e J=3--2 for HCN
($\nu_{\rm rest}$ = 265.85 GHz), 
HCO$^{+}$ ($\nu_{\rm rest}$ = 267.42 GHz), and 
HNC ($\nu_{\rm rest}$ = 271.92 GHz) are so close to the v=0 J=3--2 line
for each molecule that it is very difficult to separate the v$_{2}$=1e
J=3--2 lines from the much stronger v=0 J=3--2 emission lines. 
The fluxes of the v$_{2}$=1e emission lines of HCN/HCO$^{+}$/HNC will
not be discussed in this paper.  

We create integrated intensity (moment 0) maps of individual molecular
lines by combining channels with clear signal signs. 
Figure 3 displays these maps. 
Table 3 summarizes their properties.
The achieved spatial resolution is $\sim$0$\farcs$5 ($\sim$420 pc)
for all molecular emission lines (Table 3, column 6), and so our request
of $<$0$\farcs$6 is fulfilled.
The molecular line peak positions agree with the continuum emission peak
within 1 pixel (0$\farcs$1) in RA and DEC directions for all lines,
suggesting that several serendipitously detected faint emission lines
are real.
The estimated deconvolved, intrinsic emission sizes for the
brightest HCN, HCO$^{+}$, and HNC v=0 J=3--2 lines, using the CASA 
task $''$imfit$''$, are 0$\farcs$22 $\times$ 0$\farcs$20 (185 pc
$\times$ 170 pc), 0$\farcs$27 $\times$ 0$\farcs$26 (225 pc $\times$ 220
pc), and  0$\farcs$21 $\times$ 0$\farcs$17 (175 pc $\times$ 145 pc), 
with the uncertainties of $<$15 pc, respectively.  
HCO$^{+}$ v=0 J=3--2 emission may show slightly larger spatial
extension than HCN v=0 J=3--2 and HNC v=0 J=3--2 emission.

Figure 4 shows spectra around individual molecular lines of our interest
at the continuum peak position within the beam size. 
We fit detected molecular emission lines in the spectra with
Gaussian profiles.   
Table 3 summarizes the fitting results.
The flux estimates based on the Gaussian fits will be used in our
discussion, except for the non-detected HCO$^{+}$ v$_{2}$=1f J=3--2
emission line for which the upper limit in the integrated intensity
(moment 0) map is adopted. 
The molecular line luminosities are summarized in Table 4.

Intensity-weighted mean velocity (moment 1) and intensity-weighted
velocity dispersion (moment 2) maps for HCN, HCO$^{+}$, and HNC J=3--2
(v=0) emission lines are shown in Figure 5.  
In all moment 1 maps, the northeastern region is redshifted and 
the southwestern region is blueshifted, with respect to the nucleus,
suggesting a rotating gas motion.
A similar gas motion was seen in HCN and HCO$^{+}$ J=4--3 (v=0) emission 
lines \citep{ima13b}, suggesting that the spatial origin of these
detected molecular emission lines is similar.
A similar rotational gas motion is seen also at CO J=1--0 emission 
\citep{ued14}, although the detected CO J=1--0 emission is distributed
in a wider area than HCN, HCO$^{+}$, and HNC, due to its
lower critical density and higher brightness.

\section{Discussion}

\subsection{Molecular line flux ratio}

In Figure 6, the observed HCN-to-HCO$^{+}$ J=3--2 and HCN-to-HNC J=3--2
flux ratios of IRAS 20551$-$4250 are added to the plot for J=4--3
\citep{ima14}. 
The AGN-hosting ULIRG, IRAS 20551$-$4250, shows similar HCN-to-HCO$^{+}$
flux ratios both at J=4--3 and J=3--2, and the ratios are higher than
starburst-dominated regions. 

Both the HCN-to-HCO$^{+}$ and HCN-to-HNC flux ratios can change,
depending on 
(a) abundance ratio resulting from various chemical reactions, 
(b) excitation condition, and (c) flux attenuation by line opacity.
The flux comparison between J=4--3 \citep{ima13b} and J=3--2 (Table 5)
for HCN, HCO$^{+}$, and HNC shows that the ratios of J=4--3 to J=3--2
flux in [Jy km s$^{-1}$] are 1.6$\pm$0.2 for 
HCN, 1.7$\pm$0.2 for HCO$^{+}$, and 1.8$\pm$0.2 for HNC, where maximum
$\sim$10\% absolute calibration uncertainty in individual ALMA
observations is included. 
These ratios are comparable to the expected ratio for thermally excited
optically thick gas (= 16/9), which suggests that the excitation
temperatures of HCN, HCO$^{+}$, and HNC for J=3--2 and J=4--3 are
comparable to the kinetic temperature of molecular gas in the nuclear
region of IRAS 20551$-$4250.
In general, when molecular gas density and temperature are not
sufficiently high, collisional excitation at high-J levels can become
sub-thermal more quickly and deviate from thermal condition more
strongly for a molecule with higher critical density than another
molecule with lower critical density.
While HCN and HNC have comparable critical densities at individual 
J-transitions, HCO$^{+}$ has a slightly lower critical density than HCN
and HNC by a factor of $\sim$5 for the same J-transition \citep{mei07,gre09}. 
Thus, the HCN-to-HCO$^{+}$ flux ratio can be smaller at higher 
J-transition (e.g. J=4--3) than at lower J-transition (e.g., J=3--2) in
some galaxies. 
The observed similar HCN-to-HCO$^{+}$ and HCN-to-HNC flux ratios between
J=4--3 and J=3--2 suggest that the different excitation condition for
different molecules (item (b)above) is not an important factor for the
ratios at J=3--2 and J=4--3 in IRAS 20551$-$4250. 
However, in other galaxies, sub-thermal excitation conditions can occur, 
in which case the observed HCN-to-HCO$^{+}$ flux ratio at J=4--3 (and
J=3--2) can be smaller than that at J=1--0.  
This may be partly responsible for the small observed HCN-to-HCO$^{+}$
J=4--3 flux ratios in starburst-dominated regions in Figure 6
\citep{ima14}.   
If this is the case, sufficient excitation at up to J=4 in IRAS
20551$-$4250 contributes to its higher HCN-to-HCO$^{+}$ flux ratio
at J=4--3 (and J=3--2) than starburst-dominated regions.  

\subsection{Vibrationally excited HCN/HCO$^{+}$/HNC J=3--2 emission lines}

HCN, HCO$^{+}$, and HNC have three vibrational modes (v$_{1}$, v$_{2}$,
and v$_{3}$).
Since the energy level of v$_{2}$=1 is the lowest, its vibrational
excitation is the simplest \citep{tow75,ama83,fos84a,fos84b,dav84,kaw85}. 
HCN v$_{2}$=1f J=4--3 and J=3--2, and HNC v$_{2}$=1f
J=3--2 emission lines have previously been 
detected in the LIRG, NGC 4418 (L$_{\rm IR}$ $\sim$ 10$^{11}$L$_{\odot}$)
\citep{sak10,cos13,cos15}, which has observational signatures of a
luminous buried AGN \citep{spo01,eva03,ima04,sak13}. 
The HCN v$_{2}$=1f J=4--3 emission line has also been detected in the 
AGN-hosting ULIRG, IRAS 20551$-$4250 
(L$_{\rm IR}$ $\sim$ 10$^{12}$L$_{\odot}$) \citep{ima13b}.
The other AGN-hosting ULIRG, Mrk 231 
(L$_{\rm IR}$ $\sim$ 10$^{12.5}$L$_{\odot}$; Sanders et al. 2003), also
shows a detectable HCN v$_{2}$=1f J=3--2 emission line \citep{aal15a}. 
Very recently, the detection of HCN v$_{2}$=1f J=3--2 emission
lines in three LIRGs (IC 860, Zw 049.057, and Arp 220) and HCN
v$_{2}$=1f J=4--3 emission lines in two LIRGs (Arp 220W and IRAS
17208$-$0014) has been reported \citep{aal15b,mar16}.
This paper adds the detection of HCN v$_{2}$=1f J=3--2 and HNC
v$_{2}$=1f J=3--2 emission lines in IRAS 20551$-$4250.

Regarding the HCO$^{+}$ v$_{2}$=1f J=3--2 emission line, we obtain
non-detection in our ALMA data for IRAS 20551$-$4250 (Table 3).   
The HCO$^{+}$ v$_{2}$=1f emission line has never been discussed in
detail for external galaxies in the published literature.
\citet{sak10} covered this HCO$^{+}$ v$_{2}$=1f J=3--2 emission 
line in the data of NGC 4418 and reported its non-detection, with no
upper limit flux provided.
In the present paper, we provide the first stringent constraint on the
upper limit of its flux in external galaxies.

In the case of optically thin conditions, 
the column density of the level {\it u}
(N$_{u}$) is written as 
\begin{eqnarray}
N_u &=& \frac{4\pi}{hc A_{ul} \Omega} \times flux \\
 &\propto& \frac{flux}{A_{ul}}, 
\end{eqnarray}
where the line flux from the upper (u) to lower (l) transition level is
in units of [Jy km s$^{-1}$], $h$ is the Planck constant, $c$ is the
speed of light, $\Omega$ is the source solid angle, and $A_{ul}$ is the
Einstein A coefficient for spontaneous emission from the upper (u) to
lower (l) level \citep{gol99,izu13}.
We assume that HCN, HCO$^{+}$, and HNC surround energy sources
(buried AGN and stars) with almost full covering factor in the same
geometry, because (a) the optical spectroscopic non-Seyfert
classification of IRAS 20551$-$4250 \citep{duc97,yua10} suggests that the
putative AGN detected in infrared and X-ray observations ($\S$1) is
obscured by dust and molecular gas in virtually all directions
(=buried), and (b) newly formed stars are also deeply embedded in dust
and molecular gas. Namely, emission structure inside the ALMA
synthesized beam is assumed to be the same among HCN, HCO$^{+}$, and HNC. 
Adopting the Einstein A coefficients from the J=3 to J=2 level at
v$_{2}$=1f shown in Table 5, the ratio of the column density at the
v$_{2}$=1f J=3 level among HCN, HCO$^{+}$, and HNC is 
\begin{eqnarray}
HCN:HCO^{+}:HNC (N_{v_{2}=1f, J=3})(observed) & = & 1:<0.20:0.69, 
\end{eqnarray}
where the fluxes based on Gaussian fits (Table 3, column 10) are used
for the detected HCN v$_{2}$=1f J=3--2 and HNC v$_{2}$=1f J=3--2 lines,
whereas the upper limit ($<$3$\sigma$) in the integrated intensity 
(moment 0) map (Table 3, column 3) is adopted for the undetected
HCO$^{+}$ v$_{2}$=1f J=3--2 line. 
This will be compared to theoretical calculations in the
following subsection.  
If HC$_{3}$N v$_{7}$=1f J=30--29 emission blends with HNC v$_{2}$=1f
J=3--2 \citep{cos13}, the actual HNC v$_{2}$=1f J=3--2 emission line
flux could be smaller by 25--50\% (Table 3).

The ratios of observed v$_{2}$=1f J=3--2 to v=0 J=3--2 flux in units 
of [Jy km s$^{-1}$] for HCN, HCO$^{+}$, and HNC are summarized in 
Table 6.
The ratios are 
\begin{eqnarray}
HCN:HCO^{+}:HNC (v_{2}=1f/v=0)(observed) &=& 1:<0.25:1.5. 
\end{eqnarray}
This is the second observed quantity to indicate the efficiency of
vibrational excitation to the v$_{2}$=1 level. 

\subsection{Comparison with infrared radiative pumping model}

\subsubsection{Abundance ratio among HCN, HCO$^{+}$, and HNC}

The energy levels of v$_{2}$=1 are $\sim$1030 K (HCN), $\sim$1200 K
(HCO$^{+}$), and $\sim$670 K (HNC). 
These energy levels are too high to be collisionally excited, but 
could be excited by infrared radiative pumping, by absorbing infrared
photons at $\sim$14 $\mu$m (HCN), $\sim$12 $\mu$m (HCO$^{+}$), and 
$\sim$21.5 $\mu$m (HNC). 
The ratio in Equation (3) reflects the product of the infrared
radiative pumping rate to v$_{2}$=1 and column density at v=0 of each
molecule.
To estimate the effect of the infrared radiative pumping, we assume that 
(a) HCN, HCO$^{+}$, and HNC emission lines at both v=0 and 
v$_{2}$=1f come from the same regions, 
relative to the central buried AGN of IRAS 20551$-$4250, and 
(b) the column density at v=0 (N$_{v=0}$) is comparable among HCN,
HCO$^{+}$, and HNC. 
The first assumption comes from the fact that HCN, HCO$^{+}$, and HNC
have similar dipole moments ($\mu$ = 3.0, 3.9, and 3.1 Debye,
respectively), and so their emission lines at the same J-transition 
at v=0 probe similarly dense molecular gas in a galaxy, compared
to the widely-used CO J-lines ($\mu$ = 0.1 Debye), which trace 
much less dense molecular gas at the same J-transition.
Here, it is assumed that v$_{2}$=1f emission also comes from the 
same regions.
These assumptions will be discussed later. 

The infrared radiative pumping rate (P$_{\rm IR}$) is 
\begin{eqnarray}
P_{IR} & \propto & B_{v2=0-1,vib} \times F_{\nu (IR)} \times N_{v=0}, 
\end{eqnarray}
where B$_{v2=0-1,vib}$ is the Einstein B coefficient from v=0 to
v$_{2}$=1, F$_{\nu (IR)}$ is infrared flux in F$_{\nu}$ [Jy] used for the
infrared radiative pumping of HCN/HCO$^{+}$/HNC, and N$_{v=0}$ is the
column density at the v=0 level. 
The B$_{v2=0-1,vib}$ is related to the Einstein A coefficient from
v$_{2}$=1 to v=0 (A$_{v2=1-0,vib}$) in the form of 
\begin{eqnarray}
B_{v2=0-1,vib} &\propto& \lambda^3 \times A_{v2=1-0,vib} 
\end{eqnarray}
\citep{ryb79}.

For v$_{2}$=1, the A$_{v2=1-0,vib}$ values are estimated to be 
$\sim$1.7 s$^{-1}$ (HCN) \citep{deg86,aal07a}, 
$\sim$3.0 s$^{-1}$ (HCO$^{+}$) \citep{mau95}, 
and $\sim$5.2 s$^{-1}$ (HNC) \citep{aal07a}.
Thus, the ratio of the B$_{v2=0-1,vib}$ value among HCN, HCO$^{+}$, and
HNC is 
\begin{eqnarray}
HCN:HCO^{+}:HNC (B_{v2=0-1,vib}) &=& 1:1.1:11
\end{eqnarray}

The infrared flux for the infrared radiative pumping of HCN (14 $\mu$m),
HCO$^{+}$ (12 $\mu$m), and HNC (21.5 $\mu$m) is estimated from the 5--35
$\mu$m spectrum of IRAS 20551$-$4250 taken by Spitzer IRS (Fig. 7). 
The 5--35 $\mu$m spectrum in Figure 7 shows strong silicate dust
absorption features at $\lambda_{\rm rest}$ $\sim$ 9.7 $\mu$m and 18
$\mu$m (rest-frame), which means that mid-infrared (3--20 $\mu$m)
continuum emission originated in energy sources (buried AGN and/or
stars) is obscured by dust at the foreground.
We need to correct for these silicate dust absorption features, if we
are to obtain the intrinsic mid-infrared spectral energy distribution
which is produced in the close vicinity of the obscured energy sources
and is illuminated to their nearby molecular gas. 

The HCN v$_{2}$=1f J=3--2 or J=4--3 emission was detected in some
Galactic sources \citep{ziu86,mil13,vea13,nag15}.   
Since the emission can be produced in the region where infrared 14
$\mu$m photons are available and since Galactic sources emit some
amount of infrared 14 $\mu$m photons, the detection in nearby Galactic
sources is reasonable, as long as observations are sensitive enough to
detect low luminosity emission.   
However, IRAS 20551$-$4250 at z=0.043 (luminosity distance is 188 Mpc)
is much further away from our Galaxy and the detected v$_{2}$=1f
emission line luminosity in IRAS 20551$-$4250 (Table 4) is more than
several orders of magnitude higher than the Galactic sources. 
Since an AGN is a much stronger mid-infrared continuum emitter than a
starburst under the same bolometric luminosity, due to the larger
amount of hot dust in the former ($\S$1), an AGN can produce strong 
v$_{2}$=1f emission more easily than stars.
\citet{nar08,nar09} estimated that $\sim$90\% of the observed 5--8
$\mu$m flux of IRAS 20551$-$4250 comes from AGN-heated hot dust emission. 
We also investigate the rest-frame equivalent widths of the 6.2 $\mu$m, 
7.7 $\mu$m, and 11.3 $\mu$m polycyclic aromatic hydrocarbon (PAH)
emission features from the spectrum of IRAS 20551$-$4250 in Figure 7,
using the same definition as that employed by \citet{ima07b}.
It is widely accepted that a starburst shows strong PAH emission features,
while a pure AGN does not \citep{moo86,gen98,imd00}, because PAHs are
destroyed in the close vicinity of an AGN \citep{voi92}. 
The derived rest-frame equivalent widths of the 6.2 $\mu$m, 7.7 $\mu$m,
and 11.3 $\mu$m PAH emission features are $\sim$45 [nm], $\sim$190 [nm],
and $\sim$185 [nm], respectively.   
These values are much smaller than the typical values found in
starburst-dominated galaxies ($\sim$540 [nm], $\sim$690 [nm], and
$\sim$600 [nm] for the 6.2 $\mu$m, 7.7 $\mu$m, and 11.3 $\mu$m PAH
emission features, respectively) \citep{ima07b}, which again supports 
the scenario that the observed mid-infrared flux of this galaxy is
dominated by AGN-heated hot dust emission. 
\citet{far03} also modeled that the observed 10--25 $\mu$m
emission of IRAS 20551$-$4250 is dominated by an AGN-origin dust
continuum emission component.
Thus, we regard that the observed mid-infrared continuum from IRAS 
20551$-$4250 is dominated by the AGN-heated hot dust emission.
Since the AGN-heated hot dust emission is spatially very compact and is
more centrally-concentrated than surrounding dust \citep{ima07b}, a
simple foreground screen dust model is applicable. 

Following \citet{ima07b}, we adopt power-law-shaped intrinsic continua, 
using data unaffected by the silicate dust absorption features and
strong PAH emission features.
The derived intrinsic power law continua, after correcting for the
silicate dust absorption features using the foreground screen dust
model, are shown in Figure 7. 
The ratio of the infrared flux used for the infrared radiative pumping
among HCN (14 $\mu$m), HCO$^{+}$ (12 $\mu$m), and HNC (21.5 $\mu$m) is 
\begin{eqnarray}
HCN:HCO^{+}:HNC (F_{\nu (IR)}) &=& 1:0.8:2.5,
\end{eqnarray}
where we compare the power-law continuum fluxes at the redshifted
wavelengths with $z=$ 0.043.
The infrared spectral energy distribution of the hidden energy source of 
IRAS 20551$-$4250 (in F$_{\nu}$) shows higher fluxes with increasing
wavelength at 5--35 $\mu$m.
Strictly speaking, the continuum emission, outside the strong silicate
dust absorption features, could be flux-attenuated by foreground dust
absorption, and its shape could change slightly.
However, as we will show below, this change is insignificant.
The optical depth of the 9.7 $\mu$m silicate dust absorption feature,
relative to the adopted power law continuum level in Figure 7, is 
estimated to be $\tau_{9.7}$ $\sim$ 2.7.
This corresponds to A$_{\rm V}$ = 25--50 mag, if the relation of 
$\tau_{9.7}$/A$_{\rm V}$ = 0.054--0.11, found in the Galactic
interstellar medium \citep{roc84,roc85}, is adopted.
The dust extinction at 5--20 $\mu$m, outside the 
strong silicate dust absorption features (A$_{\rm MIR}$), is 
A$_{\rm MIR}$ = 0.027 $\times$ A$_{\rm V}$ \citep{nis08,nis09}, or 
A$_{\rm MIR}$ = 0.7--1.4 mag in the case of IRAS 20551$-$4250. 
Since the dust extinction curve at 5--20 $\mu$m is relatively flat
\citep{fri11}, the possible change in the intrinsic continuum spectral
shape at 5--20 $\mu$m outside the silicate dust absorption features,
through dust extinction, is too small to significantly affect our
discussion. Thus, we adopt the ratio of Equation (8). 

Adopting the above values, the ratio of the infrared radiative pumping
rate among HCN, HCO$^{+}$, and HNC is  

\begin{eqnarray}
HCN:HCO^{+}:HNC (P_{IR})(predicted) & = & 1:0.9:27. 
\end{eqnarray}

Namely, the infrared radiative pumping rate of HNC is much higher
than HCN and HCO$^{+}$, mainly due to the longer wavelength of the
infrared v$_{2}$=1 absorption line.
The Einstein A coefficient for the rotational J=3--2 transition at
v$_{2}$=1f (A$_{\rm J=3-2,rot,v2=1f}$), relative to that for the
vibrational transition from v$_{2}$=1 to v=0 (A$_{\rm v2=1-0,vib}$), is  
$\sim$4 $\times$ 10$^{-4}$, $\sim$4 $\times$ 10$^{-4}$, and 
$\sim$2 $\times$ 10$^{-4}$ for HCN, HCO$^{+}$, and HNC, respectively
(Table 5 and $\S$4.3,1, third paragraph).
Thus, most of the vibrationally excited molecules to v$_{2}$=1 quickly 
decay back to v=0, rather than decaying to a lower rotational-J
level at v$_{2}$=1.
However, the A$_{\rm J=3-2,rot,v2=1f}$ to A$_{\rm v2=1-0,vib}$ ratios 
agree within a factor of $\sim$2 among HCN, HCO$^{+}$, and HNC. 
Once vibrationally excited to J=3 at v$_{2}$=1f, a similar fixed
fraction of rotational J-transition from J=3 to J=2 at v$_{2}$=1f is
expected, which should be detectable as the v$_{2}$=1f J=3--2 emission line.
Thus, the detection of the HNC v$_{2}$=1f J=3--2 emission line in a
galaxy with detectable HCN v$_{2}$=1f J=3--2 and J=4--3 emission lines 
is quite reasonable, under our current assumption that HCN and HNC
have comparable column density at v=0.
On the other hand, the radiative pumping rates of HCN and
HCO$^{+}$ are comparable. 
The detection of the HCN v$_{2}$=1f J=3--2 emission line, without the
detection of the HCO$^{+}$ v$_{2}$=1f J=3--2 emission line, requires
some consideration.  

When the v$_{2}$=1 to v=0 column density ratio
(N$_{v_{2}=1}$/N$_{v=0}$), controlled by the infrared radiative pumping, 
is $<<$1 (see $\S$4.5), we obtain the relation 
\begin{eqnarray}
N_{v=0} \times B_{v2=0-1,vib} \times F_{\nu (IR)} = N_{v_{2}=1} \times A_{v2=1-0,vib}. 
\end{eqnarray}
Here, transitions among rotational J-levels within v=0 and v$_{2}$=1 are
neglected.
From this equation, we obtain 
\begin{eqnarray}
\frac{N_{v_{2}=1}}{N_{v=0}} &=& \frac{B_{v2=0-1,vib} \times
 F_{\nu (IR)}}{A_{v2=1-0,vib}}\\
& \propto & \lambda^{3} \times F_{\nu (IR)}, 
\end{eqnarray}
where the relation in Equation (6) is used.
Thus, the v$_{2}$=1 to v=0 column density ratios among HCN, HCO$^{+}$,
and HNC are 
\begin{eqnarray}
HCN:HCO^{+}:HNC (N_{v_{2}=1}}/{N_{v=0})(predicted) & = & 1:0.5:9. 
\end{eqnarray}
We note that \citet{ziu86} adopt a higher A$_{v2=1-0,vib}$ value of 3.7
s$^{-1}$ for HCN than our assumption (1.7 s$^{-1}$).
However, this predicted column density ratio is independent of the
absolute A$_{v2=1-0,vib}$ values of HCN, HCO$^{+}$, HNC, because of the
division of A$_{v2=1-0,vib}$ by B$_{v2=0-1,vib}$, as shown in Equations
(11) and (12). 
Thus, the predicted ratio is robust to the possible uncertainty of the
absolute A$_{v2=1-0,vib}$ values of HCN, HCO$^{+}$, and HNC. 

The observed J=4--3 to J=3--2 flux ratios in $\S$4.1 suggest that
HCN, HCO$^{+}$, and HNC are almost thermally excited and so are
similarly populated at J=3 and J=4 at v=0.   
Assuming that the fraction of J=3 level, relative to all
J-levels at v=0, is similar among HCN, HCO$^{+}$, and HNC,
Equations (3) (observation) and (13) (prediction) can be compared 
if the column density at v=0 (and thereby abundance, which is
proportional to the column density) is the same among HCN, HCO$^{+}$,
and HNC. 
The discrepancy between observation (Eqn. (3)) and prediction
(Eqn. (13)) can be reconciled if the column density at v=0
(N$_{v=0}$) for HCN is higher than HCO$^{+}$ and HNC by factors of 
$>$2.5 and 13, respectively, where HCN/HCO$^{+}$/HNC v$_{2}$=1f 
J-transition emission is assumed to be optically thin.
Namely, the HCN-to-HCO$^{+}$ abundance ratio of $>$2.5 and HCN-to-HNC
abundance ratio of 13 are suggested for the IRAS 20551$-$4250 nucleus.  
The HCN-to-HNC abundance ratio could be even higher if the HNC v$_{2}$=1f
J=3--2 emission line flux is contaminated by HC$_{3}$N v$_{7}=1f$
J=30--29 line (Table 3). 
In summary, we obtain the abundance ratio of  
\begin{eqnarray}
HCN:HCO^{+}:HNC (abundance) & = & 1:<0.4:<0.08.
\end{eqnarray}
The higher abundance (and thereby higher column density at v=0) of HCN
than HCO$^{+}$ and HNC, derived in Equation (14), can naturally
explain the following observational results that 
(a) HCN v$_{2}$=1f J=3--2 emission line is detected, without HCO$^{+}$
v$_{2}$=1f J=3--2 emission line detection, despite the similar infrared
radiative pumping rate to the v$_{2}$=1 level (Eqn. (9)),  
and (b) the v$_{2}$=1f J=3--2 emission line fluxes are similar between
HCN and HNC (Table 5), despite the much higher predicted infrared
radiative pumping rate to v$_{2}$=1 in HNC than HCN (Eqn. (9)).
Given that the bulk of the mid-infrared photons used for the
vibrational excitation of HCN/HCO$^{+}$/HNC is interpreted to
originate from AGN-heated hot dust thermal radiation, 
the properties of molecular gas derived from the v$_{2}$=1f
emission lines (i.e., higher abundance of HCN than HCO$^{+}$ and HNC)
should reflect those of AGN-affected molecular gas in the close vicinity
of an AGN.

In our Milky Way Galaxy, the HCN-to-HCO$^{+}$ abundance ratio is 
estimated to be 1.2$\pm$0.4 in infrared dark clouds, based on molecular
emission line studies \citep{liu13} and 
1.9$\pm$0.9 in diffuse interstellar medium, based on molecular 
absorption line studies \citep{god10}.  
However, it has been suggested from observations and 
their interpretations that the HCN-to-HCO$^{+}$ abundance ratio is
high in molecular gas in the close vicinity of AGNs \citep{yam07,izu16}. 
Thus, the derived HCN-to-HCO$^{+}$ abundance ratio with $>$2.5 at the 
IRAS 20551$-$4250 nucleus would be possible.

With respect to HCN and HNC, in our Galaxy, the abundance of HCN and HNC
are comparable in cold dark clouds \citep{liu13,hir98,gra14}.  
However, in high temperature clouds around active regions, 
the HNC abundance decreases by more than an order of magnitude, relative 
to the HCN abundance \citep{sch92,hir98,gra14}.
Given that the bulk of HCN and HNC J=3--2 emission at v$_{2}$=1 and v=0 
detected in the ALMA data for IRAS 20551$-$4250 come from nuclear active
regions ($\S$3), the increase in the HCN-to-HNC abundance ratio by more
than an order of magnitude could occur.   

Summarizing, our results suggest that in the nuclear region of IRAS
20551$-$4250, the HCN-to-HCO$^{+}$ and HCN-to-HNC abundance ratios are
$>$2.5 and $>$10, respectively, which is quite possible based on
previously obtained observational data from our own Galaxy and other
galaxies.  

The HCN v$_{2}$=1f J=3--2 or J=4--3 emission lines were detected in
some Galactic objects and star-forming regions, despite much lower
luminosity than IRAS 20551$-$4250 \citep{ziu86,vea13,mil13,nag15}.
However, the relative strengths of HCO$^{+}$ and HNC v$_{2}$=1f J=3--2 
or J=4--3 emission lines are not known, except for 
the luminous active star-forming region W49A, in which the
v$_{2}$=1f J=4--3 emission line was detected for HCN and HNC, but not
for HCO$^{+}$ with no flux upper limit provided \citep{nag15}.
Infrared 5--14.5 $\mu$m spectra were taken at some offset regions in
the vicinity of the W49A center \citep{sto14}, but are not available at
the W49A center, where the v$_{2}$=1f J=4--3 emission lines of HCN and
HNC are detected \citep{nag15}.
Thus, the same analysis to derive molecular abundance ratios as
made for IRAS 20551$-$4250 in the earlier part of this section is
currently not applicable to the W49A center and other Galactic sources
with detectable HCN v$_{2}$=1f emission lines. 
It might be possible that some specific Galactic stellar 
populations show an enhanced HCN abundance in their vicinity, despite
the fact that a modest HCN-to-HCO$^{+}$ abundance ratio (=1.9$\pm$0.9)
is derived, based on absorption studies of Galactic sources
\citep{god10}, which probe the ratio in a larger Galactic physical
scale, in front of the background continuum emitting sources, than
emission-based abundance measurements.

There is an observational trend that the observed HCN-to-HCO$^{+}$ flux
ratios in AGNs are higher than starburst galaxies ($\S$1).
The observed HCN-to-HCO$^{+}$ J=4--3 flux ratio at the W49A center is
also low, only $\sim$0.4 \citep{nag15}. 
High observed HCN-to-HCO$^{+}$ flux ratios at J=4--3 or J=3--2 are
produced not only by high HCN-to-HCO$^{+}$ abundance ratios, but also by
sufficient excitation to J=4 and 3, particularly for HCN, due to its
higher critical density than HCO$^{+}$ ($\S$4.1).
Since excitation to J=4 or 3 is more efficiently achieved in an AGN than
star-formation, because of AGN's much higher radiative energy generation
efficiency ($\S$1), buried AGN selection based on high HCN-to-HCO$^{+}$
flux ratios remains effective.  
Since an AGN is expected to show a higher mid-infrared 14 $\mu$m to
infrared (8--1000 $\mu$m) luminosity ratio than star-formation,
due to AGN-heated hot dust ($\S$1), the HCN v$_{2}$=1f J=4--3 to
infrared luminosity ratio in IRAS 20551$-$4250 is expected to be even
higher than that in W49A, if IRAS 20551$-$4250 indeed contains a
luminous buried AGN.  
For W49A, adopting the infrared luminosity of $>$10$^{7}$L$_{\odot}$ and
distance of 11 kpc \citep{nag15}, the HCN v$_{2}$=1f J=4--3 to
infrared luminosity ratio is $<$1.2$\times$10$^{-9}$. 
The ratio for IRAS 20551$-$4250 is estimated to be 4.9
$\times$10$^{-9}$, based on the infrared luminosity of L$_{\rm IR}$ =
10$^{12.0}$L$_{\odot}$ ($\S$1) and observed HCN v$_{2}$=1f J=4--3
luminosity (Table 4). 
This ratio is more than a factor of four higher than W49A, supporting
the strong AGN contributions to molecular gas emission at the IRAS 
20551$-$4250 nucleus.   
%{\bf We consider that the high HCN-to-HCO$^{+}$ and HCN-to-HNC abundance 
%ratios derived in molecular gas at the IRAS 20551$-$4250 nucleus are
%likely to be largely affected by the luminous buried AGN, even though
%active stars at the nuclear region may make some possible additional
%contributions.} 

\subsubsection{Flux-attenuation of HCN/HCO$^{+}$/HNC v=0 J=3--2 Emission}

In Equation (4), because the Einstein A
coefficient for spontaneous emission from J=3 to J=2 
is almost identical between v=0 and v$_{2}$=1f (Table 5), 
the ratio is roughly converted into the column density
ratio at v$_{2}$=1f J=3 and v=0 J=3 (see Eqn. (2)), provided that
the opacity of v=0 J=3--2 emission is insignificant.
In other words, the ratios in Equations (4) and (13) can be used to
estimate the opacity of v=0 J=3--2 emission for HCN, HCO$^{+}$, and
HNC.
From this comparison, we derive that the flux attenuation
of HCN v=0 J=3--2 emission is a factor of $>$2 larger than that of
HCO$^{+}$ v=0 J=3--2 emission, and is a factor of 6 larger than that of
HNC v=0 J=3--2 emission. 

The observed HCN J=3--2 to H$^{13}$CN J=3--2 flux ratio is $\sim$16
(Table 3). 
The $^{12}$C/$^{13}$C abundance ratios in starburst galaxies are 
estimated to be $\sim$50 \citep{hen93a,hen93b,hen14} or possibly even 
higher \citep{mar10}. 
In ULIRGs, the $^{12}$C/$^{13}$C abundance ratios are as high as
$\sim$100 \citep{hen14}.  
Assuming optically thin emission for H$^{13}$CN J=3--2, HCN 
J=3--2 emission is estimated to be optically thick, and its
flux can be attenuated by a factor of 3--6.
In this case, the flux attenuation of HNC v=0 J=3--2 emission
becomes 0.5 (if the flux attenuation of HCN v=0 J=3--2 emission is 3) 
or 1 (if that of HCN v=0 J=3--2 emission is 6). 
Since the flux attenuation factor by line opacity is by definition
always larger than unity, we adopt the flux attenuation of HCN J=3--2
emission to be 6, which provides the flux attenuation of HCO$^{+}$ v=0
J=3--2 emission to be $<$3.

According to the widely accepted model of molecular gas in galaxies, 
molecular clouds consist of clumpy structures, rather than spatially
smooth distribution, because observed molecular gas properties are
better explained by the former clumpy model \citep{sol79,sol87}.
Each clump can be optically thick, but has random motion inside
molecular clouds, so that emission from clumps at the other side of
molecular clouds is not significantly flux-attenuated by foreground
molecular gas clumps \citep{sol87}. 
The physical properties of individual clumps are assumed to be the
same throughout a molecular cloud and among different molecular clouds.
In this model, profiles of molecular lines with even different line
opacity are expected to be similar. 
Figure 8 shows similar v=0 J=3--2 emission line profiles of HCN,
HCO$^{+}$, and HNC, supporting the above clumpy molecular gas model 
for IRAS 20551$-$4250. 

Adopting this clumpy molecular gas model, line opacity comes mostly 
from the individual line emitting clumps, rather than foreground
molecular gas.
The flux attenuation factor is expressed by 
\begin{eqnarray}
Flux-attenuation & = & \frac{\tau}{1 - e^{-\tau}},
\end{eqnarray}
where $\tau$ is an optical depth, which is proportional to column
density. 
The flux attenuation of 6 for HCN v=0 J=3--2 emission corresponds to
$\tau$ = 6. 
In the same way, we obtain $\tau$ $<$ 2.8 for HCO$^{+}$ v=0 J=3--2
emission.  
For HNC J=3--2, the flux attenuation is estimated to be not
substantially larger than unity. 
We adopt $\tau$ $<$ 1 for HNC v=0 J=3--2 emission, in which case the
flux attenuation is $<$1.6. 
Consequently, we obtain the following optical depth ratio among
HCN, HCO$^{+}$, and HNC J=3--2 at v=0,  
\begin{eqnarray}
HCN:HCO^{+}:HNC (\tau) & = & 1:<0.5:<0.2.
\end{eqnarray}
If the optical depth in thermally excited gas is small, the J=4--3
to J=3--2 flux ratio can be higher than that of thermally excited
optically thick gas.
The possibly higher J=4--3 to J=3--2 flux ratio for HNC (1.8$\pm$0.2)
than HCN (1.6$\pm$0.2) ($\S$4.1) might be related to the small optical
depth of HNC. 

The optical depth ratio among v=0 J=3--2 emission lines of HCN,
HCO$^{+}$, and HNC can be roughly estimated independently from the
abundance ratio in Equation (14).
Considering the transitions between J=3 and J=2 (v=0), the optical depth
value is expressed as   
\begin{eqnarray}
\tau & \propto & \nu \times N_2 \times B_{J=2-3,rot} \times 
(1-g_{2}n_{3}/g_{3}n_{2}) \\
& \propto &  \nu^{-2} \times N_2 \times A_{J=3-2,rot} \times 
(1-g_{2}n_{3}/g_{3}n_{2})
\end{eqnarray}
where N$_{2}$ is the column density at J=2 (v=0), 
n$_{J}$ is the number density at J=2 or J=3 (v=0), 
B$_{\rm J=2-3,rot}$ is the Einstein B coefficient for
rotational transition from J=2 to J=3 (v=0), A$_{\rm J=3-2,rot}$ is
the Einstein A coefficient for spontaneous emission for rotational
transition from J=3 to J=2 (v=0), $\nu$ is frequency, and 
g$_{J}$ is the statistical weight that is related to the J-level in the
form of 2J+1 \citep{ryb79}.  
In IRAS 20551$-$4250, since HCN, HCO$^{+}$, and HNC are found to be
almost thermally excited at J=3 and J=4, 
the g$_{2}$n$_{3}$/g$_{3}$n$_{2}$ term is expressed with 
exp($-$E/k$_{\rm B}$T$_{\rm kin}$), where E is the energy level of the 
J-transition of interest, T$_{\rm kin}$ is the kinetic temperature of
molecular gas, and  k$_{\rm B}$ is the Boltzmann constant.  
For HCN, HCO$^{+}$, and HNC, since energy levels and frequencies are 
almost the same for the same rotational J-transitions at v=0, the term 
of $\nu^{-2}$ $\times$ (1$-$g$_{2}$n$_{3}$/g$_{3}$n$_{2}$) is also
nearly identical.
Adopting the A$_{J=3-2,rot}$ values shown in Table 5, we obtain the 
following optical depth ratio, from the derived abundance
ratio in Equation (14),    
\begin{eqnarray}
HCN:HCO^{+}:HNC (\tau_2) & = & 1:<0.7:<0.1.
\end{eqnarray}
The two independently derived optical depth ratios in Equations
(16) and (19) show consistent results to each other.
The ratios of various parameters among HCN, HCO$^{+}$, and HNC used for
our discussion are summarized in Table 7.

We now correct for the flux attenuation by line opacity, to derive the
intrinsic HCN-to-HCO$^{+}$ and HCN-to-HNC flux ratios from the observed
ratios in Figure 6. 
For J=3--2, we adopt the flux attenuation with factors of 6, $<$3, and
$\sim$1 for HCN, HCO$^{+}$, and HNC, respectively ($\S$4.3.2, second
paragraph). 
For J=4--3, the observed HCN-to-HCO$^{+}$ and HCN-to-HNC flux ratios are 
almost identical to those at J=3--2.
Since excitation condition does not significantly change the flux ratios
between J=3--2 and J=4--3 in IRAS 20551$-$4250 ($\S$4.1) and since the
abundance effect should be the same between J=3--2 and J=4--3, the
observed flux ratios at J=4--3 can be naturally interpreted with the
similar intrinsic flux ratios and flux attenuation to J=3--2.   
We thus apply the same flux attenuation correction for J=4--3 as
employed above for J=3--2.  

The intrinsic molecular line flux ratio of IRAS 20551$-$4250 at J=3--2
and J=4--3, after the correction for flux attenuation by line opacity, 
are plotted as open symbols in Figure 6, which are further deviating from 
the distribution of starburst-dominated regions.
Other data points could also move if opacity-corrected intrinsic 
flux ratios are used.  
If HCN flux enhancement is indeed a good signature of
AGNs, and if this is due in part to HCN abundance enhancement, then 
the line opacity of HCN emission and thereby the opacity correction
factor of HCN flux, relative to HCO$^{+}$ and HNC, could be higher in
AGNs than in starburst galaxies. This will move AGNs to the right and
upper direction, even deviating from the distribution of starburst
galaxies. This could make the HCN-flux-based separation between AGNs and 
starbursts more solid.
Furthermore, if higher HCN opacity is the case in AGNs, while some AGNs
show enhanced observed HCN fluxes, other AGNs may not.
This could explain the result that some fraction of AGNs do not
necessarily show observed HCN flux enhancements \citep{pri15}, 
although it is possible that the result is partly due to the dilution of
the signatures of AGN-affected molecular gas emission, by the large
contamination from spatially-extended starburst-affected molecular gas
emission in their large-aperture single-dish telescope data.
Correction for flux attenuation through isotopologue observations 
(e.g., H$^{13}$CN, H$^{13}$CO$^{+}$, HN$^{13}$C, HC$^{15}$N, H$^{15}$NC)
is necessary not only for IRAS 20551$-$4250, but 
also for other galaxies, to derive intrinsic flux ratios in Figure 6.

Finally, in Figure 6, IRAS 22491$-$1808 shows no clear
infrared-identified buried AGN signatures, and yet shows a high
HCN-to-HCO$^{+}$ flux ratio at J=4--3 \citep{ima13b}. 
For IRAS 22491$-$1808, the HCN v$_{2}$=1f J=3--2 emission line has been
detected in our ALMA Cycle 2 data \citep{ima16}, suggesting
the presence of a strongly 14 $\mu$m emitting energy source, which is
naturally explained by a luminous buried AGN, due to AGN-heated hot dust
($\S$1 and $\S$4.3.1).   
IRAS 22491$-$1808 may contain an extremely deeply buried AGN whose
signatures are not detectable in infrared energy diagnostic methods, but
can be detected in (sub)millimeter observations for the first time, due 
to reduced effects of dust extinction. 
The presence of similar sources, infrared-classified-starbursts with
elevated HCN-to-HCO$^{+}$ flux ratios, is found also at J=1--0 
\citep{cos11,pri15}.
ALMA (sub)millimeter observations of these sources are interesting to
scrutinize the presence of infrared-elusive, but
(sub)millimeter-detectable, extremely deeply buried AGNs.

\subsection{Possible caveats}

We here comment a few caveats for our discussion above. 
First, transitions among rotational J-levels within v=0 and
v$_{2}$=1 are not treated in our calculations (Eqns. (10)--(12)), because 
the number of currently available molecular J-transition data is not
large enough to do this in a reliable manner. 
This could be the largest uncertainty.
However, the similar excitation up to J=4 at v=0 among
HCN, HCO$^{+}$, and HNC, estimated from their similar observed J=4--3 to
J=3--2 flux ratios ($\S$4.1), suggests that the effects of this
uncertainty for our discussion at J=3 and J=4 are expected to be
relatively limited, unless excitation at higher-J is considerably
different among HCN, HCO$^{+}$, and HNC.
When more J-transition data become available, this point will be tested
more.

Second, we assumed that HCN, HCO$^{+}$, and HNC emission both at
v=0 and v$_{2}$=1 come from the same region ($\S$4.3.1). This is clearly
too simplistic. 
Even though the dipole moments are comparable among HCN, HCO$^{+}$,
and HNC ($\S$4.3.1), their critical densities are not perfectly the
same.
HCO$^{+}$ has a lower critical density than HCN and HNC by a
factor of $\sim$5 for the same J-transition at v=0 \citep{mei07,gre09},
so that the spatial extent of collisionally-excited HCO$^{+}$ v=0 J=3--2 
emission line can be larger than those of HCN/HNC v=0 J=3--2 emission
lines, if molecular gas has a gradually decreasing radial
density distribution from the very center to the outer region in
a galaxy \citep{big12}, due to the decrease of typical density of
individual molecular gas clumps.
In fact, it is sometimes seen in a galaxy nuclear scale that HCO$^{+}$
emission is spatially more extended than HCN emission at the same
J-transition \citep{ima07a,sai15}. 
This is also the case for IRAS 20551$-$4250, because HCO$^{+}$ v=0
J=3--2 emission is estimated to be slightly more extended spatially than
HCN/HNC v=0 J=3--2 emission ($\S$3).
The size of the continuum emitting region around a buried AGN is also 
different at 14 $\mu$m (used for the vibrational excitation to v$_{2}$=1
for HCN), 12 $\mu$m (HCO$^{+}$), and 21.5 $\mu$m (HNC). 
These different geometries are shown as schematic diagrams in Figure 9.
For the v$_{2}$=1f J=3--2 emission, in the case of constant radial
density distribution of molecular gas, since the number of infrared
photons per molecule is larger in the close vicinity of an AGN than the 
outer part, the v$_{2}$=1f to v=0 flux ratios of HCN/HCO$^{+}$/HNC 
should be higher at the inner part, and the bulk of the v$_{2}$=1f
J=3--2 emission is likely to come from the inner region in Figure
9. Emission-weighted size of the v$_{2}$=1f J=3--2 line can become smaller
than that of the v=0 J=3--2 line for HCN/HCO$^{+}$/HNC. 
The difference of v$_{2}$=1f to v=0 flux ratios between the inner and
outer parts, and the emission-weighted size difference between
v$_{2}$=1f and v=0, are reduced, if molecular gas has a decreasing
radial density distribution from the galaxy center to the outer part.
We need to consider how these differences in geometry could alter our
results obtained based on the assumption that HCN/HCO$^{+}$/HNC v=0 and
v$_{2}$=1f lines come from the same volume. 

Regarding the size difference between v$_{2}$=1f J=3--2 and v=0 J=3--2
emission, the total fluxes of HCN/HCO$^{+}$/HNC 
v$_{2}$=1f J=3--2 emission lines from molecular gas in Figure 9 are
primarily determined by the number of infrared photons which are
absorbed and used for the vibrational excitation to v$_{2}$=1.   
As long as (a) the covering solid angle around the
central infrared continuum emitting energy source does not differ
a lot among HCN, HCO$^{+}$, and HNC, and (b) the bulk of the
v$_{2}$=1f J=3--2 emission is covered within our ALMA beam, then our
discussion in $\S$4.3 is essentially unchanged. 
Both of the conditions (a) and (b) are likely to be met.  

Regarding the size difference of the infrared continuum emitting
regions among HCN/HCO$^{+}$/HNC, the most important factor for our
infrared radiative pumping calculations of v$_{2}$=1f fluxes in
$\S$4.3.1 is again the number of infrared photons used for the
vibrational excitation to v$_{2}$=1.
The slight difference of the spatial extent of continuum emitting
regions at different infrared wavelengths will not alter our
results significantly, as long as (a) the infrared continuum emitting 
regions are more compact than the surrounding molecular gas, and (b) the 
covering fraction around the central energy source is similar among 
HCN, HCO$^{+}$, and HNC. Both of the conditions (a) and (b) are expected
to be fulfilled.  

Regarding the size difference of v=0 J=3--2 emission among HCN,
HCO$^{+}$, and HNC, the HCO$^{+}$ J=3--2 emission is estimated to be
spatially more extended than HCN/HNC J=3--2 emission inside the ALMA
beam ($\S$3).
In this case, the source solid angle ($\Omega$) in Equation (1) becomes
larger for HCO$^{+}$ J=3--2 than HCN/HNC J=3--2, and the
HCO$^{+}$ J=3--2 (v=0) column density is smaller than our estimate. 
This affects our discussion of column density and thereby abundance.
In fact, if both HCN J=3--2 and HCO$^{+}$ J=3--2 emission are
thermalized (excitation temperature is comparable to the kinetic
temperature of molecular gas) ($\S$4.1) and if the optical depth of HCN
J=3--2 line is larger than that of HCO$^{+}$ J=3--2 line ($\S$4.3.2),
then the HCO$^{+}$ J=3--2 flux cannot be larger than the HCN J=3--2
flux, as long as both lines are emitted from the same volume, because
their frequencies are almost identical.  
The HCO$^{+}$ v=0 J=3--2 to HCN v=0 J=3--2 flux ratio of 1.4 (Table 5)
can be explained, if the HCO$^{+}$ v=0 J=3--2 line emitting region is
$>$1.4 times larger in $\Omega$ than the HCN J=3--2 line emitting region
within our ALMA beam. 
In our ALMA data, the spatial extent of HCO$^{+}$ v=0 J=3--2
emission is estimated to be larger than those of HCN/HNC v=0 J=3--2
emission by a factor of $\sim$1.5 in $\Omega$ ($\S$3). 
In this case, the actual HCO$^{+}$ column density (abundance) relative
to HCN will decrease by $\sim$50\%.   
This will even strengthen our result that HCN-to-HCO$^{+}$ abundance ratio
is high in the IRAS 20551$-$4250 nucleus.
It is also possible that due to the lower critical density of
HCO$^{+}$ v=0 J=3--2 than HCN/HNC v=0 J=3--2, the size of individual
molecular gas clumps is larger for HCO$^{+}$ v=0 J=3--2 than HCN/HNC v=0
J=3--2, if each molecular clump has a decreasing radial density
distribution \citep{gie92,ima07a}. 
In this case, HCO$^{+}$ v=0 J=3--2 emission could occupy a larger volume
fraction inside a molecular cloud than HCN/HNC v=0 J=3--2 emission.  
This will also work to decrease the HCO$^{+}$ column density (abundance) 
relative to HCN and HNC, which again reinforces our main result of 
enhanced HCN-to-HCO$^{+}$ abundance ratio in the nucleus of IRAS
20551$-$4250.  

Third, we assumed that the observed mid-infrared continuum is 
dominated by AGN-heated hot dust emission ($\S$4.3.1).
Even though this is the case for IRAS 20551$-$4250, there may be
non-negligible contribution from starburst emission to the observed
mid-infrared flux. Since starburst-heated dust usually has cooler
temperature than AGN-heated dust, the starburst contribution to the
observed mid-infrared flux becomes relatively higher at a longer
wavelength \citep{far02,far03,pro04,veg08,rui10}. 
The relative contribution from starbursts can be larger at 21.5 $\mu$m
(for HNC) than 12 $\mu$m (HCO$^{+}$) and 14 $\mu$m (HCN).
If certain fraction of infrared 21.5 $\mu$m flux comes from starbursts,
rather than the buried AGN, the AGN-origin HNC term in Equation (13)
will decrease.
Even if this slightly modifies the HNC term in Equation (13), our main
result of very high HCN-to-HNC abundance ratio ($\S$4.3.1) will not change. 

Summarizing, even though neglecting the transitions among
rotational J-levels within v=0 and v$_{2}$=1 needs to be tested in
future data, our main result is robust to possible ambiguity of the
assumed geometry of HCN, HCO$^{+}$, and HNC emission regions around
energy sources.

\subsection{Role of infrared radiative pumping}

The excitation temperature (T$_{\rm ex}$) is related to the column
density at level $u$ (N$_{u}$), in the form of   
\begin{equation}
N_u = \frac{N_{\rm{mol}}}{Q(T)} g_u \exp \Biggl(-\frac{E_u}{k_{\rm B} T_{\rm{ex}}}\Biggr),
\end{equation}
where $N_{\rm{mol}}$ is the total column density of a given
molecule, $Q(T)$ is a partition function, $E_u$ is an energy at
level $u$ above the ground state, k$_{\rm B}$ is the Boltzmann constant, 
and g$_{\rm u}$ is the statistical weight. 
The excitation temperatures (T$_{\rm ex}$) in IRAS 20551$-$4250, based on
the observed fluxes in this paper and \citet{ima13b}, are shown in
Table 8. 
If v$_{2}$=1 emission is optically thin and v=0 emission is optically
thick, the intrinsic vibrational excitation temperature 
(T$_{\rm ex-vib}$) after correction for line opacity will be smaller 
than that based on the observed flux ratio between v$_{2}$=1 and v=0.
For HCN, the T$_{\rm ex-vib}$ values corrected for the flux attenuation
with the factor of 6 for HCN v=0 J=3--2 emission ($\S$4.3.2) are also
shown in Table 8. 
The same opacity correction is applied for HCN v=0 J=4--3 emission 
(see $\S$4.3.2).

In Table 6, we also show the HCN v$_{2}$=1f to v=0 flux ratios at 
J=4--3 and/or J=3--2 in NGC 4418 and Mrk 231, for which flux attenuation
of v=0 emission is quantitatively estimated \citep{sak10,aal15a}.
The observed ratios are similar between IRAS 20551$-$4250 and Mrk 231. 
For NGC 4418, the observed HCN v$_{2}$=1f to v=0 flux ratios at J=3--2
and J=4--3 are higher by a factor of 4--5 than IRAS 20551$-$4250 and 
Mrk 231.
However, \citet{sak10} estimated a factor of $\sim$20 flux attenuation
of HCN v=0 J=4--3 emission in NGC 4418. 
If we apply this correction, the intrinsic HCN v$_{2}$=1f to v=0 flux
ratios for NGC 4418 become $\sim$0.01 at J=3--2 and J=4--3.
For HCN emission from IRAS 20551$-$4250, assuming flux
attenuation by a factor of 6 (see $\S$4.3.2), 
the intrinsic HCN v$_{2}$=1f to v=0 flux ratios at J=3--2 and J=4--3
become $\sim$0.01.  
For Mrk 231, if we adopt a factor of $\sim$10 flux attenuation of HCN v=0
J=3--2 emission \citep{aal15a}, the intrinsic HCN v$_{2}$=1f to v=0 flux
ratio at J=3--2 becomes $\sim$0.005. 
Given the almost identical Einstein A coefficient at v$_{2}$=1f and
v=0 for the same J transition lines (Table 5), we can derive that the
population ($\propto$flux/A) at v$_{2}$=1 (J=3,4) is approximately
0.5--1\% of that at v=0 (J=3,4) in IRAS 20551$-$4250, NGC 4418, and Mrk
231. 
The observed HCN v$_{2}$=1f to v=0 flux ratios at J=4--3 or 
J=3--2 shown by \citet{aal15b} range from $\sim$0.1 (Zw 049.057 and IRAS
17208$-$0014) to $\sim$0.2 (IC 860 and Arp 220W).
The information of flux attenuation for v=0 emission is needed to 
derive the intrinsic ratio.
These four sources are not included in our quantitative discussion 
about intrinsic flux ratios, but it is quite possible that these
sources have similarly low intrinsic values to the above three LIRGs,
because large line opacities for the v=0 emission are argued
\citep{aal15b,mar16}. 

Although the HCN population at v$_{2}$=1 is only $\sim$1\% of that at v=0
for the same J-level, the Einstein A coefficient from 
v$_{2}$=1f J=4 to v=0 J=3 (1.7--3.7 s$^{-1}$)
\citep{deg86,ziu86,aal07a} is nearly three orders of magnitude larger
than that from v=0 J=4 to v=0 J=3 (0.002 s$^{-1}$; Table 5).
The photon number of the transition from v$_{2}$=1f J=4 to v=0 J=3 is
estimated to be an order of magnitude higher than that from v=0 J=4 to
v=0 J=3.  
Therefore, we can conclude that infrared radiative pumping 
affects rotational excitation at v=0, and thus could alter the v=0 
J-transition flux ratios compared to collisional excitation alone  
as the excitation mechanism. 

The condition where infrared radiative pumping can have a significant 
influence on the HCN rotational excitation at v=0 is
defined as 
\begin{equation}
T_{ex-vib} > \frac{T_{0}}{ln\frac{A_{v2=1-0,vib}}{A_{rot}}},
\end{equation} 
where T$_{ex-vib}$ is the v$_{2}$=1 vibrational excitation temperature, 
T$_{0}$ is the energy level of HCN v$_{2}$=1f J=4 or J=3, 
A$_{v2=1-0,vib}$ is the Einstein A coefficient from v$_{2}$=1 to v=0, 
and A$_{rot}$ is the Einstein A coefficient between J-levels at v=0 
\citep{car81,sak10,mil13}. 
For J=3--2, adopting T$_{0}$=1050.0 [K] (Table 5), A$_{v2=1-0,vib}$
= 1.7 s$^{-1}$ ($\S$4.3.1), and A$_{3-2,rot}$ = 8.4$\times$10$^{-4}$
(Table 5), the right-hand column becomes $\sim$140 [K]. 
For J=4--3, adopting T$_{0}$=1067.1 [K] (Table 5), A$_{v2=1-0,vib}$
= 1.7 s$^{-1}$ ($\S$4.3.1), and A$_{4-3,rot}$ = 20.6$\times$10$^{-4}$
(Table 5), the right-hand column becomes $\sim$160 [K]. 
The T$_{ex-vib}$ in the left-hand column is $>$275 [K], if we neglect
line opacity for HCN v=0 emission (Table 8).
Even if flux attenuation by a factor of 6 for HCN v=0 emission is 
included, the T$_{ex-vib}$ value is still $>$185 [K] (Table 8).
Thus, the condition in Equation (21) is fulfilled for HCN at J=4 and 3
in IRAS 20551$-$4250. 

For HNC J=3--2, the same logic is applicable. 
The value in the right-hand column is 80 [K] for HNC J=3--2.
However, the T$_{ex-vib}$ value in the left-hand column is 
250 [K], where the observed v$_{2}$=1f and v=0 fluxes are used, because
the flux attenuation of HNC v=0 J=3--2 emission is estimated to be
not significantly larger than unity ($\S$4.3.2). 
We can therefore conclude that infrared radiative pumping affects the
HNC rotational excitation at v=0 as well.

For HCO$^{+}$ J=3--2, the values in the left-hand and right-hand columns
are $<$270 [K] and 160 [K], respectively, where the observed 
v$_{2}$=1f and v=0 fluxes are used, because the estimated flux
attenuation for HCO$^{+}$ v=0 J=3--2 emission is $<$3 ($\S$4.3.2).  
Although the HCO$^{+}$ v$_{2}$=1f J=3--2 flux is an upper limit, if 
the actual flux is larger than 5\% of the upper limit, the left-hand
column will be higher than the right-hand column, and thus it is still
possible that infrared radiative pumping plays some role for HCO$^{+}$. 

If infrared radiative pumping is at work to excite HCN/HCO$^{+}$/HNC
molecules to the v$_{2}$=1 level, a sufficient number of infrared
photons have to be absorbed. 
Figure 10 displays the Spitzer IRS high-resolution (R $\sim$ 600) spectra
of IRAS 20551$-$4250 at $\lambda_{\rm rest}$ $\sim$ 14 $\mu$m  
(corresponding to HCN v$_{2}$=1 absorption), 12 $\mu$m (HCO$^{+}$),
and 21.5 $\mu$m (HNC) \citep{leb15}.
Although possible signatures relating to the HCN v$_{2}$=1 absorption
features may be present, the absorption features are not clear at the
expected wavelengths of HCO$^{+}$ and HNC v$_{2}$=1 absorption. 
Assuming that HCN/HCO$^{+}$/HNC v$_{2}$=1 absorption features in the
infrared spectra have the same velocity profile as the HCN/HCO$^{+}$/HNC 
v$_{2}$=1f J=3--2 emission features at 1 mm ($\sim$270 GHz), the
required condition for infrared radiative pumping to work is 

\begin{equation}
\sum_{i=P,Q,R} f_{i}S_{IR}/\lambda_{IR} > S_{rot}/\lambda_{rot} 
\end{equation}
\citep{sak10}.
Here, $S_{\rm IR}$ is the continuum flux density (in [Jy]) around the
absorption wavelength $\lambda_{\rm IR}$, $f_i$ is the fractional
absorption depth at the absorption peak, and $S_{\rm rot}$ is the peak
flux density (in [Jy]) of the v$_{2}$=1f J-transition line emission at
the wavelength $\lambda_{\rm rot}$. 
The ``1/$\lambda$'' term is included, because a unit frequency width 
for the given same velocity width, $\Delta \nu$, is inversely proportional
to $\lambda$ (= proportional to $\nu$).
In the left-hand term, infrared absorption by the P, Q, and R branches
is summed.  
For HCN absorption at $\lambda_{\rm rest}$ $\sim$ 14.0 $\mu$m, 
the absorption dip, f$_{\rm i}S_{\rm IR}$, is at most $<$0.03 [Jy] in
the infrared spectrum (Fig. 10). 
The peak flux of the HCN v$_{2}$=1f J=3--2 emission line in the
millimeter wavelength is $\sim$1.5 [mJy] (Fig. 4). 
The upper limit in the left-hand term of Equation (22) is approximately
three orders of magnitude larger than the right-hand term.  
For HCO$^{+}$ and HNC, the dips of infrared absorption features, 
f$_{\rm i}S_{\rm IR}$, are at most $<$0.01 [Jy] and $<$0.1 [Jy], 
respectively (Fig. 10).
The peak fluxes of the HCO$^{+}$ v$_{2}$=1f J=3--2 and HNC v$_{2}$=1f
J=3--2 emission lines in the millimeter wavelength are $<$1.0 [mJy] 
and $\sim$0.8 [mJy], respectively (Fig. 4). 
The upper limit and value in the left-hand term of Equation (22) is
about three orders of magnitude larger than the right-hand term also for
HCO$^{+}$ and HNC, respectively. 
Although the energy absorbed by the infrared photons is used for other
J-transition lines at v$_{2}$=1f for HCN/HCO$^{+}$/HNC, 
there is a large enough margin for infrared radiative pumping as the
source of vibrational excitation of HCN/HCO$^{+}$/HNC, unless the actual 
strengths of the infrared v$_{2}$=1 absorption features of
HCN/HCO$^{+}$/HNC are more than three orders of magnitude smaller than
the current upper limits. 
The observed velocity widths of HCN and HNC v$_{2}$=1f J=3--2 emission
lines are $\sim$200 km s$^{-1}$ in FWHM (Table 3).
To detect absorption features with this level of velocity width, a
spectral resolution with R $>$ 1500 is necessary.  
The minimum required infrared absorption features of HCN/HCO$^{+}$/HNC
are too weak to be detected with the current quality Spitzer IRS
R $\sim$ 600 spectrum.
A higher S/N and spectral resolution spectrum is therefore required to
unambiguously detect the infrared v$_{2}$=1 absorption features of
HCN/HCO$^{+}$/HNC, and to better constrain the energetics between
absorbed infrared photons and v$_{2}$=1f J-transition
emission lines more quantitatively.

In summary, it is very likely that infrared radiative pumping is at work 
in IRAS 20551$-$4250, and therefore the infrared radiative pumping
effect must be properly taken into account for HCN and HNC, and possibly
for HCO$^{+}$ too, if we are to understand rotational excitation at v=0
and J-transition flux ratio at v=0 in IRAS 20551$-$4250. 
However, based on the currently available data, it is not yet
quantitatively clear what percentages of the observed molecular line
fluxes at v=0 are altered by the infrared radiative pumping, compared to 
collisional excitation.
Molecular gas rotational J-level populations and fluxes of rotational
J-transitions at v=0 in IRAS 20551$-$4250 are determined by (a)
collisional excitation, which is dependent on molecular gas physical
parameters (density, kinetic temperature etc), and (b) contribution from 
infrared radiative pumping.   
Obtaining further molecular line data at other J-transition lines at
both v=0 and v$_{2}$=1 will help to better disentangle these two factors
and to better understand the physical origin of the observed
HCN/HCO$^{+}$/HNC molecular line flux ratios in IRAS 20551$-$4250. 

\section{Summary}

We conducted ALMA Cycle 2 observations of the AGN-hosting ULIRG,
IRAS 20551$-$4250, at HCN/HCO$^{+}$/HNC J=3--2 (rotational transition)
at both vibrational-ground (v=0) and vibrationally excited (v$_{2}$=1)
levels. 
Since our ALMA Cycle 0 observations showed a tentatively detectable
($\sim$5$\sigma$) vibrationally excited HCN v$_{2}$=1f J=4--3 emission
line from this source, our new data are useful for investigating 
(a) whether the vibrationally excited (v$_{2}$=1f) HCN emission line is
visible at J=3--2 as well, and 
(b) the strength of the vibrationally excited (v$_{2}$=1f) J=3--2 emission
lines for HCO$^{+}$ and HNC. 
The main results of these observations are as follows:

\begin{enumerate}
\item 
HCN, HCO$^{+}$, and HNC J=3--2 emission lines at vibrational-ground
level were clearly detected at the continuum peak position.

\item 
We detected HCN v$_{2}$=1f J=3--2 and HNC v$_{2}$=1f J=3--2 emission
lines, based on Gaussian fits of the spectra and integrated intensity
(moment 0) maps, supporting the scenario that IRAS 20551$-$4250 shows
detectable vibrationally excited v$_{2}$=1f J-transition emission lines
for both HCN and HNC. 

\item 
The vibrationally excited HCO$^{+}$ v$_{2}$=1f J=3--2 emission 
line was not clearly detected, but a stringent upper limit was in place. 
The observed v$_{2}$=1f J=3--2 to v=0 J=3--2 flux ratio for HCO$^{+}$ is 
more than a factor of four smaller than those for HCN and HNC.

\item 
Since collisional excitation is very difficult to significantly populate
the v$_{2}$=1 level, due to its high energy level ($>$600 K), we 
calculated how infrared radiative pumping can achieve the vibrational
excitation, using the available infrared 5--35 $\mu$m spectrum. 
It was estimated that the infrared radiative pumping rate to v$_{2}$=1
of HCN is comparable to HCO$^{+}$, but is smaller than HNC by more than
an order of magnitude. 
Our observational results about the v$_{2}$=1f J=3--2 emission line
fluxes of HCN, HCO$^{+}$, and HNC can naturally be explained, if the HCN
abundance is greater than a factor of few higher than HCO$^{+}$, and an
order of magnitude higher than HNC. 
Because of the low observed equivalent widths of the PAH emission
features at 5--25 $\mu$m, it was interpreted that infrared photons used
for the vibrational excitation of HCN/HCO$^{+}$/HNC mostly come from
AGN-heated hot dust emission near an AGN.
Thus, the properties of molecular gas derived from the v$_{2}$=1f
emission largely reflect those of AGN-affected molecular gas in the
close vicinity of an AGN.
In fact, HCN abundance enhancement in the close vicinity of AGNs was
argued in several previous works.

\item 
The observed v$_{2}$=1f J=3--2 to v=0 J=3--2 flux ratios for
HCN/HCO$^{+}$/HNC and the observed HCN v=0 J=3--2 to H$^{13}$CN v=0
J=3--2 flux ratio suggest that HCN v=0 J=3--2 emission has a significant
line opacity and is flux-attenuated by a factor of 6, whereas the flux
attenuation of HCO$^{+}$ v=0 J=3--2 and HNC v=0 J=3--2 emission lines
was estimated to be $<$3 and $\sim$1, respectively.
The higher flux attenuation by line opacity for HCN than HCO$^{+}$ and
HNC is as expected from the estimated higher HCN abundance than
HCO$^{+}$ and HNC. 

\item 
The observed flux ratio and excitation temperature between v$_{2}$=1f
and v=0 suggest that infrared radiative pumping has an effect 
to rotational excitation at v=0 for HCN and HNC. 

\item 
The HCN-to-HCO$^{+}$ J=3--2 flux ratio at v=0 is similar to the 
previously obtained HCN-to-HCO$^{+}$ J=4--3 flux ratio at v=0 in IRAS
20551$-$4250, and both are higher than starburst-dominated regions.
The enhanced HCN-to-HCO$^{+}$ flux ratio in this AGN-hosting ULIRG 
was interpreted to be originated mainly in the enhanced HCN-to-HCO$^{+}$
abundance ratio and sufficient HCN excitation, rather than the
substantially higher infrared radiative pumping efficiency of HCN than
that of HCO$^{+}$. 
When the estimated line opacity for HCN/HCO$^{+}$/HNC v=0 emission
is corrected, the HCN-to-HCO$^{+}$ J=3--2 and HCN-to-HNC J=3--2 flux
ratio at v=0 for IRAS 20551$-$4250 will increase, even deviating from
the ratios found in starburst-dominated regions.
If larger HCN line opacity due to enhanced HCN abundance is common in
AGNs, AGNs and starburst galaxies will be separated even more in the
plot of line-opacity-corrected, intrinsic HCN-to-HCO$^{+}$ and
HCN-to-HNC flux ratios, than in the plot of observed flux ratios. 
Line opacity correction of other sources is also necessary to 
refine these molecular line flux ratios for use as an even more
solid energy diagnostic method of dust/gas-rich ULIRGs. 

\item 
Various emission lines, other than the targeted HCN/HCO$^{+}$/HNC, were 
also detected, including HOC$^{+}$, SO$_{2}$, SO, H$^{13}$CN, 
HC$_{3}$N, and CH$_{3}$CCH, demonstrating the high sensitivity of ALMA and
the abundant presence of excited molecular gas by the nuclear energy
source in IRAS 20551$-$4250. 

\end{enumerate}

\acknowledgments

We are grateful to the anonymous referee for his/her useful comments.
We thank Dr. H. Nagai for his kind advice regarding ALMA data analysis,
and Dr. K. Sakamoto for useful discussions. 
M.I. was supported by JSPS KAKENHI Grant Number 23540273 and 15K05030,
and the ALMA Japan Research Grant of the NAOJ Chile Observatory,
NAOJ-ALMA-0001 and -0023.  
T. I. is thankful for the fellowship received from the Japan Society for
the Promotion of Science (JSPS).
This paper makes use of the following ALMA data:
ADS/JAO.ALMA\#2013.1.00033.S. 
ALMA is a partnership of ESO (representing its member states), NSF (USA)
and NINS (Japan), together with NRC (Canada), NSC and ASIAA (Taiwan),
and KASI (Republic of Korea), in cooperation with the Republic of Chile.  
The Joint ALMA Observatory is operated by ESO, AUI/NRAO and NAOJ.
This research has made use of NASA's Astrophysics Data System and the
NASA/IPAC Extragalactic Database (NED) which is operated by the Jet
Propulsion Laboratory, California Institute of Technology, under
contract with the National Aeronautics and Space Administration. 

%\clearpage

%\appendix

\clearpage

%%%%%%%%%% Table 1 %%%%%%%%%
\begin{deluxetable}{llcccccc}
%\tabletypesize{\small}
\tabletypesize{\scriptsize}
\tablecaption{Log of our ALMA Cycle 2 observations \label{tbl-1}}
\tablewidth{0pt}
\tablehead{
\colhead{Line} & \colhead{Date} & \colhead{Antenna} & \colhead{Baseline}
& \colhead{Integration} & \multicolumn{3}{c}{Calibrator} \\ 
\colhead{} & \colhead{(UT)} & \colhead{Number} & \colhead{(m)} &
\colhead{time (min)} & \colhead{Bandpass} & \colhead{Flux} &
\colhead{Phase}  \\ 
\colhead{(1)} & \colhead{(2)} & \colhead{(3)} & \colhead{(4)} &
\colhead{(5)} & \colhead{(6)} & \colhead{(7)} & \colhead{(8)} 
}
\startdata 
HCN/HCO$^{+}$ J=3--2 & 2014 May 17 & 35 & 15--650 & 14 & J2056$-$4714 &
J2056$-$4714 & J2056$-$4714 \\ 
HNC J=3--2 & 2014 May 17 & 35 & 15--650 & 36 & J2056$-$4714 & J2056$-$4714 &
J2056$-$4714 \\ 
\enddata

\tablecomments{
Col.(1): Observed line.
Col.(2): Observation date (UT). 
Col.(3): Number of antennas used for observations.
Col.(4): Baseline length in meter. Minimum and maximum baseline lengths
are shown.
Col.(5): Net on-source integration time in minutes. 
Cols.(6), (7), and (8): Bandpass, flux, and phase calibrator for the
target source, respectively.  
The flux density of J2056$-$4714 is estimated from ALMA monitoring
observations.
}

\end{deluxetable}

%%%%%%%%%% Table 2 %%%%%%%%%
\begin{deluxetable}{lrccl}
%\tabletypesize{\small}
\tabletypesize{\scriptsize}
\tablecaption{Continuum emission \label{tbl-2}}
\tablewidth{0pt}
\tablehead{
\colhead{Frequency} & \colhead{Flux} & 
\colhead{Peak Coordinate} & \colhead{rms} & \colhead{Synthesized beam} \\
\colhead{[GHz]} & \colhead{[mJy beam$^{-1}$]} & 
\colhead{(RA,DEC)J2000} & \colhead{[mJy beam$^{-1}$]} & 
\colhead{[arcsec $\times$ arcsec] ($^{\circ}$)} \\  
\colhead{(1)} & \colhead{(2)} & \colhead{(3)}  & \colhead{(4)}  &
\colhead{(5)} 
}
\startdata 
a: 244.8--246.6, 247.1--249.0, 259.8--263.5  & 3.8 (59$\sigma$) & 
(20 58 26.80, $-$42 39 00.3) & 0.065 & 0.50$\times$0.46 (74$^{\circ}$) \\
b: 254.0--258.5 &  3.6 (88$\sigma$) & (20 58 26.80, $-$42 39 00.3) & 0.041
 & 0.52$\times$0.47 (66$^{\circ}$) \\  
\enddata

\tablecomments{
Col.(1): Frequency range in [GHz] used for extraction of continuum-a 
(taken simultaneously with HCN and HCO$^{+}$ J=3--2 observations), 
and -b (obtained at the same time as HNC J=3--2 observations). 
Frequencies at obvious emission line features are removed.
Col.(2): Flux in [mJy beam$^{-1}$] at the emission peak.
Value at the highest flux pixel (0$\farcs$1 pixel$^{-1}$) is extracted.
The detection significance relative to the rms noise is shown in
parentheses. 
Possible systematic uncertainty is not included. 
Col.(3): The coordinate of the continuum emission peak in J2000.
Col.(4): The rms noise level (1$\sigma$) in [mJy beam$^{-1}$].
Col.(5): Synthesized beam in [arcsec $\times$ arcsec] and position angle
in [degree]. 
The position angle is 0$^{\circ}$ along the north--south direction
and increases in the counterclockwise direction. 
}

\end{deluxetable}

\clearpage

%%%%%%%%%% Table 3 %%%%%%%%%
\begin{deluxetable}{ll|llcl|cccc}
%\rotate
%\tabletypesize{\small}
\tabletypesize{\scriptsize}
\tablecaption{Molecular Line Flux \label{tbl-3}} 
\tablewidth{0pt}
\tablehead{
\colhead{Line} & \colhead{$\nu_{\rm rest}$} & 
\multicolumn{4}{c}{Integrated intensity (moment 0) map} & 
\multicolumn{4}{c}{Gaussian line fit} \\  
\colhead{} & \colhead{[GHz]} & \colhead{Peak} & \colhead{rms} & 
\colhead{Elements} & 
\colhead{Beam} & \colhead{Velocity} & \colhead{Peak} & \colhead{FWHM} & 
\colhead{Flux} \\ 
\colhead{} & \colhead{} & \multicolumn{2}{c}{[Jy beam$^{-1}$ km s$^{-1}$]} &
\colhead{Summed} & 
\colhead{[$''$ $\times$ $''$] ($^{\circ}$)} &
\colhead{[km s$^{-1}$]} & \colhead{[mJy]} & \colhead{[km s$^{-1}$]} & 
\colhead{[Jy km s$^{-1}$]} \\ 
\colhead{(1)} & \colhead{(2)} & \colhead{(3)} & \colhead{(4)} & 
\colhead{(5)} & \colhead{(6)} & \colhead{(7)} & \colhead{(8)} &
\colhead{(9)} & \colhead{(10)}    
}
\startdata 
HCN J=3--2 & 265.89 & 5.5 (85$\sigma$) & 0.064 & 16 & 0.49$\times$0.46
(73$^{\circ}$) & 12894$\pm$1 & 31$\pm$1 & 187$\pm$2 & 5.9$\pm$0.1  \\  
HCO$^{+}$ J=3--2 & 267.56 & 8.0 (87$\sigma$) & 0.092 & 18 &
0.49$\times$0.46 (75$^{\circ}$) & 12889$\pm$1 & 43$\pm$1 & 193$\pm$2 &
8.4$\pm$0.1 \\   
HCN J=3--2, v$_{2}$=1f & 267.20 & 0.22 (7.6$\sigma$) & 0.029 & 9 &
0.49$\times$0.46 (75$^{\circ}$) & 12883$\pm$12 & 1.5$\pm$0.2 &
165$\pm$34 & 0.25$\pm$0.07 \\   
HOC$^{+}$ J=3--2 & 268.45 & 0.18 (6.4$\sigma$) & 0.028 & 8 & 0.49$\times$0.46
(74$^{\circ}$) & 12884$\pm$21 & 1.4$\pm$0.4 & 131$\pm$55 & 0.19$\pm$0.10 \\
HCO$^{+}$ J=3--2, v$_{2}$=1f & 268.69 & $<$0.088 ($<$3$\sigma$) & 0.029
& 9 & 0.49$\times$0.46 (74$^{\circ}$) & --- & --- & --- & --- \\ \hline 
HNC J=3--2 & 271.98 & 3.0 (86$\sigma$) & 0.035 & 14 & 0.50$\times$0.47
(74$^{\circ}$) & 12895$\pm$2 & 19$\pm$1 & 164$\pm$3 & 3.2$\pm$0.1 \\  
HNC J=3--2, v$_{2}$=1f \tablenotemark{a} & 273.87 & 0.17 (6.5$\sigma$)
& 0.027 & 12 & 0.51$\times$0.44 (66$^{\circ}$) & 12891$\pm$29 &
0.80$\pm$0.17 & 243$\pm$68 & 0.20$\pm$0.07 \\   
SO$_{2}$ 5(3,3)--5(2,4) & 256.25 & 0.48 (20$\sigma$) & 0.024 & 12
& 0.54$\times$0.47 (65$^{\circ}$) & 12805$\pm$5 & 3.1$\pm$0.2 &
153$\pm$12 & 0.49$\pm$0.05 \\   
SO 6(6)--5(5) & 258.26 & 0.48 (22$\sigma$) & 0.021 & 11 &
0.53$\times$0.47 (63$^{\circ}$) & 12912$\pm$6 & 2.8$\pm$0.2 &
188$\pm$13 & 0.54$\pm$0.05 \\  
H$^{13}$CN J=3--2 & 259.01 & 0.34 (12$\sigma$) & 0.028 & 15 &
0.53$\times$0.47 (63$^{\circ}$) & 12884$\pm$9 & 1.7$\pm$0.2 & 205$\pm$22
& 0.37$\pm$0.05 \\  
HC$_{3}$N J=30--29 & 272.89 & 0.16 (9.2$\sigma$) & 0.017 & 8 &
0.50$\times$0.47 (74$^{\circ}$) & 12900$\pm$13 & 1.1$\pm$0.2 &
166$\pm$32 & 0.19$\pm$0.05 \\  
CH$_{3}$CCH 16(2)--15(2) & 273.40 & 0.59 (17$\sigma$) & 0.034 & 17 &
0.51$\times$0.44 
(66$^{\circ}$) & 12866$\pm$7 &  2.8$\pm$0.2 &  203$\pm$17 & 0.58$\pm$0.06 \\  
\enddata

\tablenotetext{a}{
HNC v$_{2}$=1f J=3--2 emission line could possibly be contaminated by 
HC$_{3}$N v$_{7}$=1f J=30--29 line ($\nu_{\rm rest}$=273.94 GHz)
\citep{cos13}. 
Although the peak velocity of the HNC v$_{2}$=1f J=3--2 emission line based
on Gaussian fit is comparable to those of HCN, HCO$^{+}$, and HNC v=0 
J=3--2 emission lines, as well as the HCN v$_{2}$=1f J=3--2 emission line, 
the line width is 25--50\% larger, possibly indicating the contamination.}

\tablecomments{ 
Col.(1): Observed molecular line. 
Molecular lines above the solid horizontal line were obtained
simultaneously with continuum-a. 
Those below the line were taken at the same time as continuum-b. 
Col.(2): Rest-frame frequency of each molecular line in [GHz]. 
Col.(3): Integrated intensity in [Jy beam$^{-1}$ km s$^{-1}$] at the 
emission peak. 
Value at the highest flux pixel (0$\farcs$1 pixel$^{-1}$) is extracted.
Detection significance relative to the rms noise (1$\sigma$) in the 
moment 0 map is shown in parentheses. 
Possible systematic uncertainty is not included. 
Col.(4): rms noise (1$\sigma$) level in the moment 0 map in 
[Jy beam$^{-1}$ km s$^{-1}$], derived from the standard deviation 
of sky signals in each moment 0 map. 
Col.(5): The number of spectral elements summed to create moment 0 maps.
Each spectral element ($\sim$20 MHz width) consists of 40 correlator
channels binning (see $\S$2).
Col.(6): Beam size in [arcsec $\times$ arcsec] and position angle in
[degree].  
Position angle is 0$^{\circ}$ along the north-south direction, 
and increases counterclockwise. 
Cols.(7)--(10): Gaussian fits of emission lines in the spectra at the 
continuum peak position, within the beam size. 
Col.(7): Optical LSR velocity (v$_{\rm opt}$) of emission peak in
[km s$^{-1}$].  
Col.(8): Peak flux in [mJy]. 
Col.(9): Observed FWHM in [km s$^{-1}$] in Figure 4.  
Col.(10): Flux in [Jy km s$^{-1}$]. 
Possible systematic uncertainty is not included.
The observed FWHM in [km s$^{-1}$] in column 9 is divided by ($1+z$) to
obtain the intrinsic FWHM in [km s$^{-1}$]. 
}

\end{deluxetable}

%%%%%%%%%% Table 4 %%%%%%%%%
\begin{deluxetable}{lcr}
%\rotate
%\tabletypesize{\small}
%\tabletypesize{\scriptsize}
\tablecaption{Molecular line luminosity for IRAS 20551$-$4250 \label{tbl-4}}
\tablewidth{0pt}
\tablehead{
\colhead{Line} & \colhead{[10$^{4}$ L$_{\odot}$]} &
\colhead{[10$^{7}$ K km s$^{-1}$ pc$^{2}$]} \\   
\colhead{(1)} & \colhead{(2)} & \colhead{(3)} 
}
\startdata 
HCN J=3--2 & 5.5$\pm$0.1 & 9.2$\pm$0.2 \\
HCO$^{+}$ J=3--2 & 7.9$\pm$0.1 & 12.9$\pm$0.2 \\
HNC J=3--2 & 3.1$\pm$0.1 & 4.7$\pm$0.1 \\
HCN J=3--2 v$_{2}$=1f & 0.23$\pm$0.07 & 0.35$\pm$0.10 \\
HCO$^{+}$ J=3--2 v$_{2}$=1f & $<$0.084 & $<$0.13 \\
HNC J=3--2 v$_{2}$=1f & 0.19$\pm$0.07 & 0.27$\pm$0.09 \\ \hline
HCN J=4--3 & 11.8$\pm$0.3 & 8.3$\pm$0.2 \\
HCO$^{+}$ J=4--3 & 17.5$\pm$1.3 & 12.1$\pm$0.9 \\
HNC J=4--3 & 7.4$\pm$0.3 & 4.8$\pm$0.2 \\
HCN J=4--3 v$_{2}$=1f & 0.49$\pm$0.09 & 0.34$\pm$0.06 \\
\enddata

\tablecomments{
Col.(1): Molecular line.
Col.(2): Luminosity in units of [10$^{4}$ L$_{\odot}$], calculated with
equation (1) of \citet{sol05}.
Col.(3): Luminosity in units of [10$^{7}$ K km s$^{-1}$ pc$^{2}$],
calculated with equation (3) of \citet{sol05}.
Luminosities at J=4--3 are also calculated from data by 
\citet{ima13b}.  
}

\end{deluxetable}

%%%%%%%%%% Table 5 %%%%%%%%%
\begin{deluxetable}{lcrrc}
%\tabletypesize{\small}
%\tabletypesize{\scriptsize}
\tablecaption{Parameters for molecular transition lines \label{tbl-5}}
\tablewidth{0pt}
\tablehead{
\colhead{Line} & \colhead{Frequency} & \colhead{E$_{\rm u}$/k$_{\rm B}$} & 
\colhead{A$_{\rm ul}$} & \colhead{Flux} \\
\colhead{} & \colhead{[GHz]} & \colhead{[K]} & 
\colhead{[10$^{-4}$ s$^{-1}$]} & \colhead{[Jy km s$^{-1}$]} \\
\colhead{(1)} & \colhead{(2)} & \colhead{(3)}  & \colhead{(4)}  &
\colhead{(5)}  
}
\startdata 
HCN J=3--2  & 265.89 & 25.5 & 8.4 & 5.9$\pm$0.1 \\  
HCN J=3--2, v$_{2}$=1f  & 267.20 & 1050.0 & 7.3 & 0.25$\pm$0.07 \\ \hline
HCO$^{+}$ J=3--2  & 267.56 & 25.7 & 14.5 & 8.4$\pm$0.1 \\  
HCO$^{+}$ J=3--2, v$_{2}$=1f  & 268.69 & 1217.4  & 13.1 & $<$0.088 \\ \hline
HNC J=3--2  & 271.98 & 26.1 & 9.3 & 3.2$\pm$0.1 \\  
HNC J=3--2, v$_{2}$=1f  & 273.87 & 692.0 & 8.5 &  0.20$\pm$0.07 \\ \hline
HCN J=4--3  & 354.51 & 42.5 & 20.6 & 9.5$\pm$0.2 \\  
HCN J=4--3, v$_{2}$=1f & 356.26 & 1067.1 & 19.0 & 0.39$\pm$0.07 \\ 
\enddata

\tablecomments{
Col.(1): Transition line.
Col.(2): Rest-frame frequency in [GHz].
Col.(3): Upper energy level in [K].
Col.(4): Einstein A coefficient for spontaneous emission in 
[10$^{-4}$ s$^{-1}$]. 
Values in Cols. (3) and (4) are from the Cologne Database of
Molecular Spectroscopy (CDMS) \citep{mul05} via Splatalogue
(http://www.splatalogue.net).
Col.(5): Flux in [Jy km s$^{-1}$] estimated based on Gaussian fits 
(Table 3, column 10).
For the undetected HCO$^{+}$ v$_{2}$=1f J=3--2 emission line, an upper
limit based on the integrated intensity (moment 0) map is adopted. 
HCN J=4--3 fluxes at v$_{2}$=1f and v=0 are from \citet{ima13b}.
}

\end{deluxetable}

%%%%%%%%%% Table 6 %%%%%%%%%
\begin{deluxetable}{llrccl}
%\tabletypesize{\small}
\tabletypesize{\scriptsize}
\tablecaption{v$_{2}$=1f to v=0 flux ratio \label{tbl-7}}
\tablewidth{0pt}
\tablehead{
\colhead{Object} & \colhead{Line} & \colhead{Observed ratio} & 
\colhead{Flux attenuation} & \colhead{Intrinsic ratio} &
\colhead{Reference} \\  
%\colhead{} & \colhead{[GHz]} & \colhead{[K]} & \colhead{[s$^{-1}$]} \\
\colhead{(1)} & \colhead{(2)} & \colhead{(3)} & \colhead{(4)} & 
\colhead{(5)} & \colhead{(6)} 
}
\startdata 
IRAS 20551$-$4250 & HCN J=3--2 & 0.042 & 6 & $\sim$0.01 & This work  \\  
 & HCO$^{+}$ J=3--2 & $<$0.011 & $<$3 & $<$0.01 & This work \\
 & HNC J=3--2 & 0.063 & $\sim$1 & $\sim$0.06 & This work \\  
 & HCN J=4--3 & 0.041 & 6 & $\sim$0.01 & \citet{ima13b} \\ \hline
NGC 4418 & HCN J=4--3 & 0.23 & $\sim$20 & $\sim$0.01 & \citet{sak10} \\
 & HCN J=3--2 & 0.17 & $\sim$20 & $\sim$0.01 & \citet{sak10} \\ \hline
Mrk 231 & HCN J=3--2 & 0.042 & $\sim$10 & $\sim$0.005 & \citet{aal15a} 
\enddata

\tablecomments{
Col.(1): Object name. Only LIRGs with available quantitative
information of flux attenuation for v=0 emission are shown.
Col.(2): Molecular transition line.
Col.(3): Ratio of the observed v$_{2}$=1f to v=0 emission line flux in 
[Jy km s$^{-1}$]. 
Col.(4): Estimated flux attenuation for v=0 emission.
The value of 6 means that flux is attenuated by line opacity
with a factor of 6.
Col.(5): Ratio of the intrinsic v$_{2}$=1f to v=0 emission line 
flux in [Jy km s$^{-1}$], after line opacity correction of v=0 emission. 
Col.(6): Reference.
}

\end{deluxetable}

%%%%%%%%%% Table 7 %%%%%%%%%
\begin{deluxetable}{lc}
%\rotate
%\tabletypesize{\small}
%\tabletypesize{\scriptsize}
\tablecaption{Summary of the ratios of various parameters for HCN,
HCO$^{+}$, and HNC \label{tbl-6}} 
\tablewidth{0pt}
\tablehead{
\colhead{Parameters} & \colhead{HCN : HCO$^{+}$ : HNC} \\
\colhead{(1)} & \colhead{(2)} 
}
\startdata 
Column density at v$_{2}$=1f, J=3 & 1 : $<$0.20 : 0.69 \\
Observed v$_{2}$=1f to v=0 flux ratio at J=3--2 & 1 : $<$0.25 : 1.5 \\
B coefficient from v=0 to v$_{2}$=1 & 1 : 1.1 : 11 \\
Infrared flux used for vibrational excitation & 1 : 0.8 : 2.5 \\
Predicted infrared radiative pumping rate & 1 : 0.9 : 27 \\
Predicted v$_{2}$=1 to v=0 column density ratio & 1 : 0.5 : 9 \\
Derived abundance & 1 : $<$0.4 : $<$0.08 \\
Line opacity derived from isotopologue flux & 1 : $<$0.5 : $<$0.2 \\
Line opacity derived from abundance & 1: $<$0.7 : $<$0.1 \\
\enddata

\tablecomments{
Col.(1): Parameters used for our discussion in $\S$ 4.3.
Col.(2): Ratio among HCN, HCO$^{+}$, and HNC.
}

\end{deluxetable}

%%%%%%%%%% Table 8 %%%%%%%%%
\begin{deluxetable}{lcc}
%\tabletypesize{\small}
%\tabletypesize{\scriptsize}
\tablecaption{Excitation temperature for IRAS 20551$-$4250 \label{tbl-8}}
\tablewidth{0pt}
\tablehead{
\colhead{Molecule} & \colhead{(v$_{2}$,J; v, J)} & 
\colhead{T$_{\rm ex}$ [K]} \\ 
\colhead{(1)} & \colhead{(2)} & \colhead{(3)} 
}
\startdata 
HCN & (1f,3; 0,3) & 340 (210) \\  
 & (1f,4; 0,4) & 330 (210) \\  
 & (1f,3; 0,4) & 430 (245) \\  
 & (1f,4; 0,3) & 275 (185) \\  
 & (1f,4; 1f,3) & 23 \\  
 & (0,4; 0,3) & 25  \\  \hline
HNC & (1f,3; 0,3) & 250 \\ \hline
HCO$^{+}$ & (1f,3; 0,3) & $<$270 \\ 
\enddata

\tablecomments{
Col.(1): Molecular line.
Col.(2): Transition.
Col.(3): Excitation temperature (T$_{\rm ex}$) in [K].
For HCN v=0 J=3--2 and J=4--3 emission, flux attenuation
with a factor of 6 is estimated ($\S$4.3.2).
In first four rows, T$_{\rm ex}$ values, after this correction by a
factor of 6, are shown in the parentheses.
}

\end{deluxetable}

%\clearpage

%--- Figure 1 ---%
\begin{figure}
\begin{center}
\includegraphics[angle=0,scale=.5]{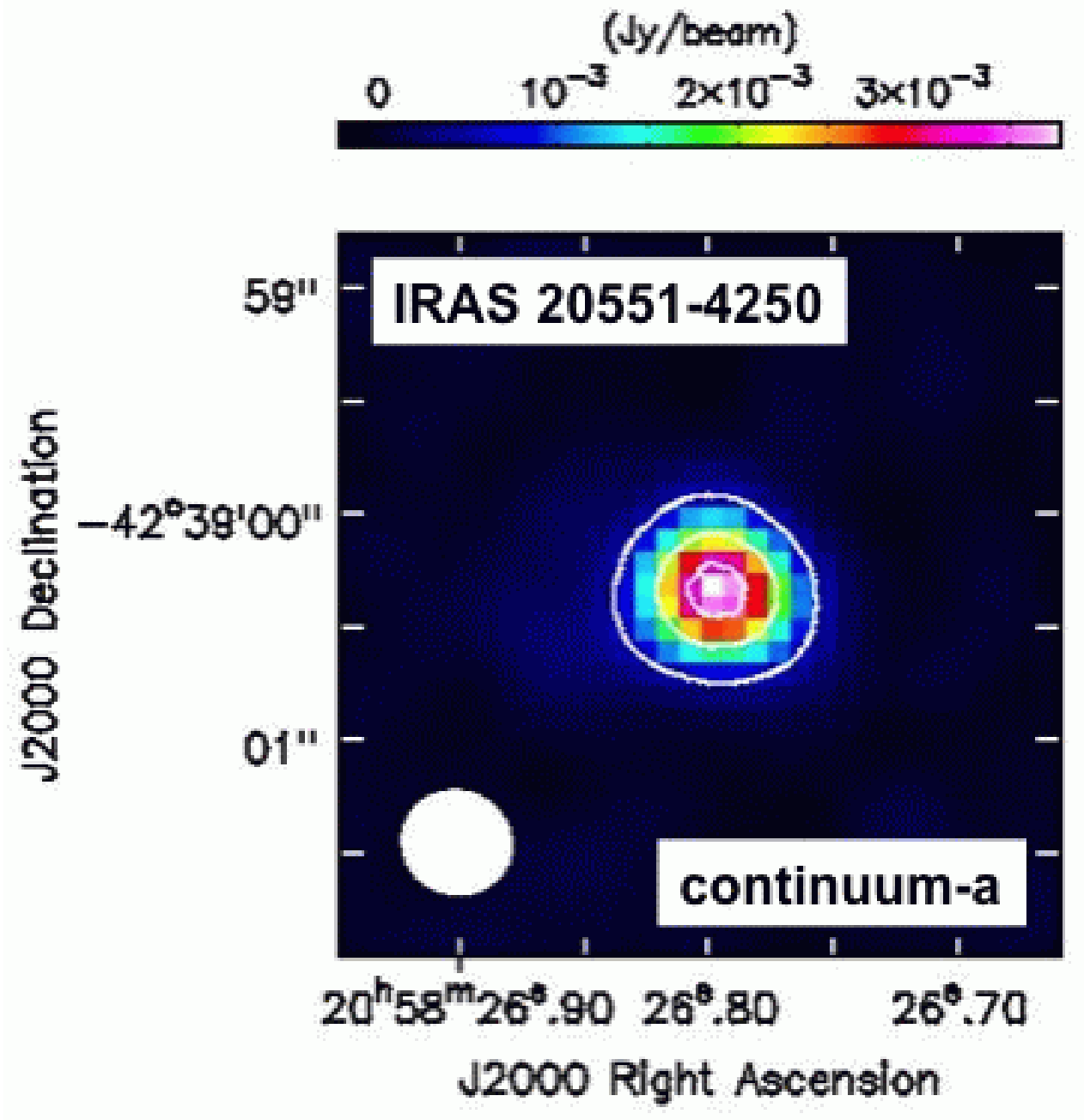} 
\includegraphics[angle=0,scale=.5]{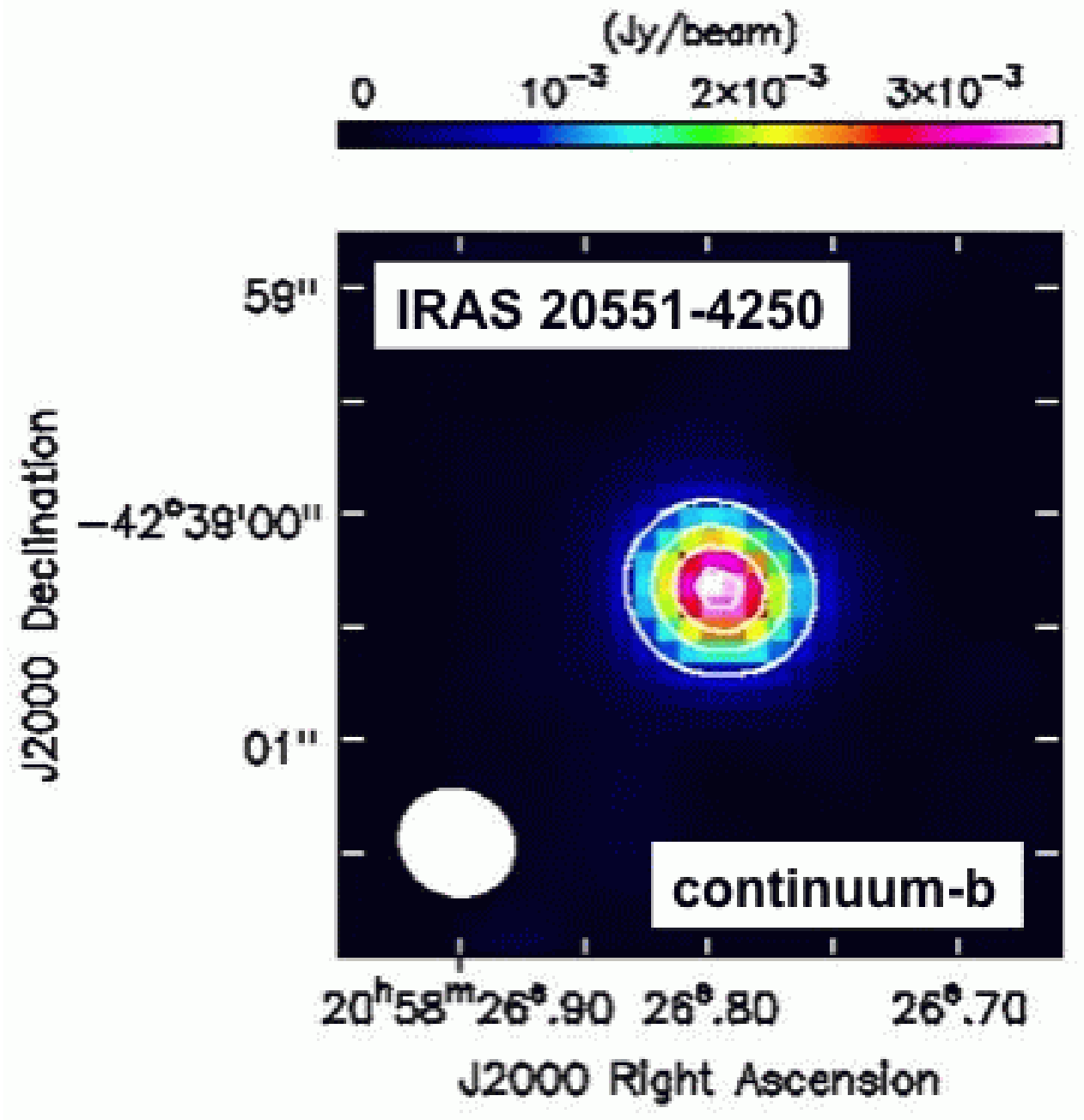} \\
\end{center}
\caption{
Continuum-a (central observed frequency $\nu_{\rm center}$ $\sim$ 256.3
GHz) and -b ($\nu_{\rm center}$ $\sim$ 254.5 GHz) data taken 
during the observations of HCN/HCO$^{+}$ J=3--2 and HNC J=3--2,
respectively.  
The contours represent the 10$\sigma$, 30$\sigma$, and 50$\sigma$ 
levels for continuum-a and 
20$\sigma$, 40$\sigma$, 60$\sigma$, and 80$\sigma$ levels for
continuum-b.  
}
\end{figure}

%--- Figure 2 ---%
\begin{figure}
\begin{center}
\includegraphics[angle=0,scale=.41]{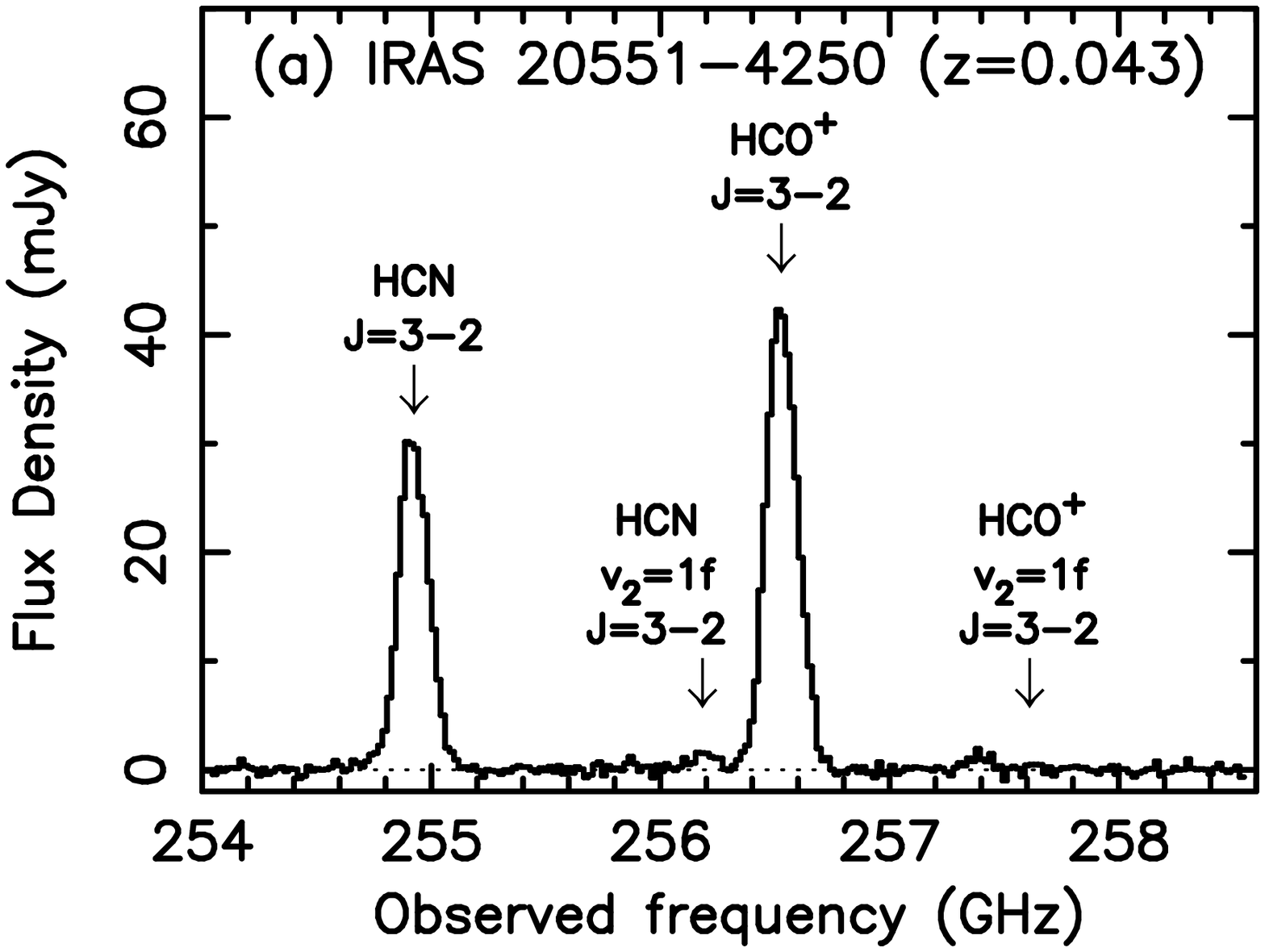} 
\includegraphics[angle=0,scale=.41]{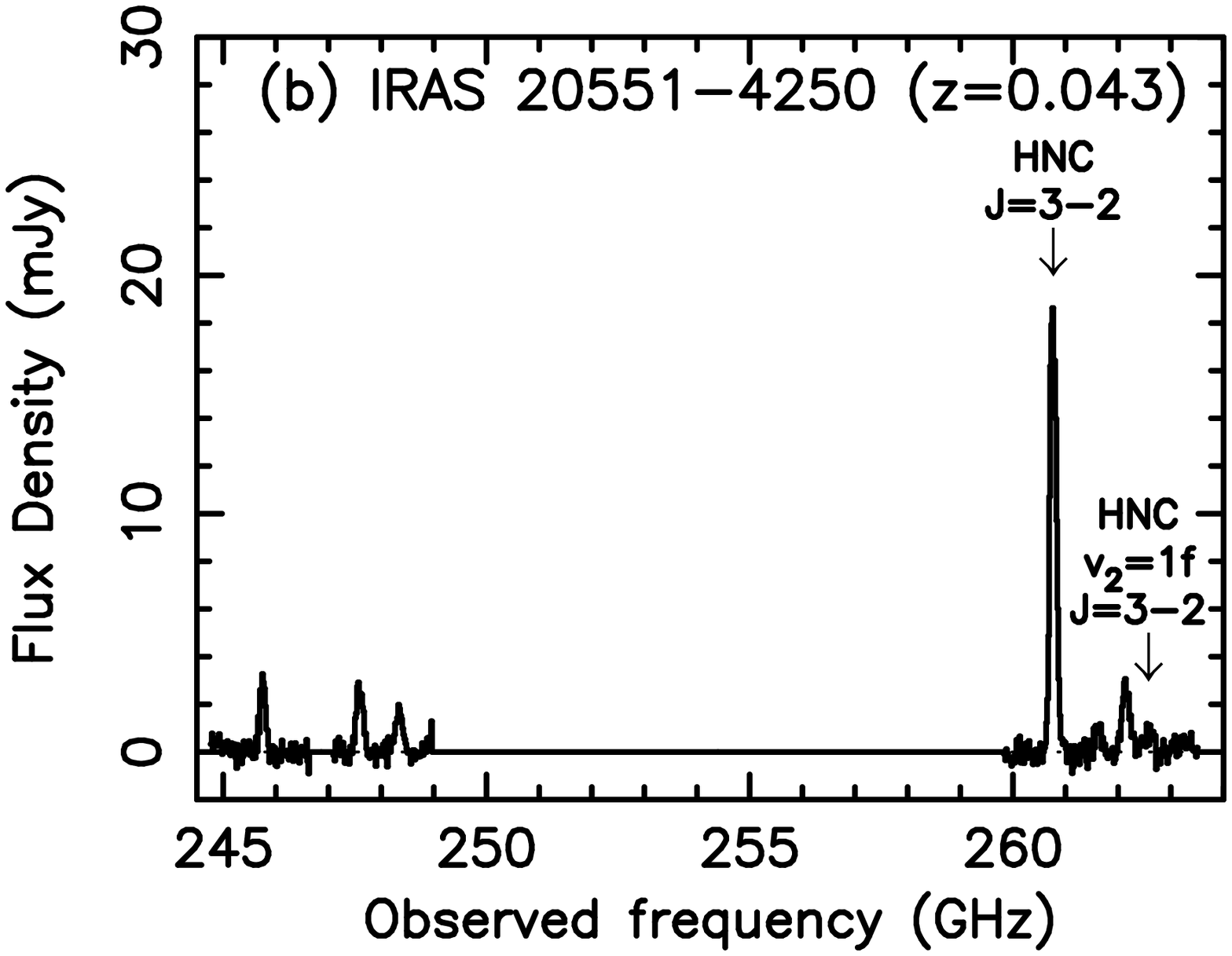} \\
\vspace{-0.5cm}
\includegraphics[angle=0,scale=.41]{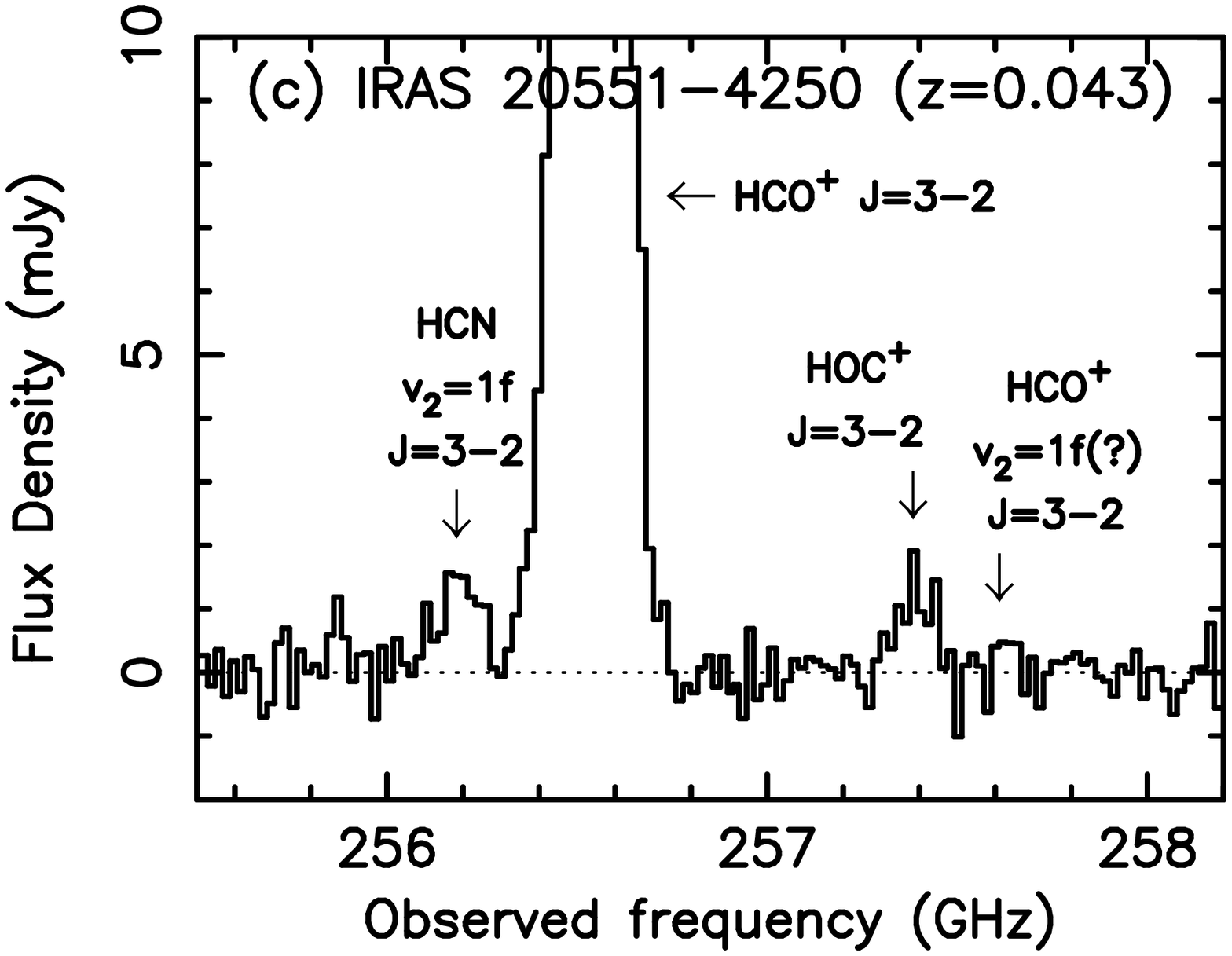} \\ 
\vspace{-0.5cm}
\includegraphics[angle=0,scale=.41]{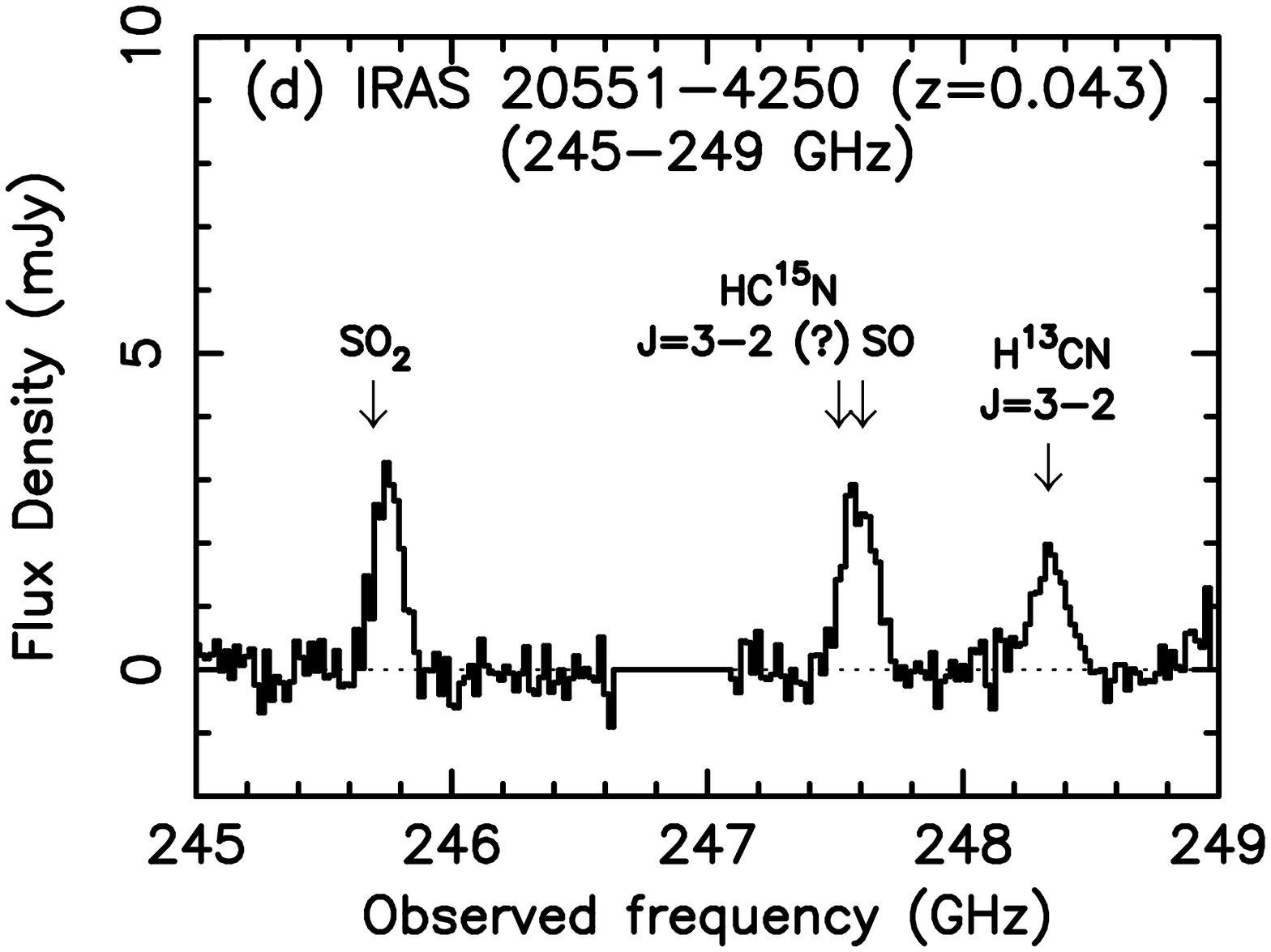}  
\includegraphics[angle=0,scale=.41]{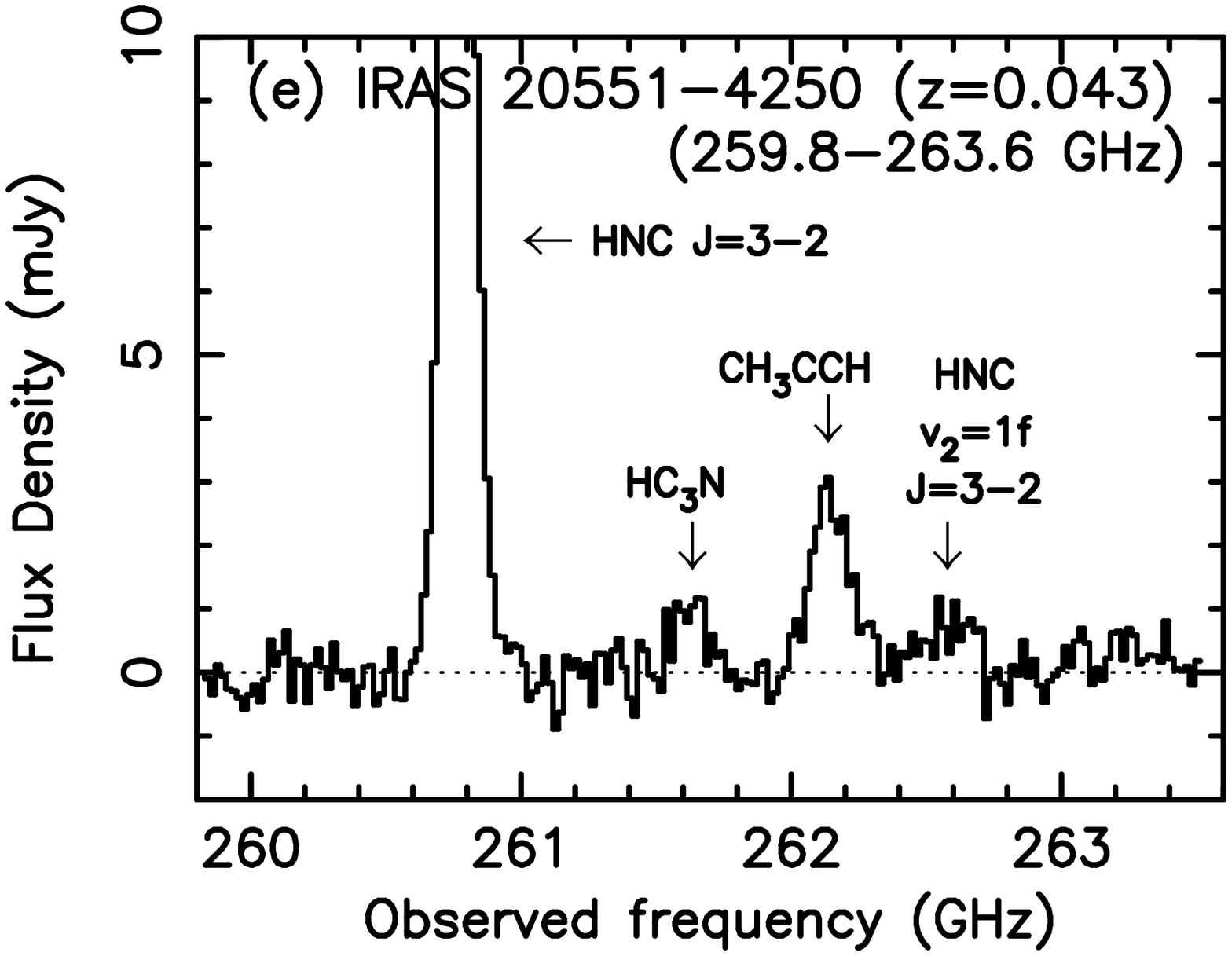}  
\end{center}
\vspace{-0.5cm}
\caption{
(a) : Full spectrum at the continuum peak position within the
beam size for HCN/HCO$^{+}$ J=3--2 observation.   
(b) : Full spectrum at the continuum peak position within the
beam size for HNC J=3--2 observations.   
(c): Magnified spectrum around the HCO$^{+}$ J=3--2 emission line,
to display HCN v$_{2}$=1f J=3--2 and HCO$^{+}$ v$_{2}$=1f J=3--2 lines.
(d) : Magnified spectrum of lower frequency part for HNC J=3--2
observations, to show serendipitously detected emission lines in more
detail.  
(e): Magnified spectrum of higher frequency part for HNC J=3--2
observations, to show serendipitously detected emission lines in more
detail.  
The abscissa is the observed frequency in [GHz] and the ordinate is flux
density in [mJy].
Down arrows are shown at the expected observed frequency for 
z = 0.0430, for individual molecular emission lines of our interest.  
}
\end{figure}

%--- Figure 3 ---%
\begin{figure}
\begin{center}
\includegraphics[angle=0,scale=.41]{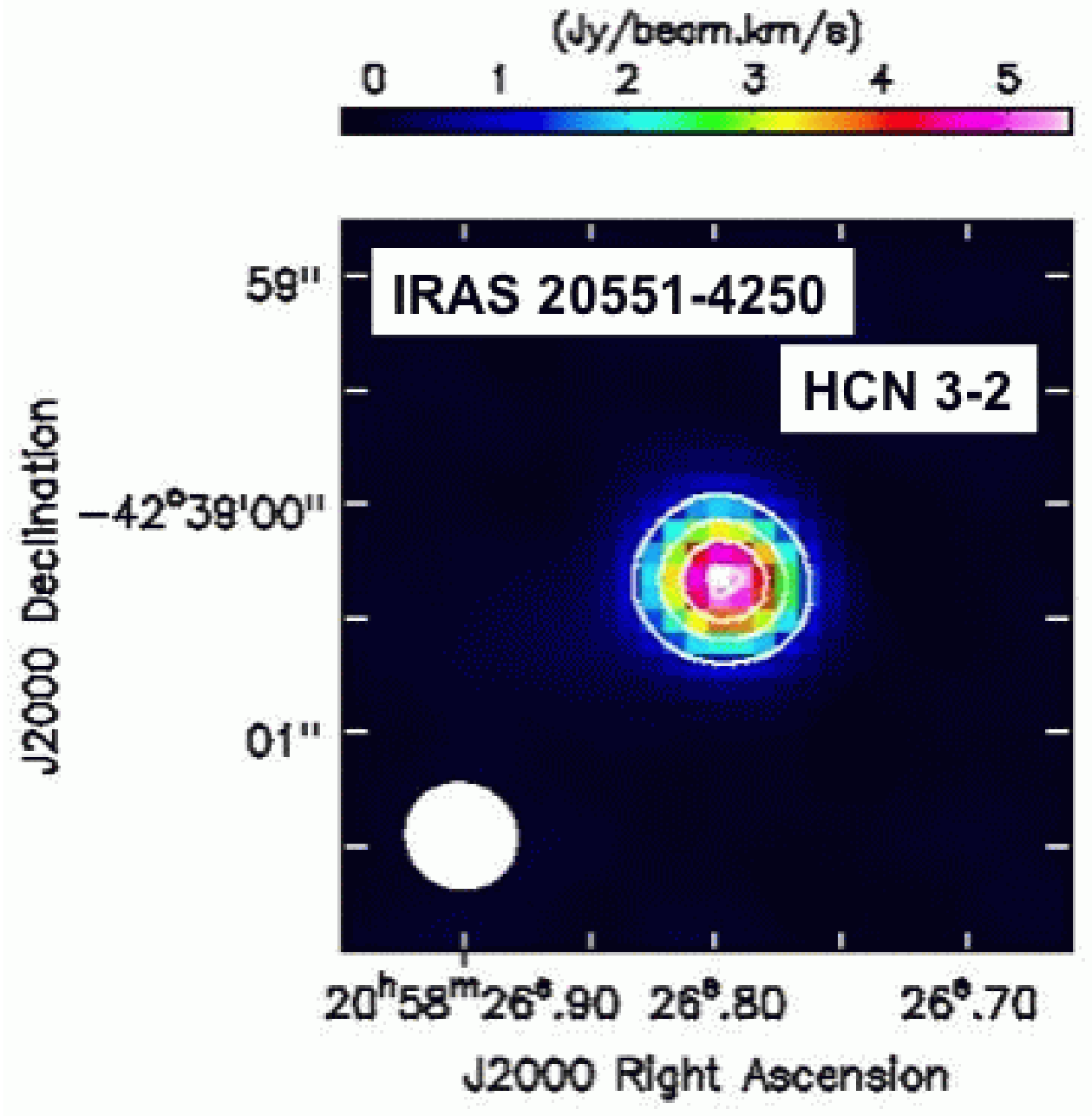} 
\includegraphics[angle=0,scale=.41]{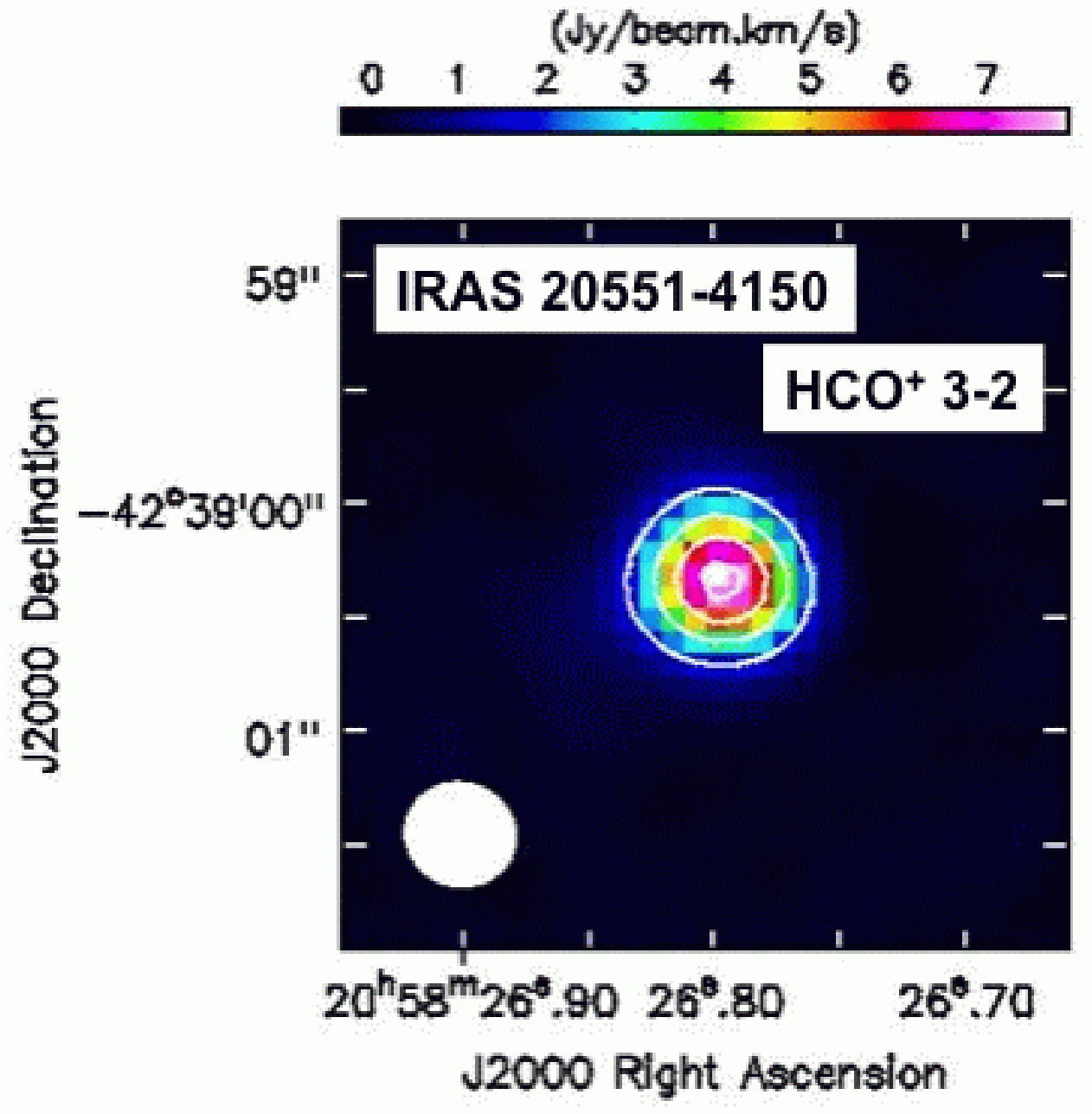} 
\includegraphics[angle=0,scale=.41]{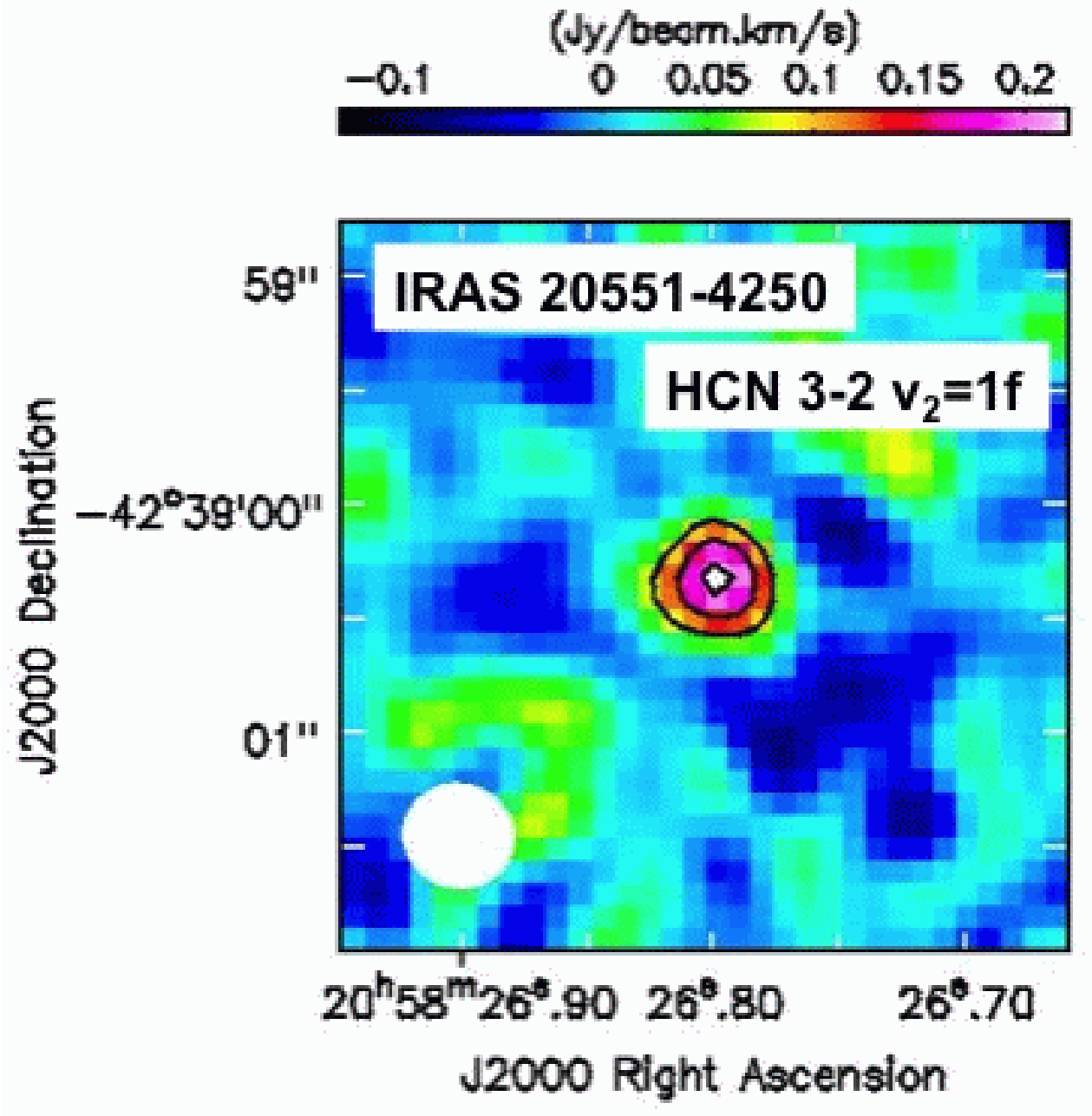} \\
\vspace{-1.1cm}
\includegraphics[angle=0,scale=.41]{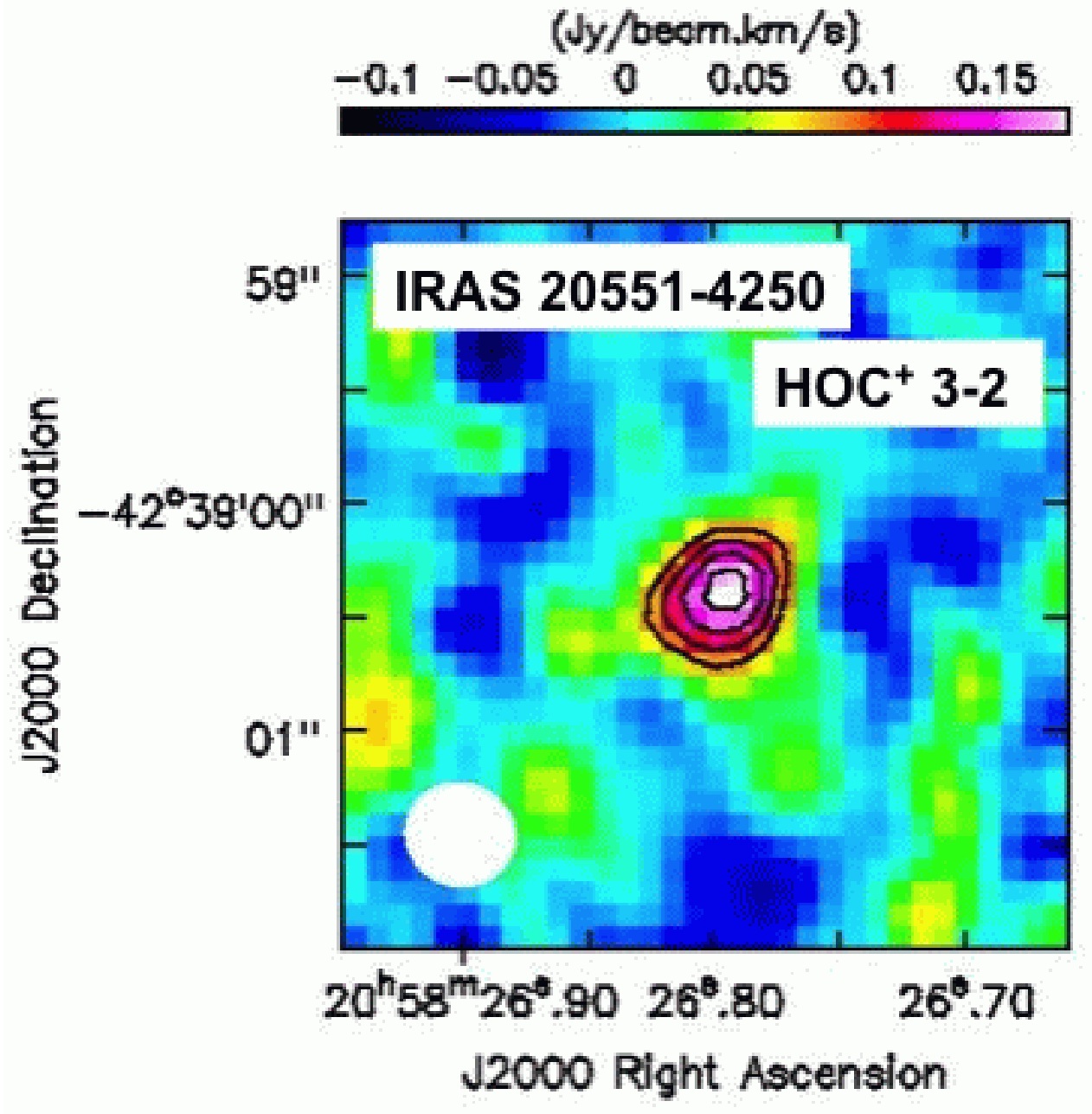} 
\includegraphics[angle=0,scale=.41]{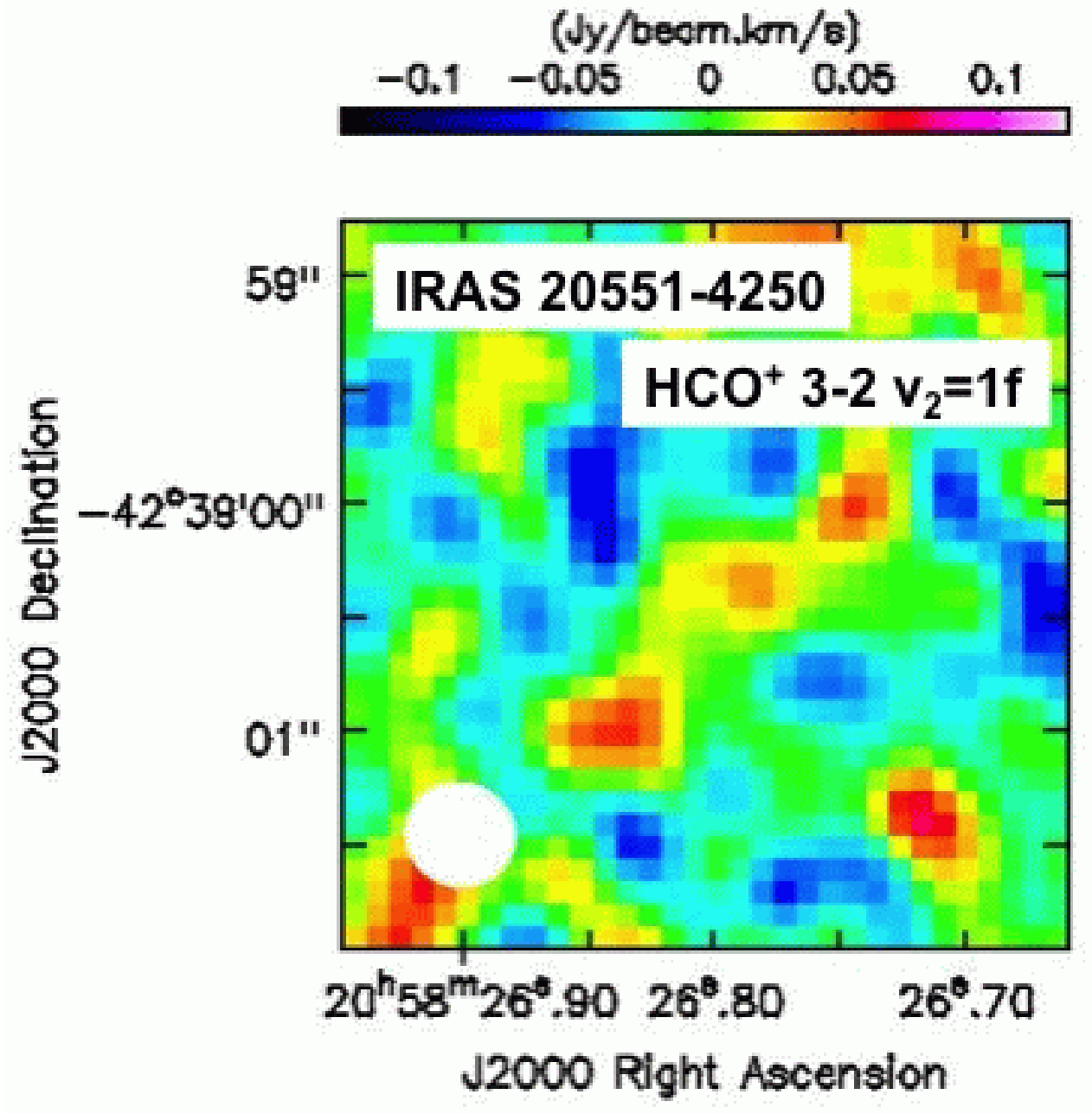} 
\includegraphics[angle=0,scale=.41]{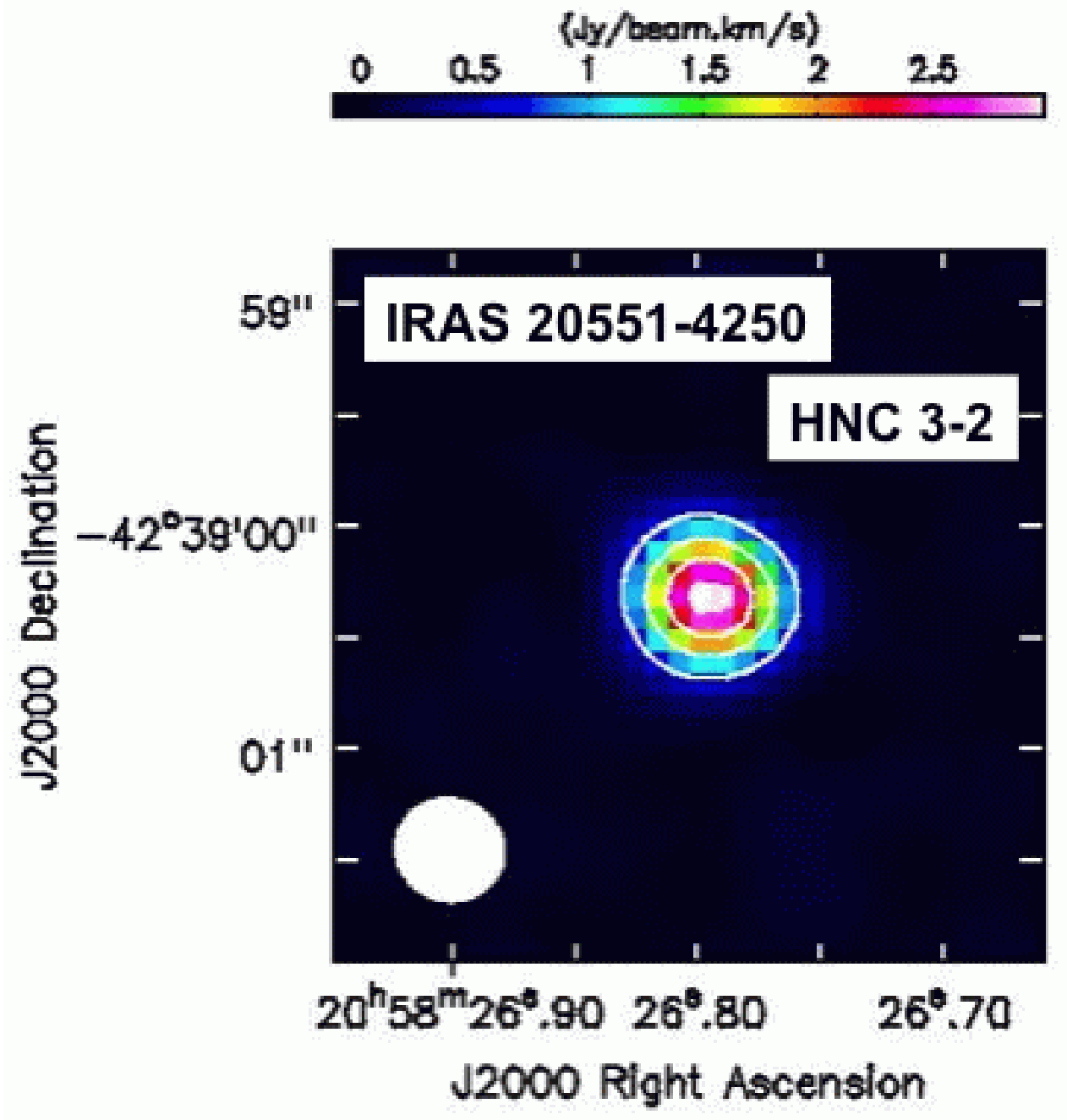} \\ 
\vspace{-1.1cm}
\includegraphics[angle=0,scale=.41]{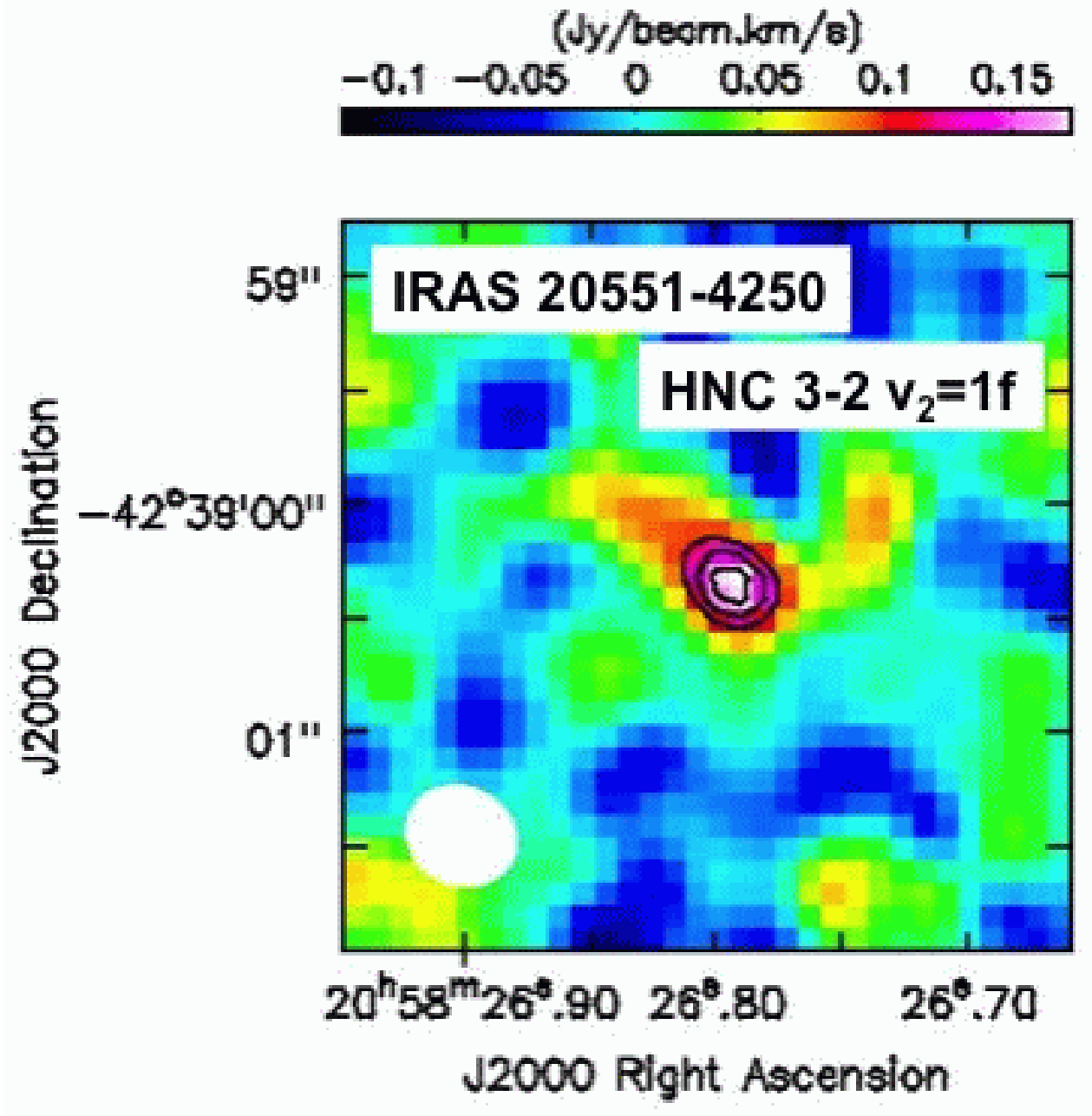} 
\includegraphics[angle=0,scale=.41]{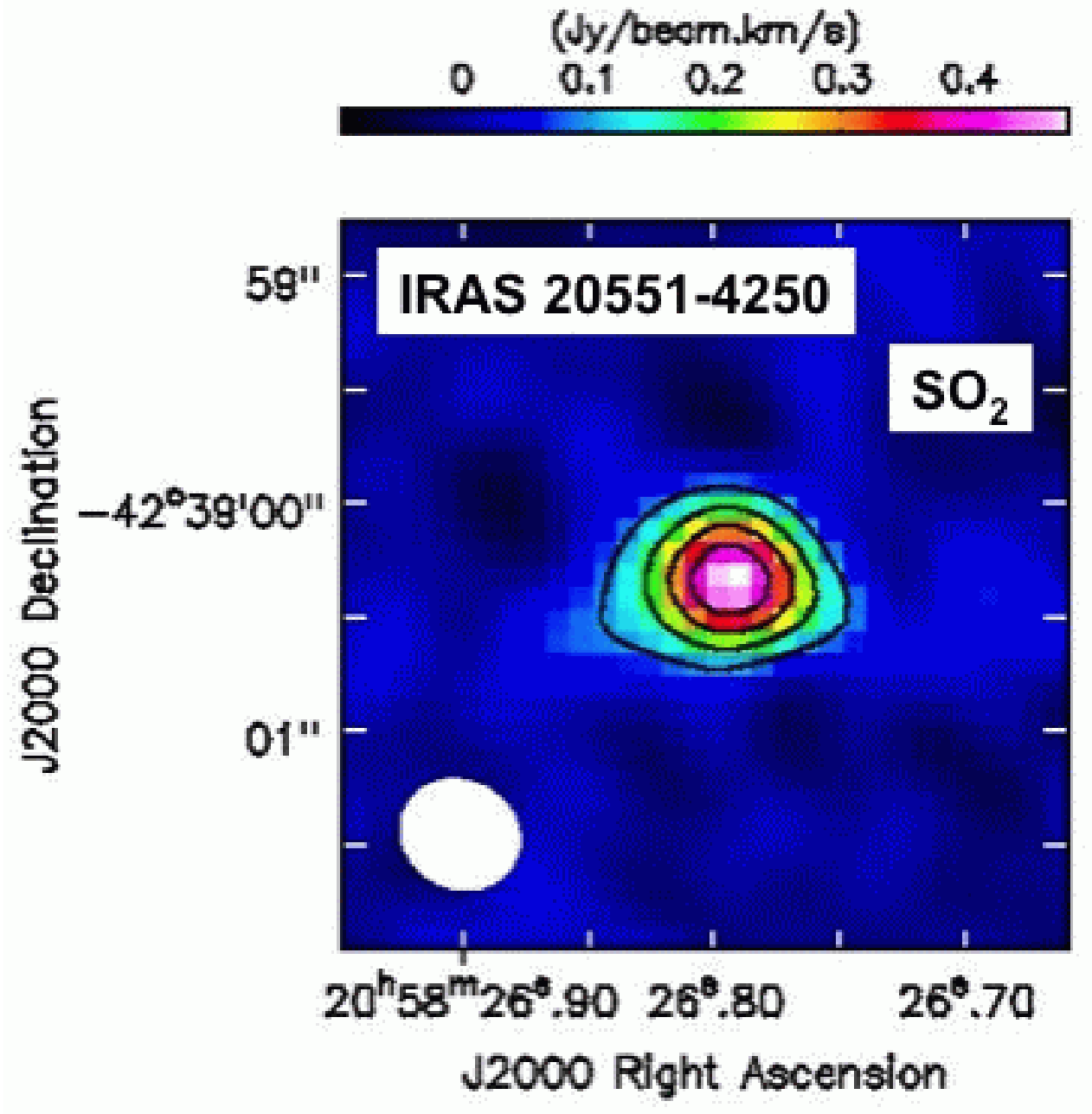} 
\includegraphics[angle=0,scale=.41]{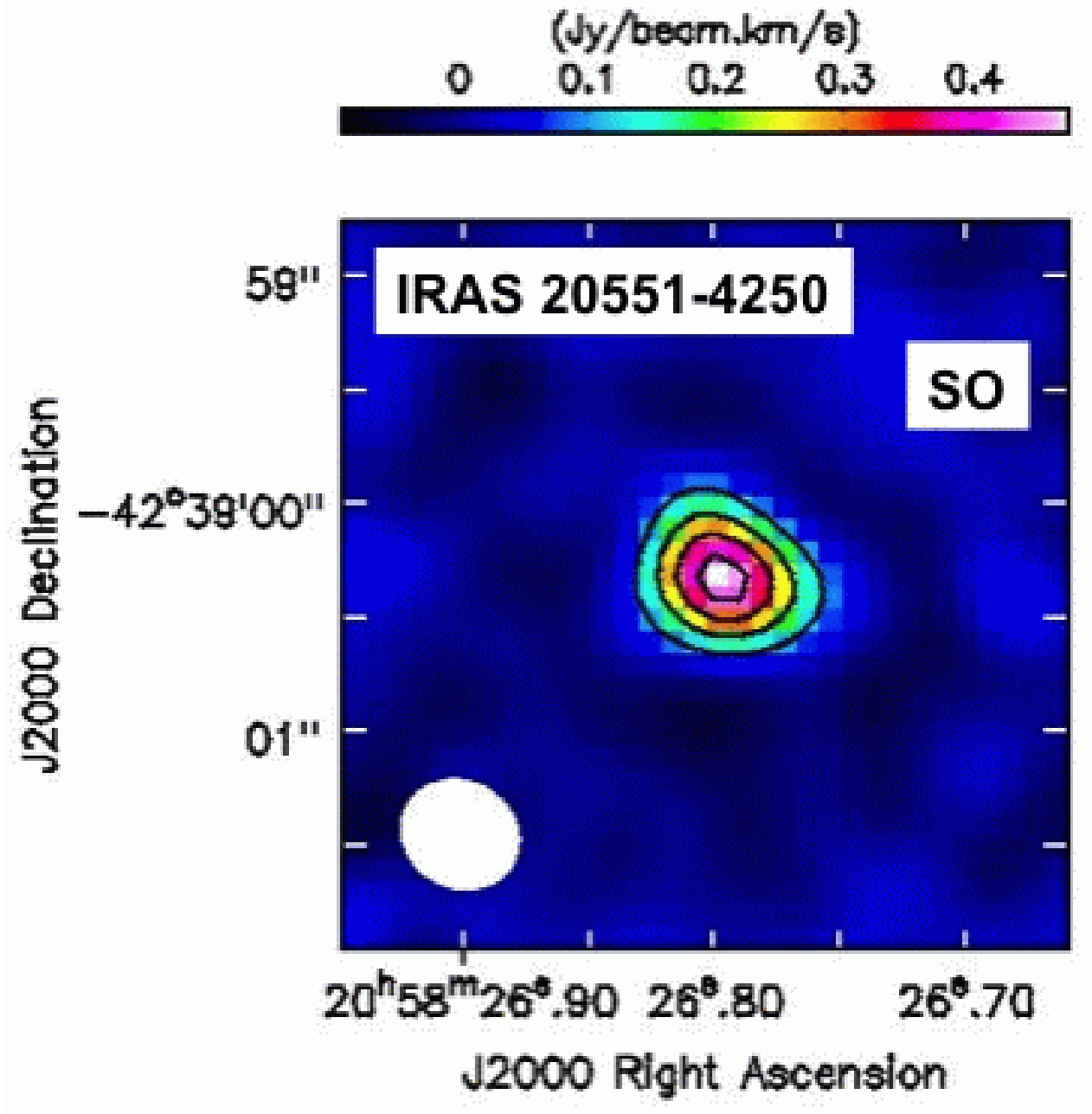} 
\end{center}
\end{figure}

%\clearpage

\begin{figure}
\begin{center}
\includegraphics[angle=0,scale=.41]{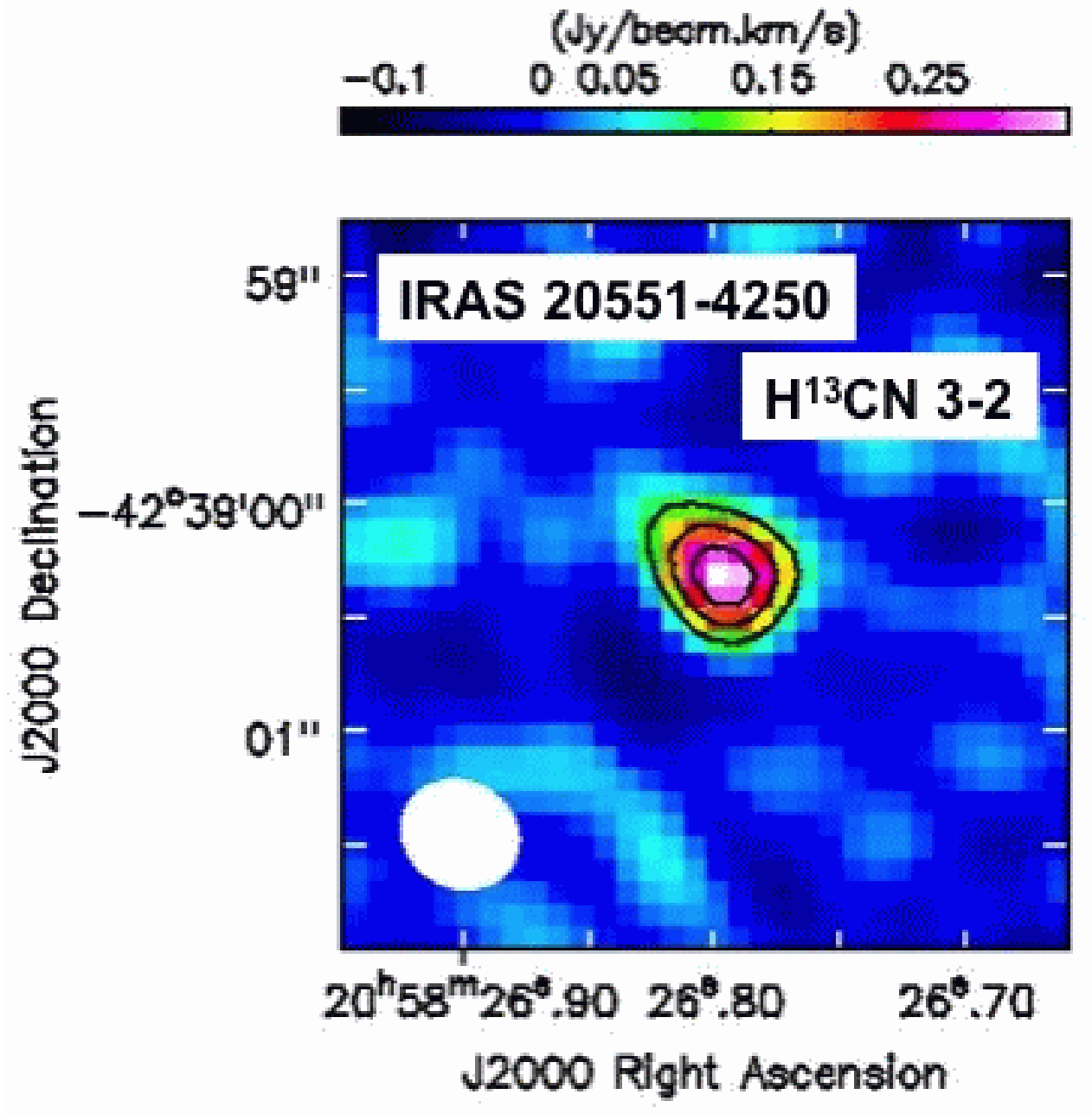} 
\includegraphics[angle=0,scale=.41]{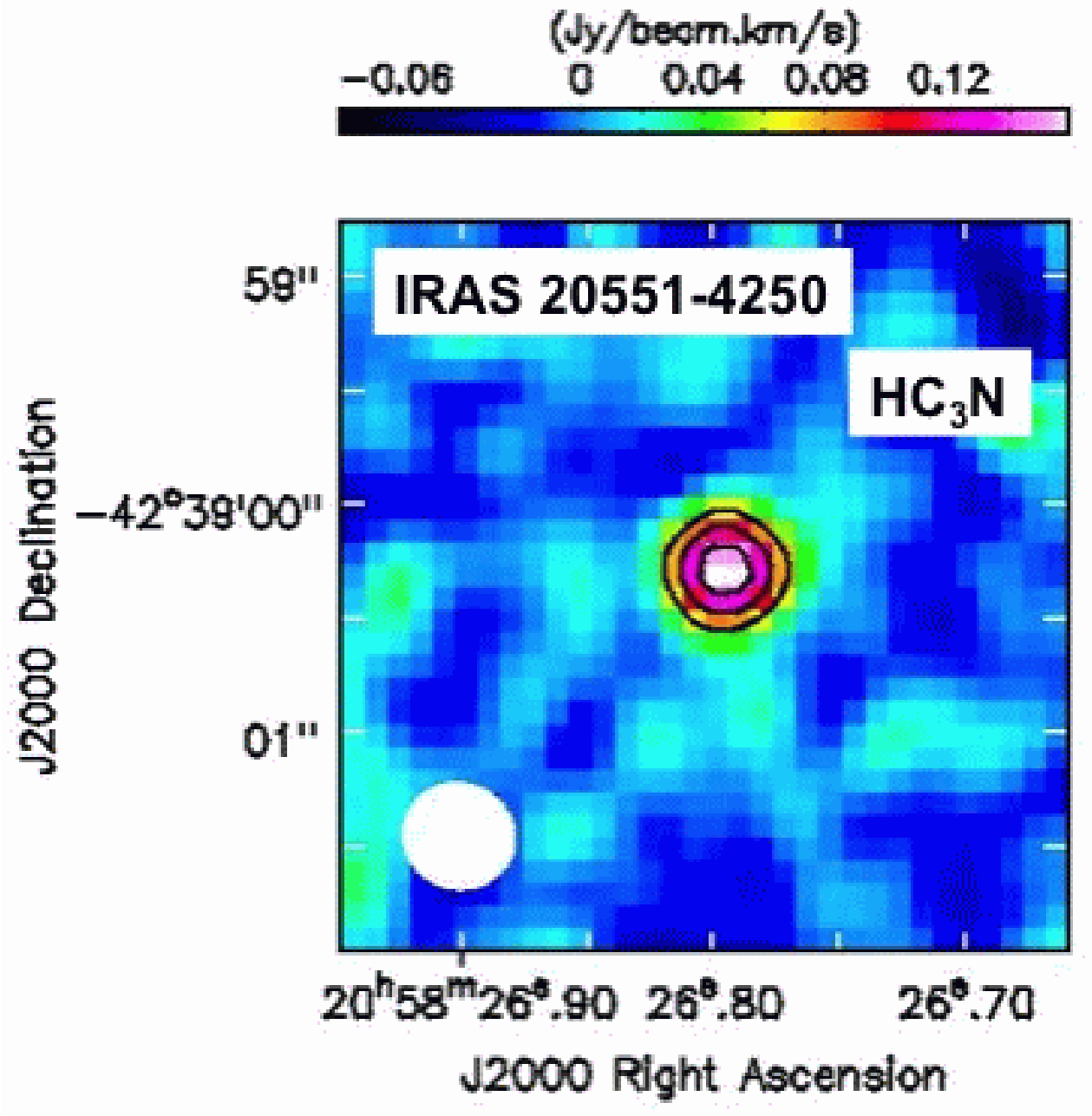}  
\includegraphics[angle=0,scale=.41]{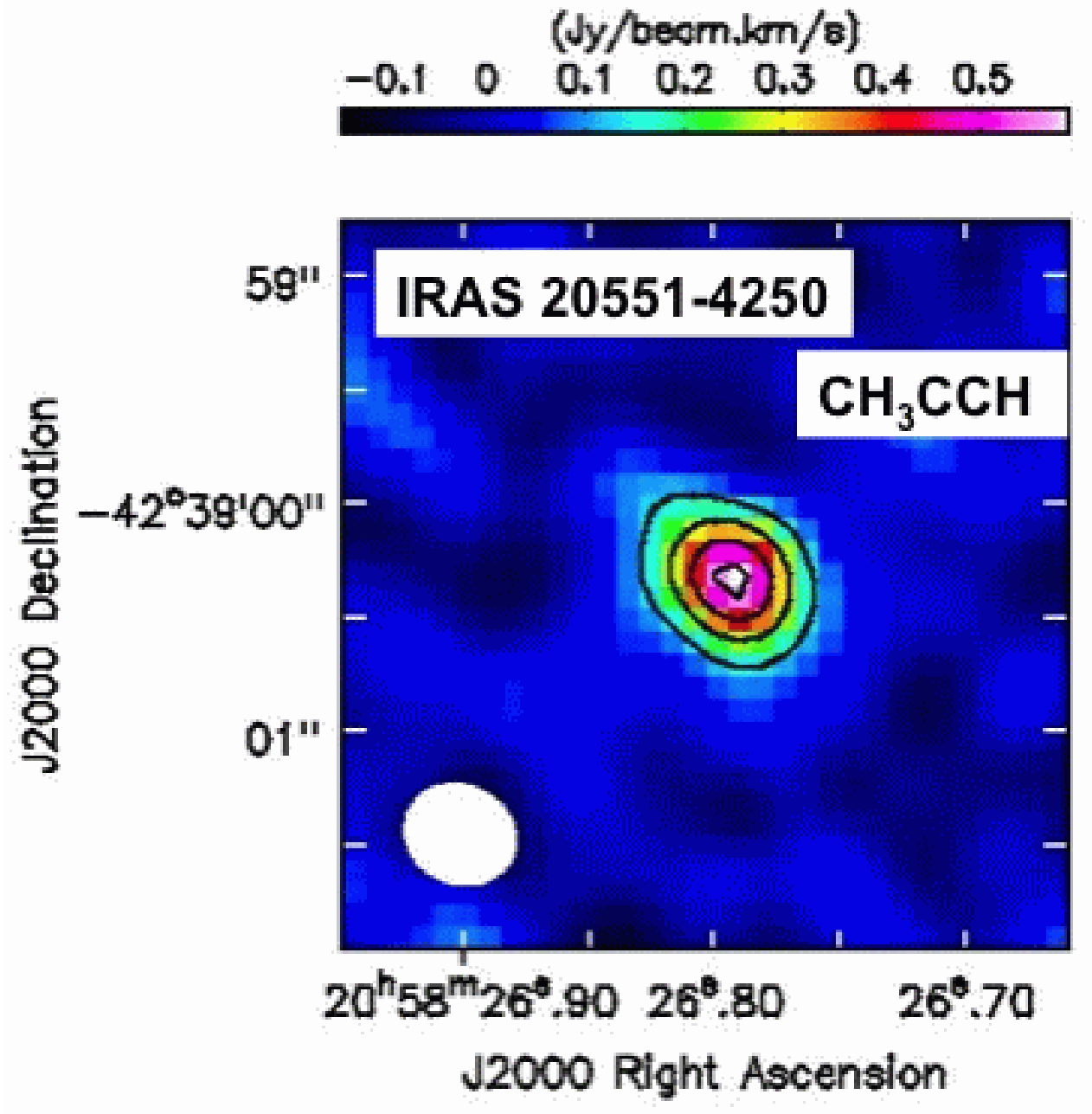} 
\end{center}
\caption{
Integrated intensity (moment 0) maps of our interesting molecular lines
in IRAS 20551$-$4250.  
The contours represent 20$\sigma$, 40$\sigma$, 60$\sigma$, 80$\sigma$ 
for HCN J=3--2, 
20$\sigma$, 40$\sigma$, 60$\sigma$, 80$\sigma$ for HCO$^{+}$ J=3--2, 
3$\sigma$, 5$\sigma$, 7$\sigma$ for HCN v$_{2}$=1f J=3--2, 
3$\sigma$, 4$\sigma$, 5$\sigma$, 6$\sigma$ for HOC$^{+}$ J=3--2, 
20$\sigma$, 40$\sigma$, 60$\sigma$, 80$\sigma$ for HNC J=3--2, 
4$\sigma$, 5$\sigma$, 6$\sigma$ for HNC v$_{2}$=1f J=3--2, 
4$\sigma$, 8$\sigma$, 12$\sigma$, 16$\sigma$ for SO$_{2}$, 
5$\sigma$, 10$\sigma$, 15$\sigma$, 20$\sigma$ for SO, 
4$\sigma$, 7$\sigma$, 10$\sigma$ for H$^{13}$CN J=3--2, 
4$\sigma$, 6$\sigma$, 8$\sigma$ for HC$_{3}$N, and 
4$\sigma$, 8$\sigma$, 12$\sigma$, 16$\sigma$ for CH$_{3}$CCH. 
For HCO$^{+}$ v$_{2}$=1f J=3--2, no emission with $\geqq$3$\sigma$ is
found. 
The 1$\sigma$ levels are different for different molecular lines.
They are summarized in Table 3.
}
\end{figure}

%--- Figure 4 ---%
\begin{figure}
\begin{center}
\includegraphics[angle=0,scale=.274]{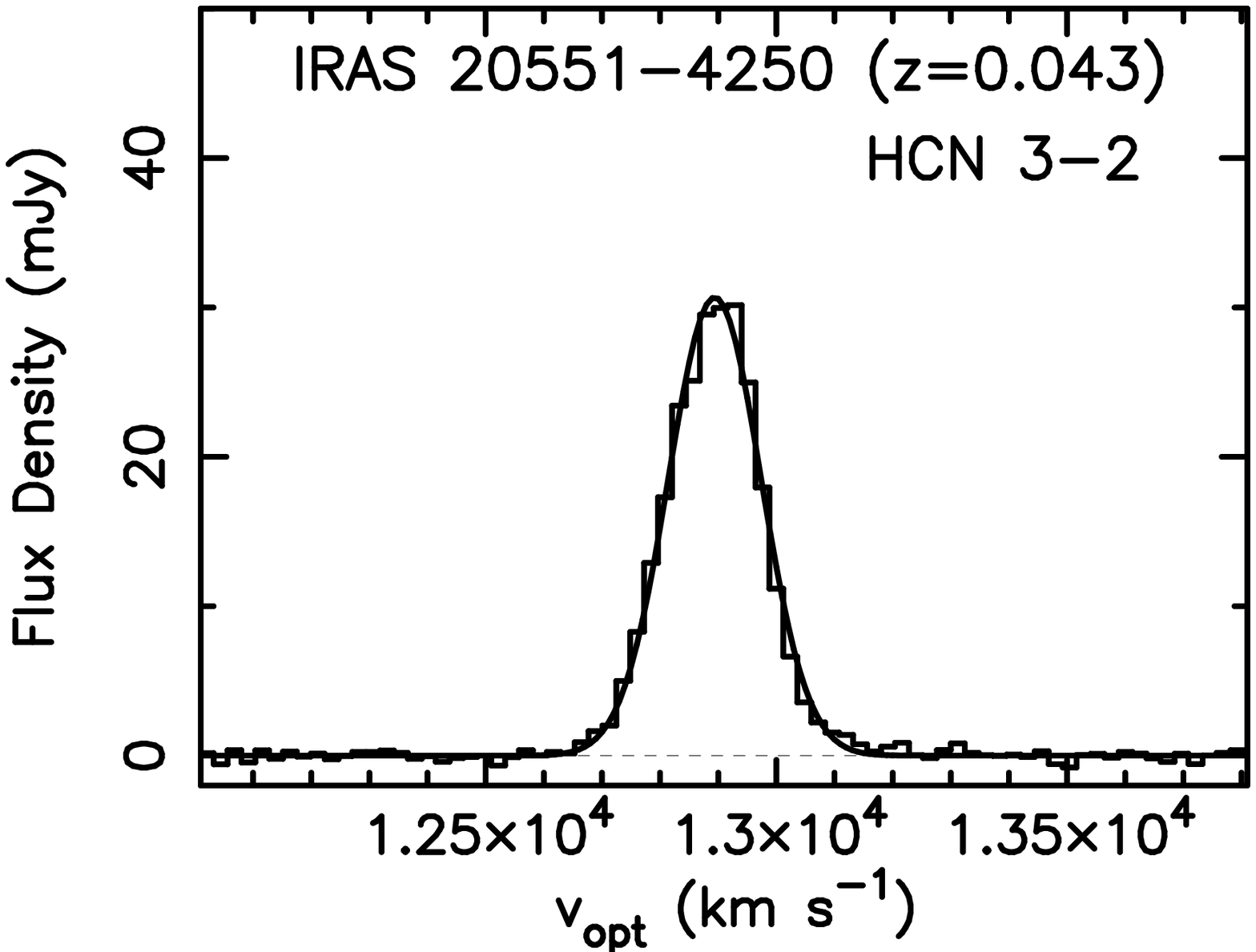}  
\includegraphics[angle=0,scale=.274]{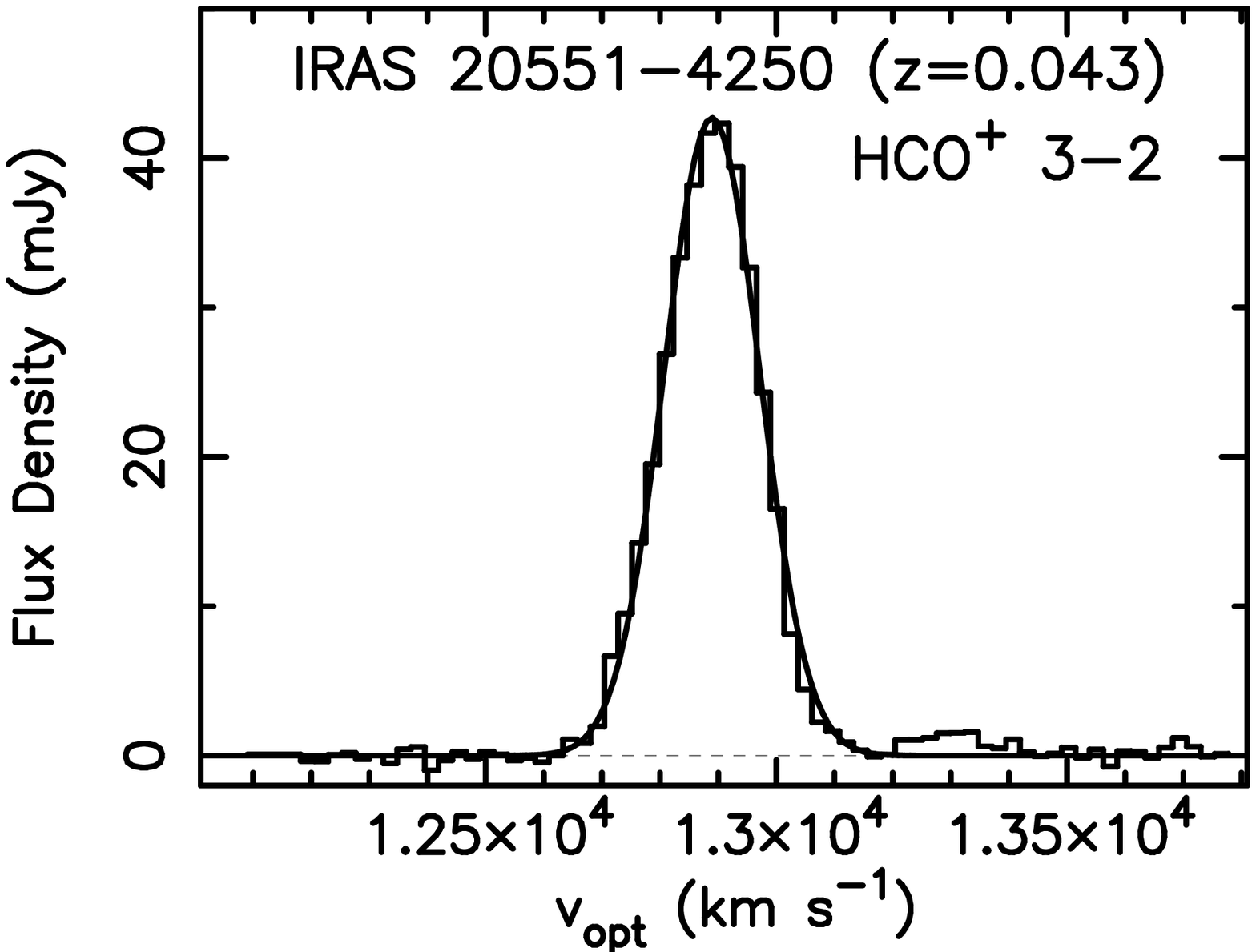}  
\includegraphics[angle=0,scale=.274]{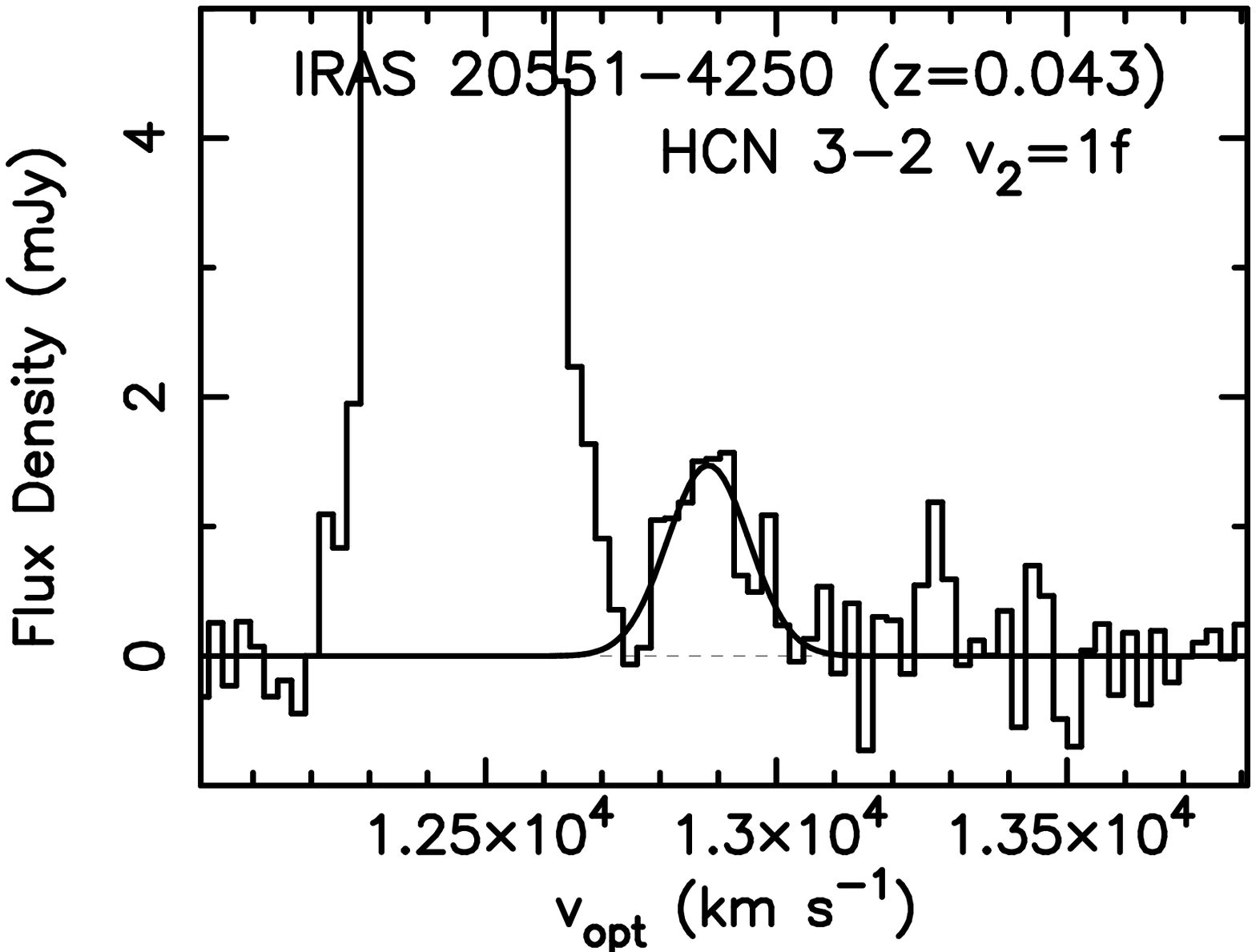} \\
\includegraphics[angle=0,scale=.274]{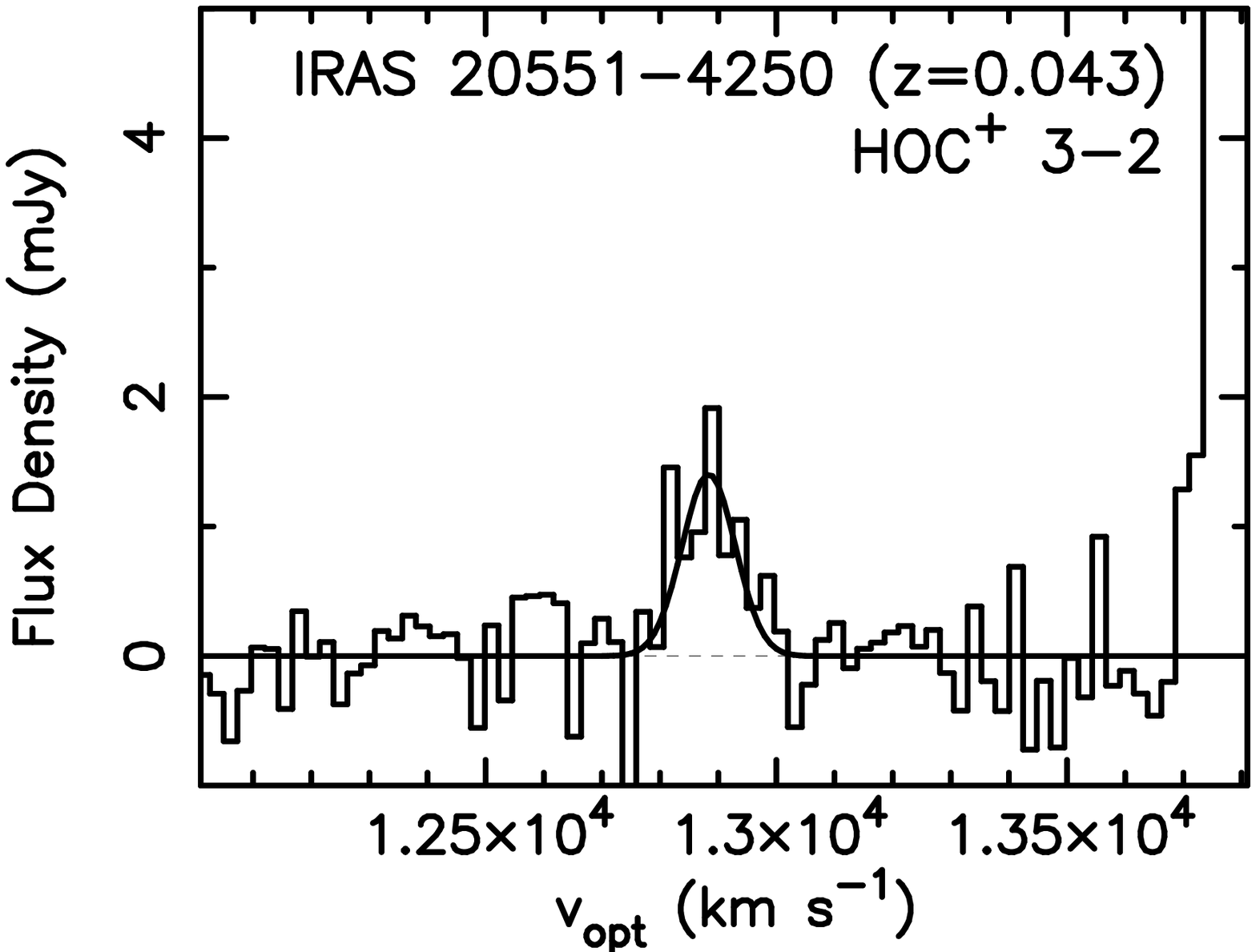}  
\includegraphics[angle=0,scale=.274]{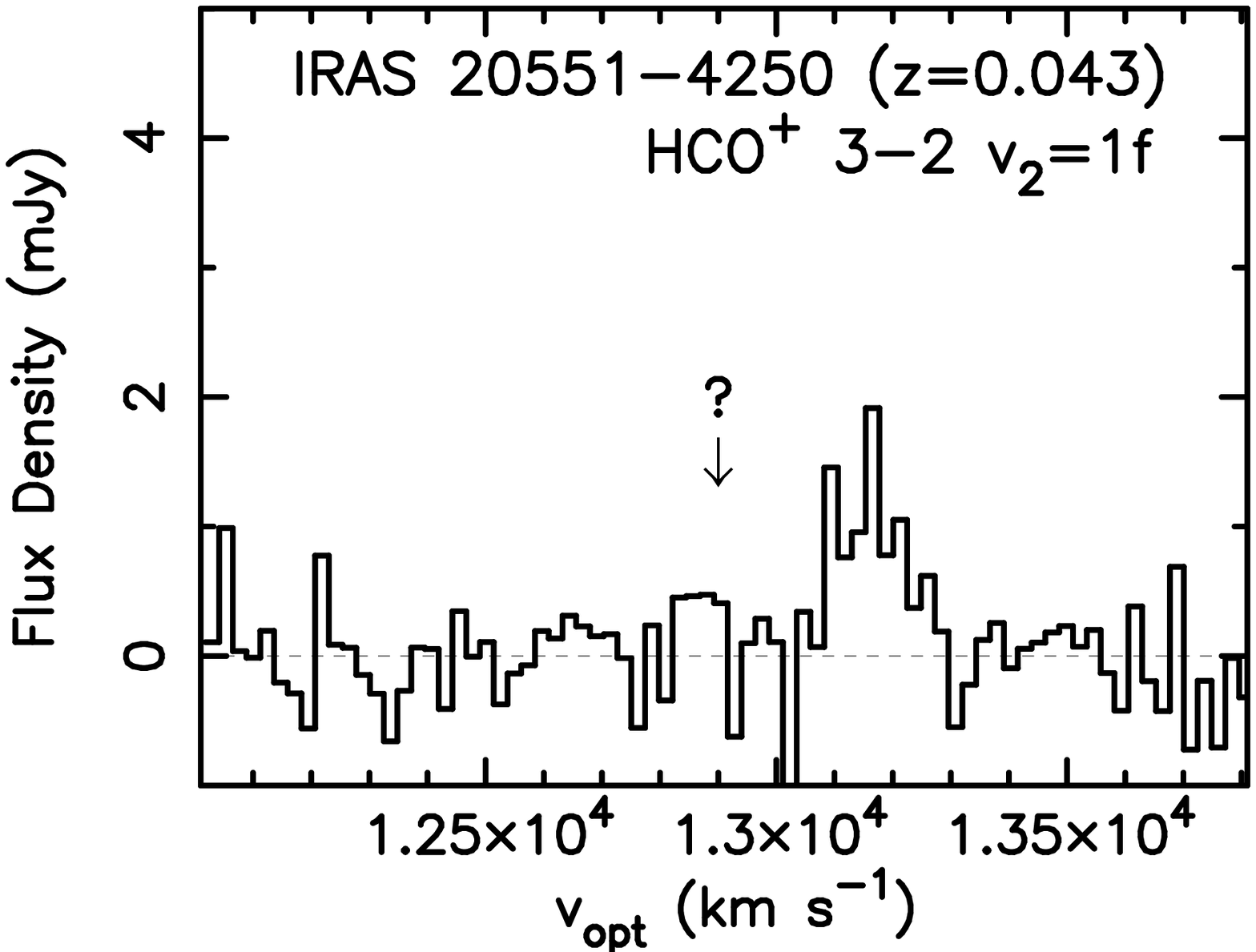}  
\includegraphics[angle=0,scale=.274]{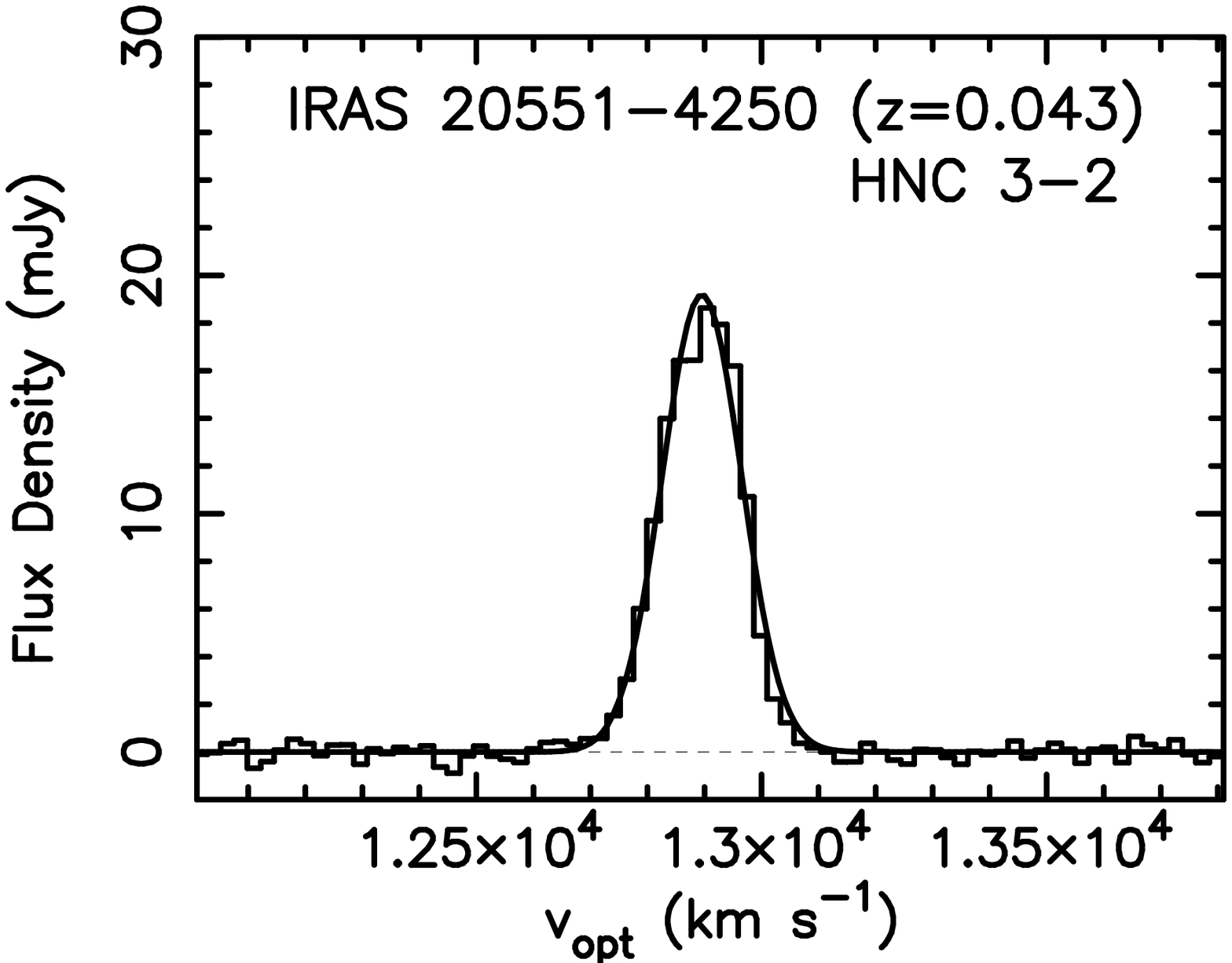}\\ 
\includegraphics[angle=0,scale=.274]{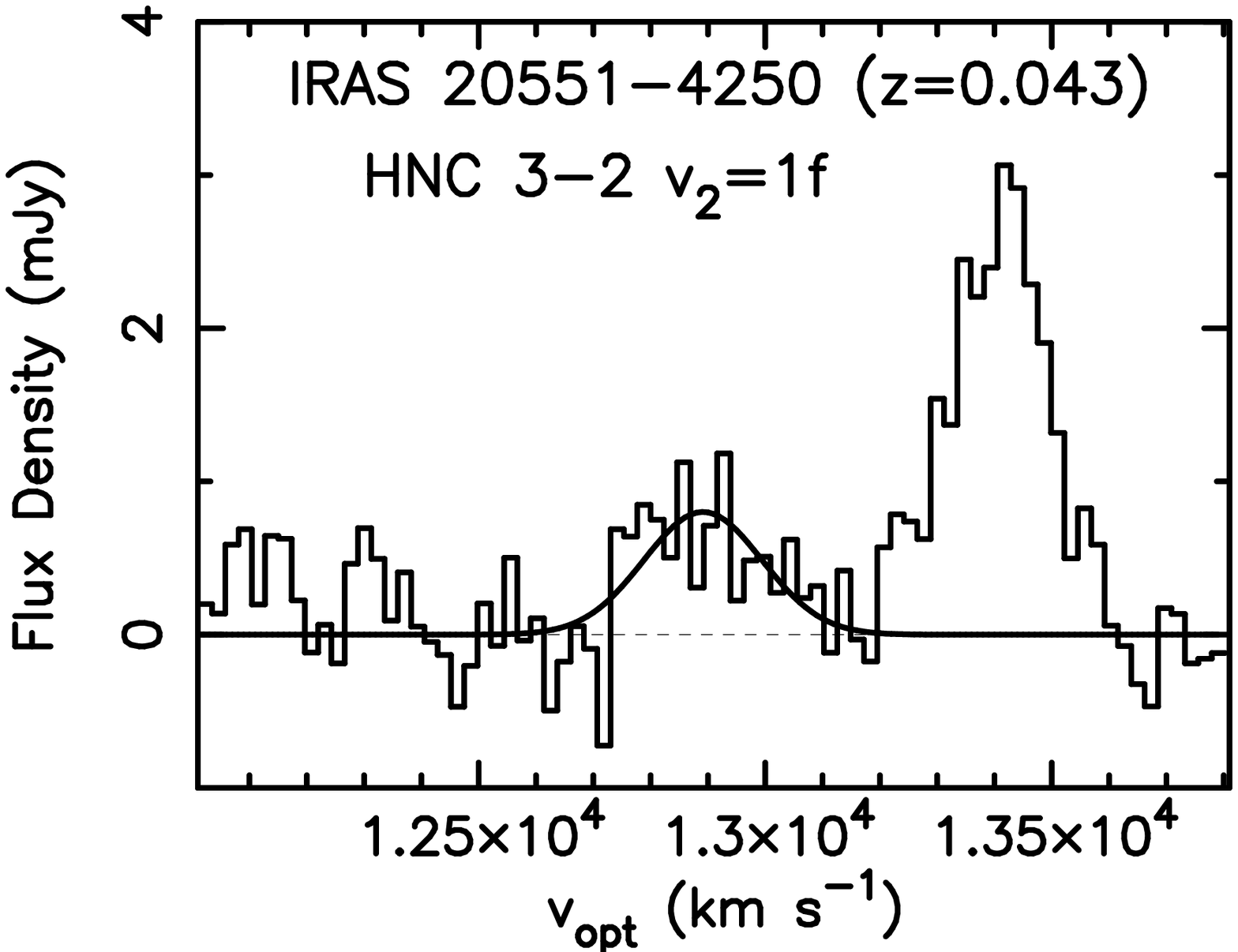}
\includegraphics[angle=0,scale=.274]{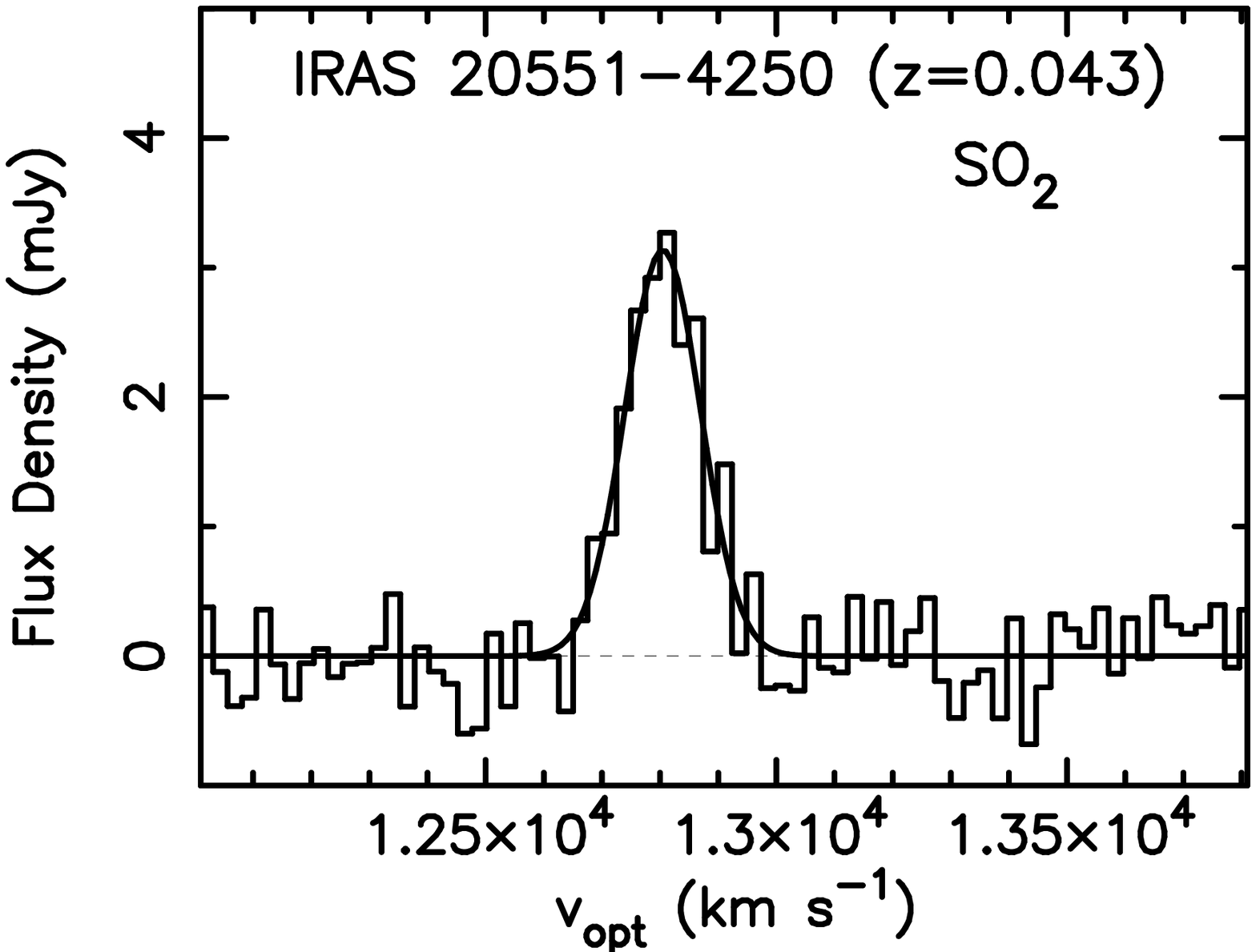}  
\includegraphics[angle=0,scale=.274]{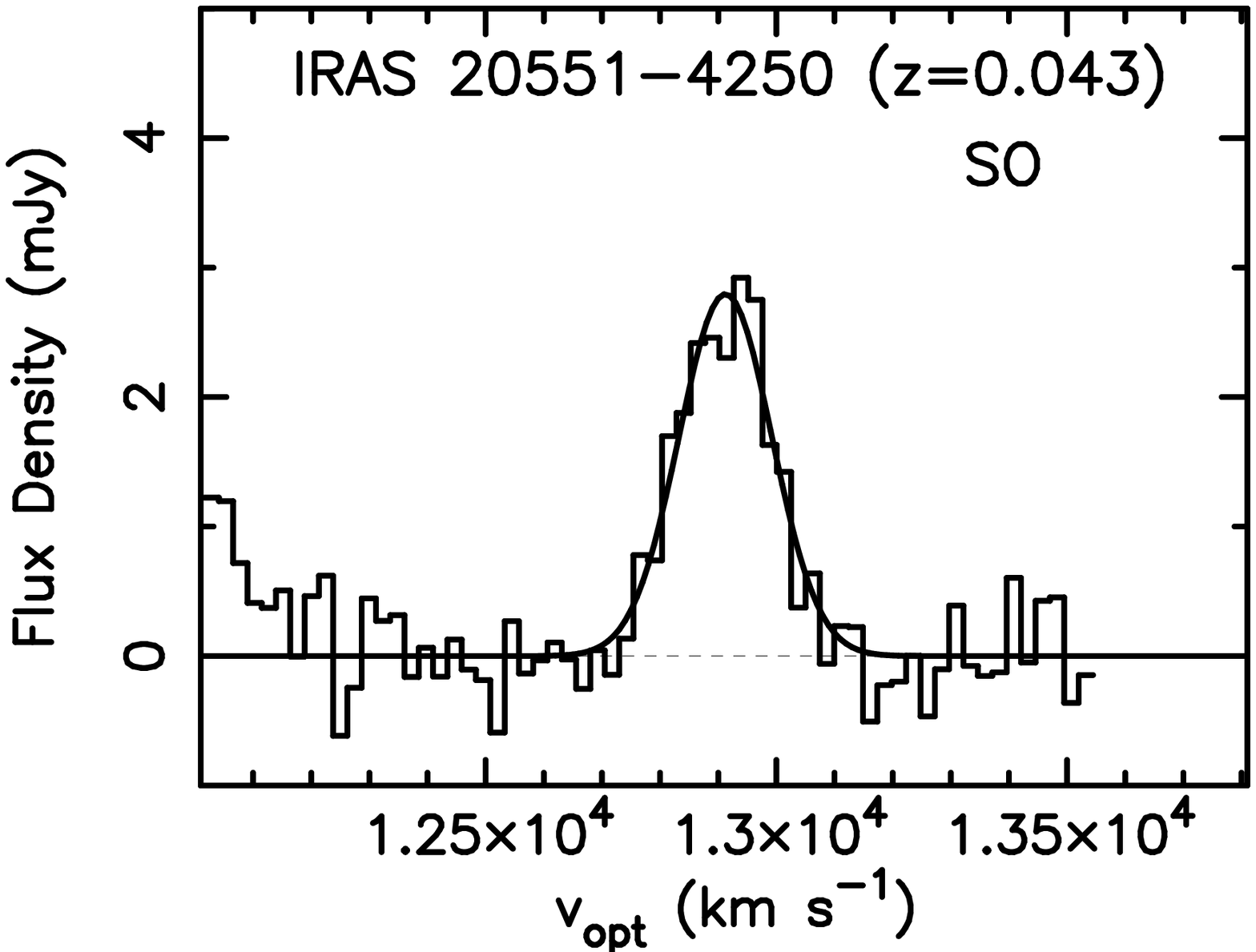}\\ 
\includegraphics[angle=0,scale=.274]{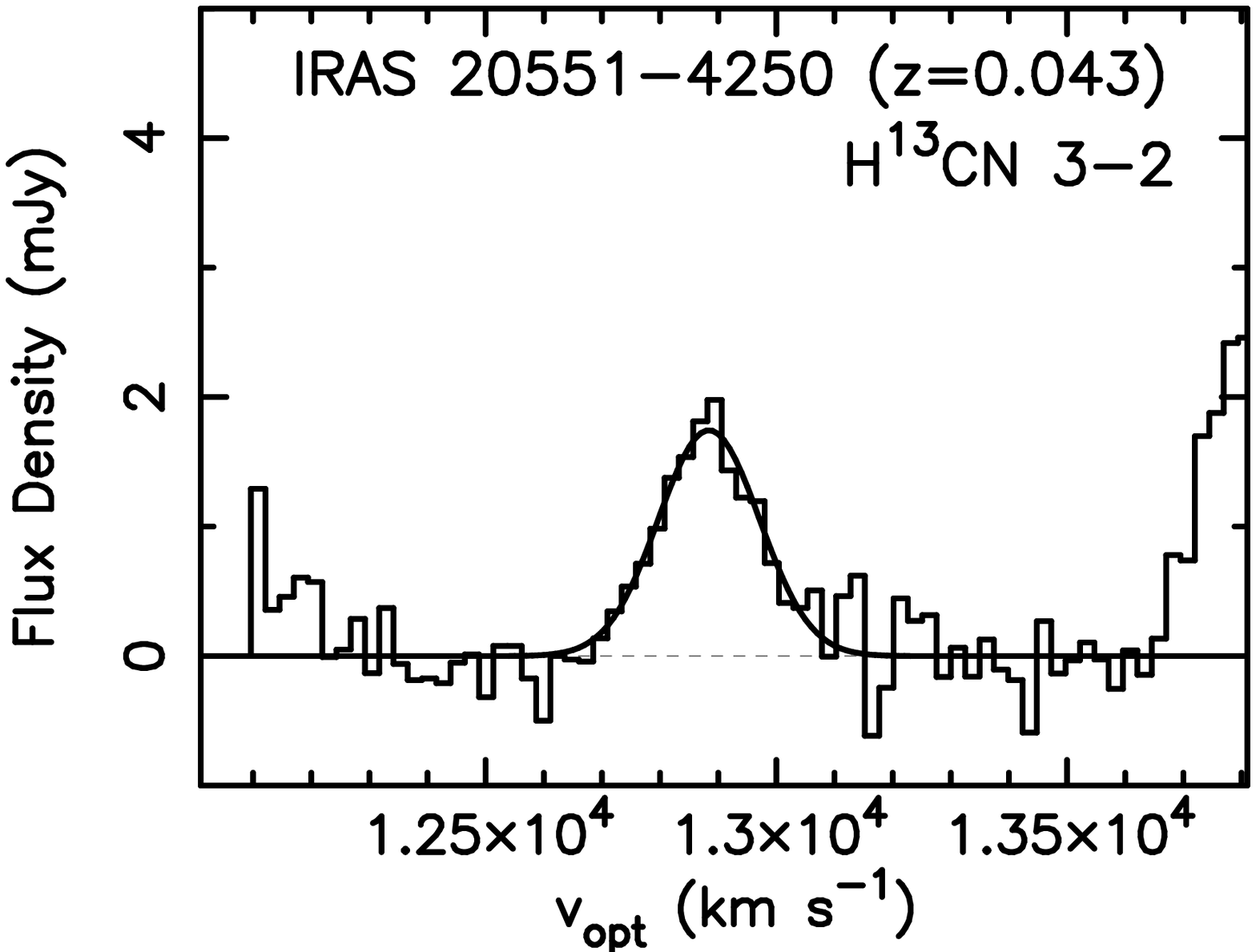} 
\includegraphics[angle=0,scale=.274]{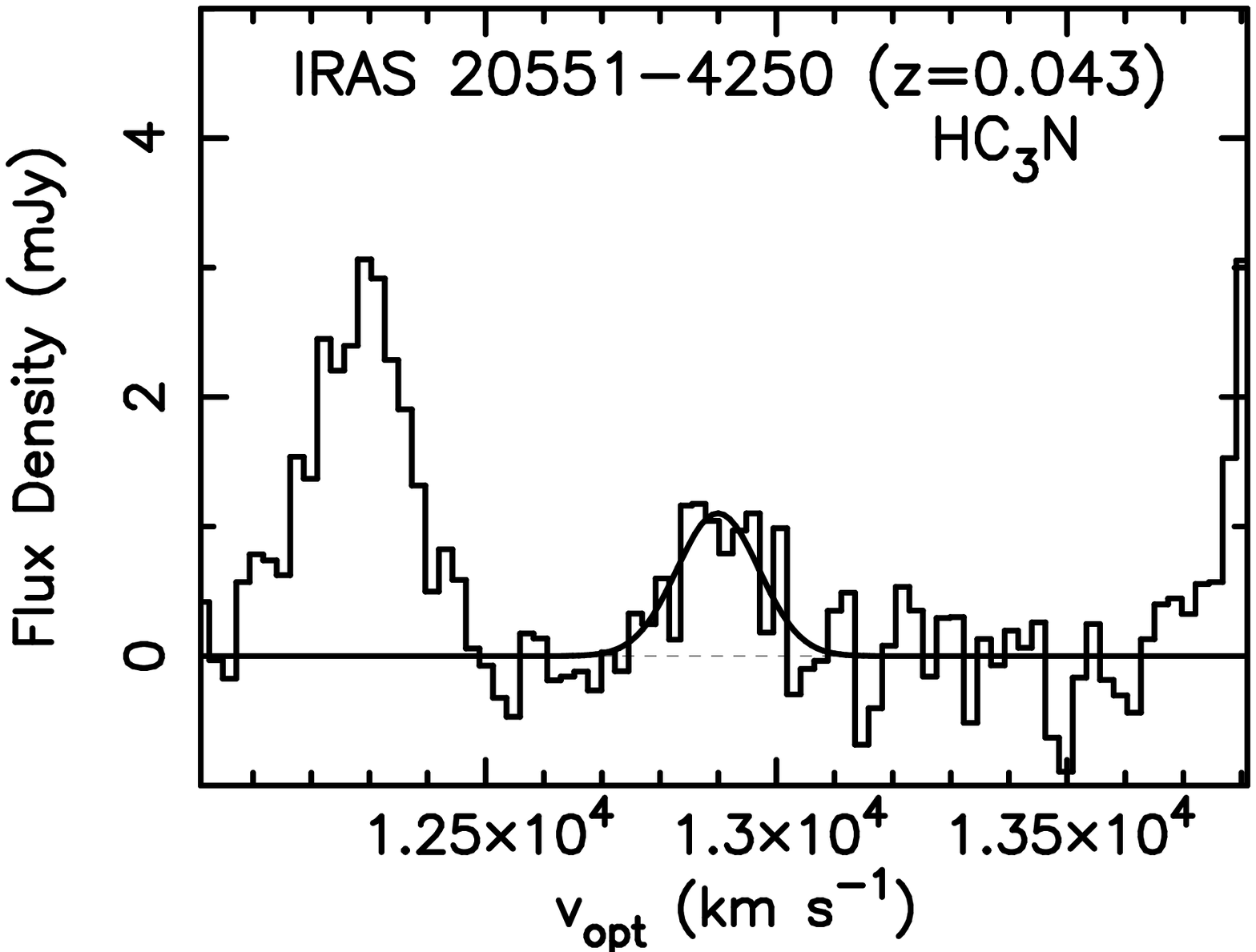} 
\includegraphics[angle=0,scale=.274]{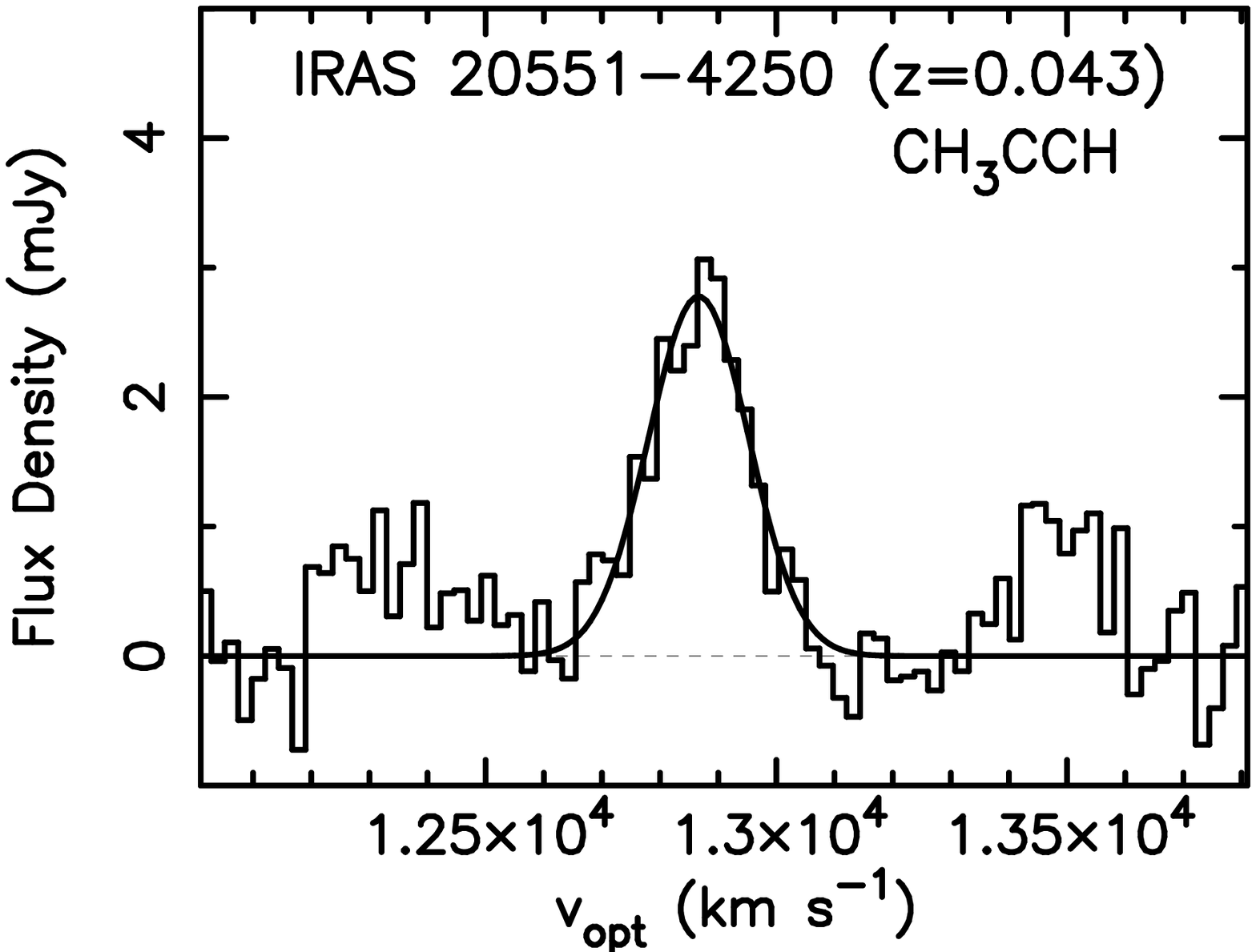}  
\end{center}
%\vspace{-0.9cm}
\caption{
Spectra at the continuum peak position, within the beam size, of our
interesting molecular lines in IRAS 20551$-$4250.  
All spectra are extracted at the continuum peak position at 
(20 58 26.80, $-$42 39 00.3)J2000.
The abscissa is optical LSR velocity (v$_{\rm opt}$ $\equiv$ 
c ($\lambda$$-$$\lambda_{\rm 0}$)/$\lambda_{\rm 0}$), and the
ordinate is flux density in [mJy].  
The best Gaussian fits (Table 3) are overplotted as solid curved lines.
}
\end{figure}

%--- Figure 5 ---%
\begin{figure}
%\begin{center}
\includegraphics[angle=0,scale=.42]{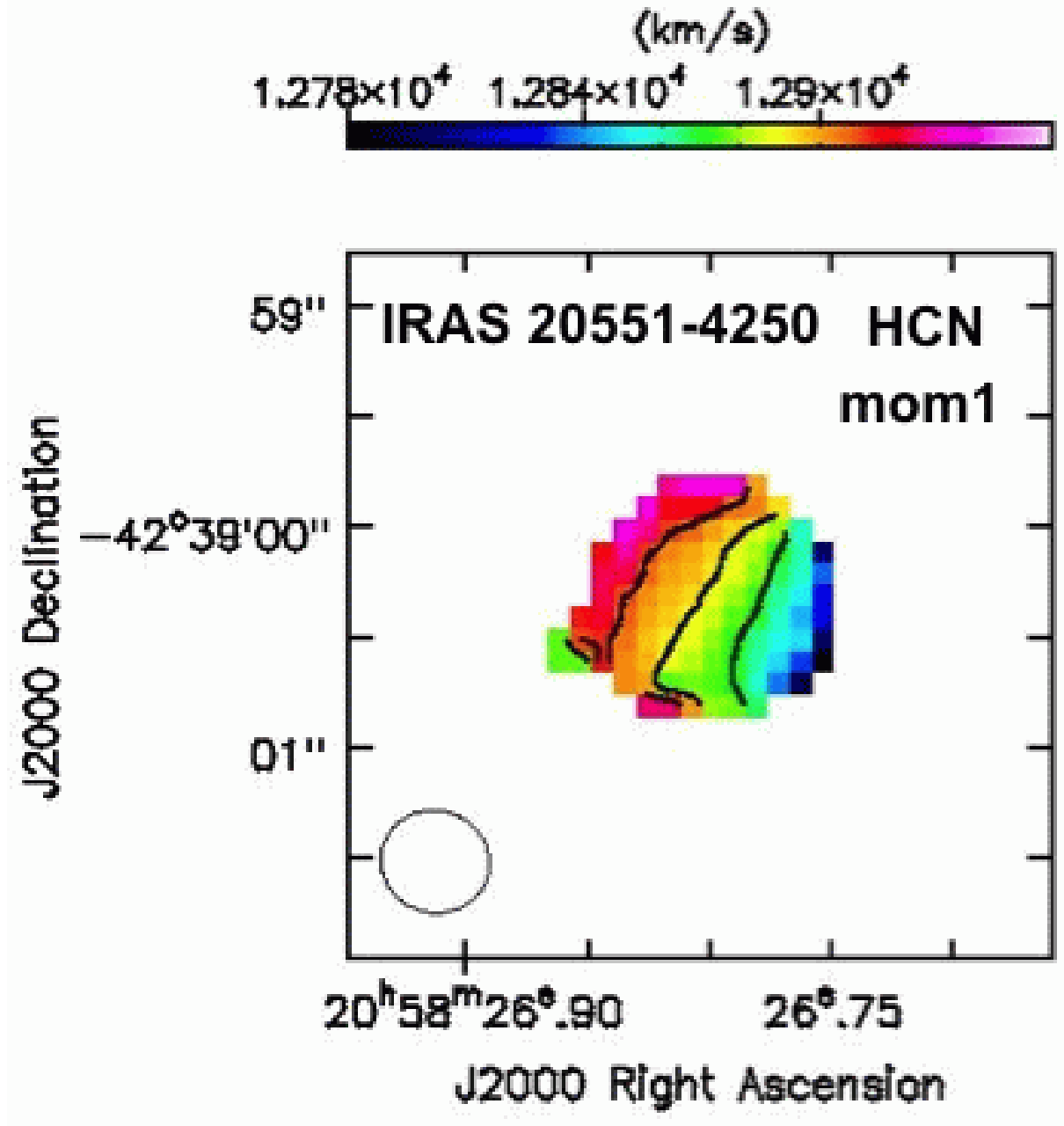} 
\includegraphics[angle=0,scale=.42]{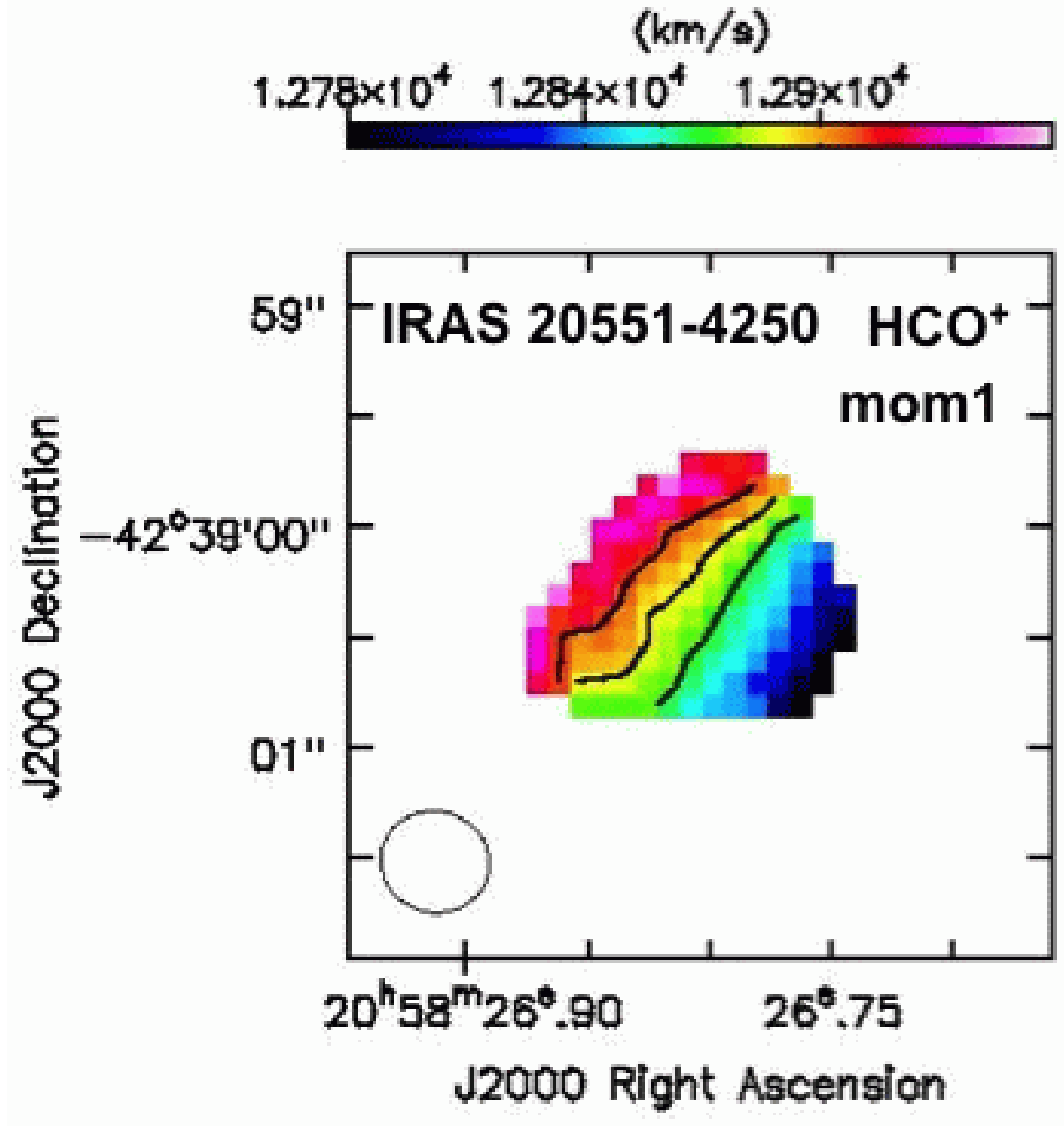} 
\includegraphics[angle=0,scale=.42]{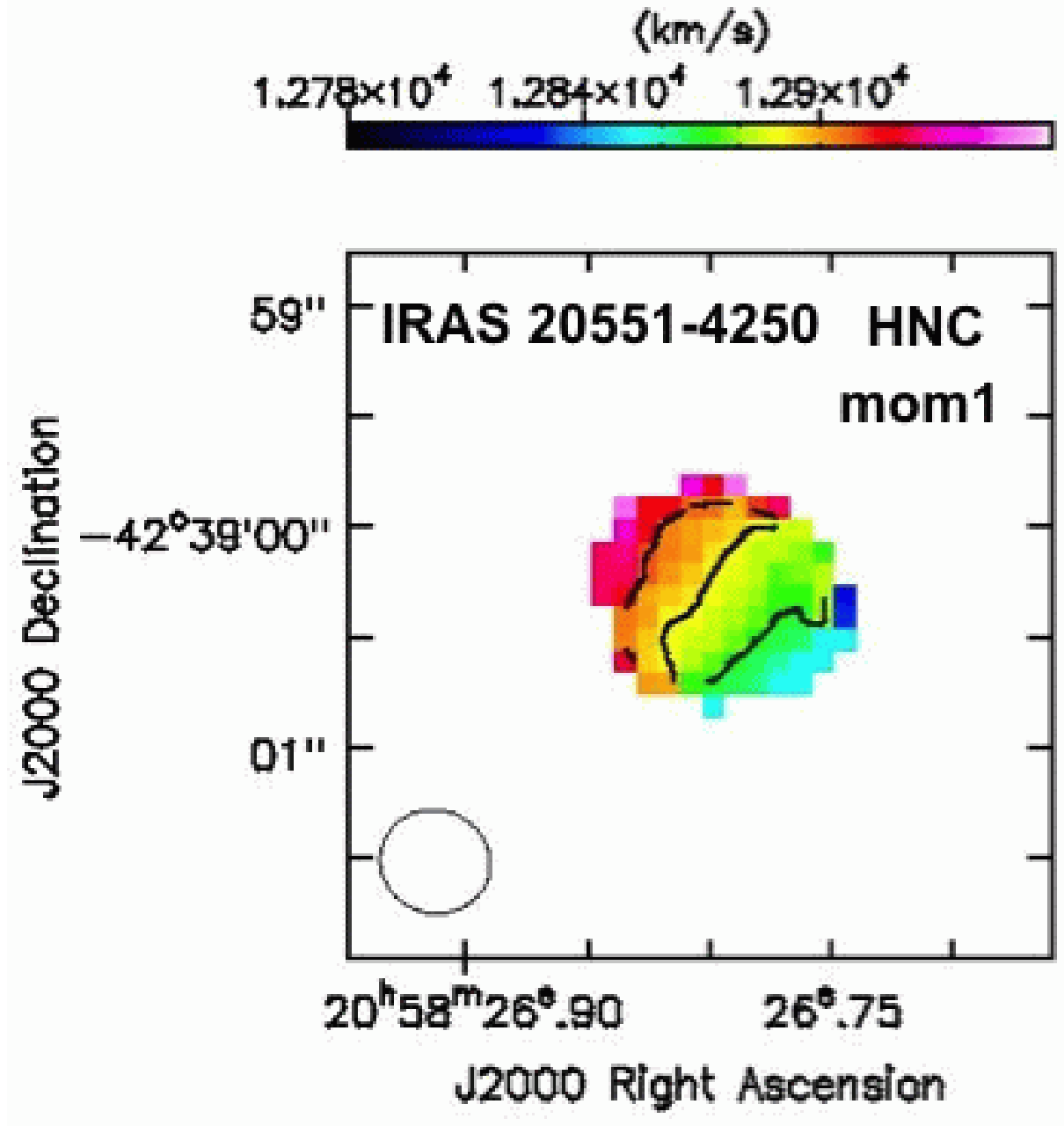} \\
\includegraphics[angle=0,scale=.42]{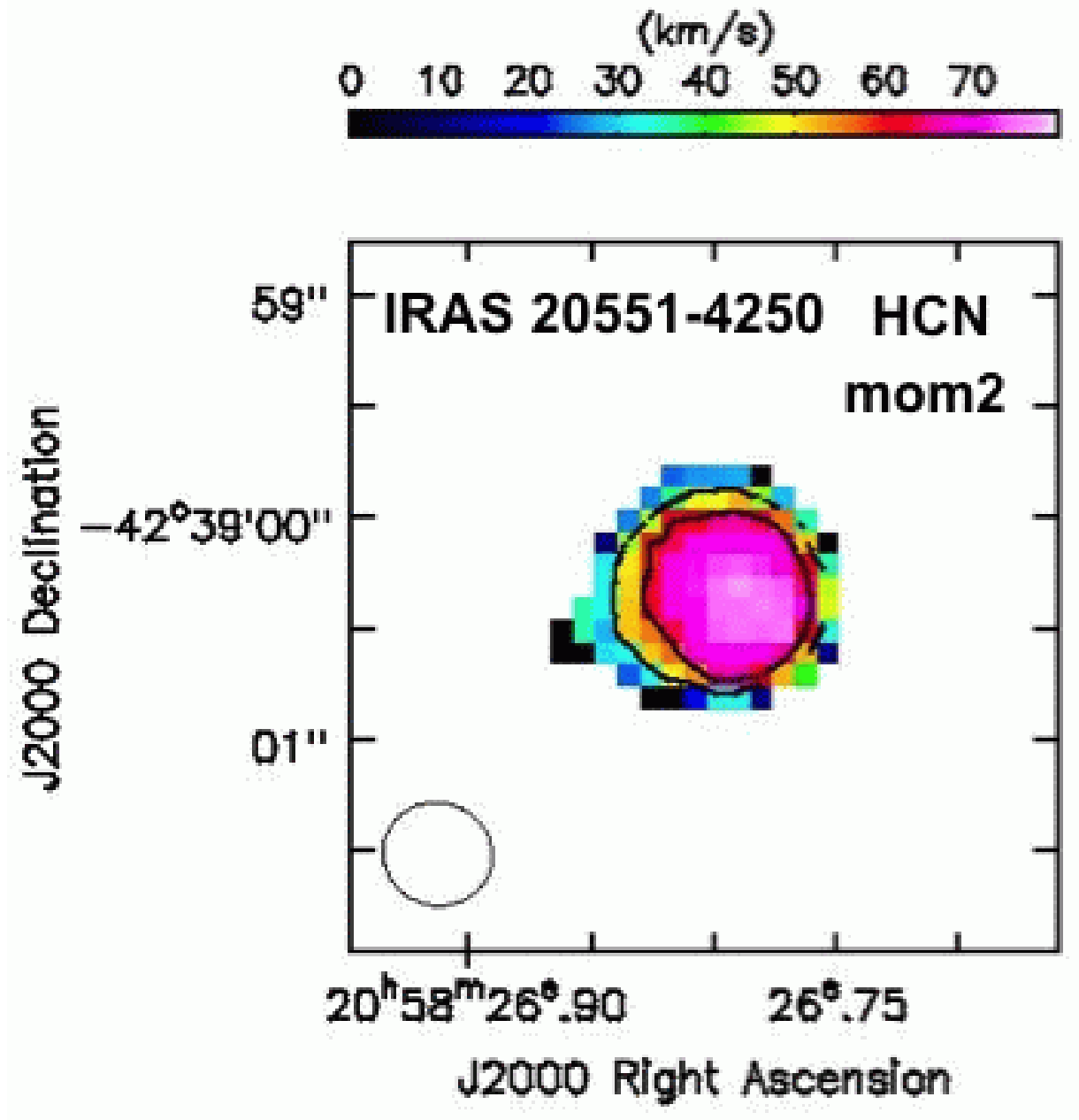} 
\includegraphics[angle=0,scale=.42]{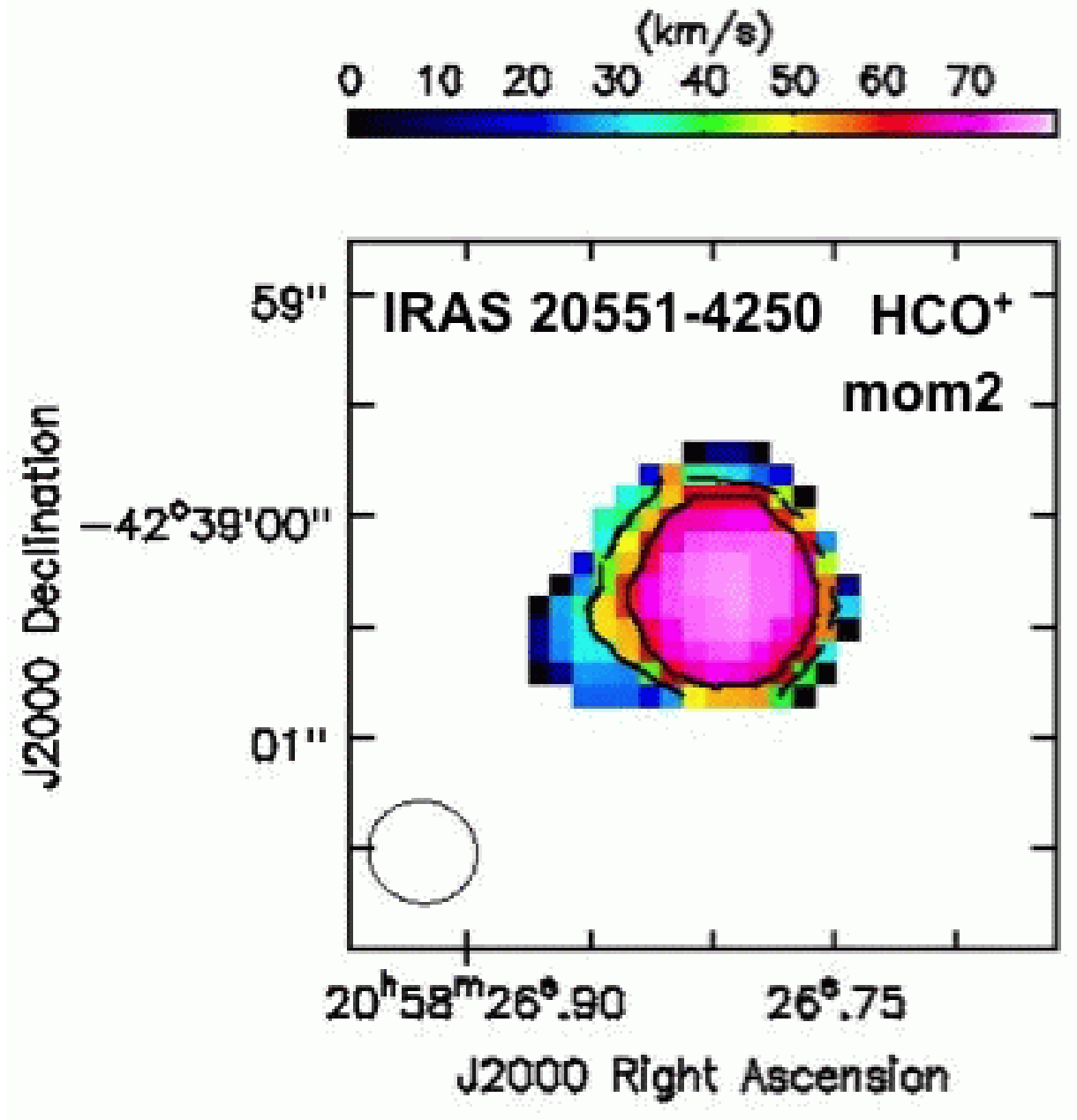} 
\includegraphics[angle=0,scale=.42]{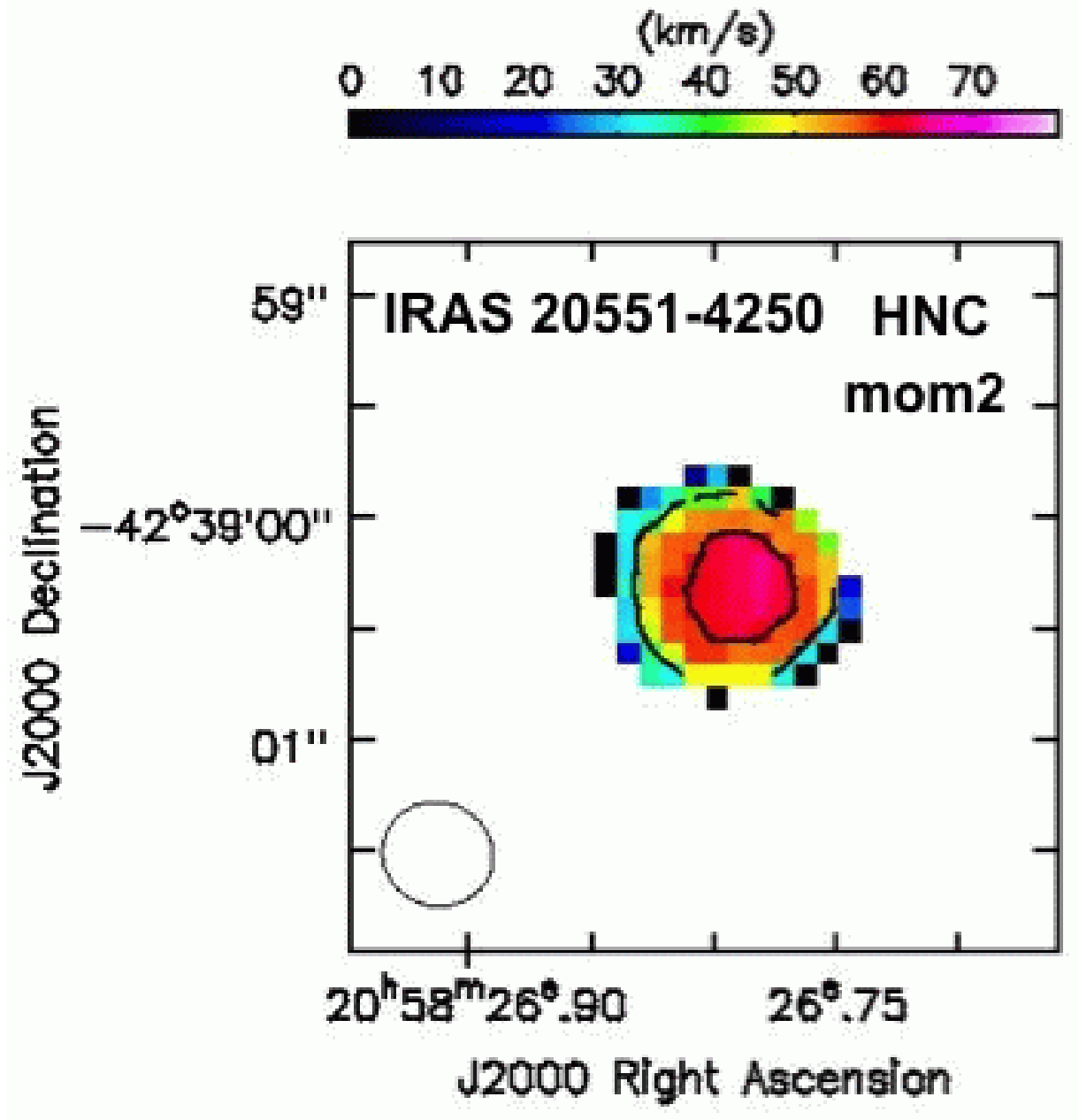} 
%\end{center}
\vspace{1cm}
\caption{
Intensity-weighted mean velocity (moment 1) and intensity-weighted
velocity dispersion (moment 2) maps of HCN/HCO$^{+}$/HNC J=3--2 (v=0)
emission lines for IRAS 20551$-$4250.  
For moment 1 maps (top panels), the velocity is in optical LSR velocity 
(v$_{\rm opt}$ $\equiv$ c ($\lambda$$-$$\lambda_{\rm 0}$)/$\lambda_{\rm 0}$). 
The contours in moment 1 maps are 12870, 12890, and 12910 km s$^{-1}$
for HCN/HCO$^{+}$/HNC J=3--2.
The contours in moment 2 maps (bottom panels) are 40 and 60 km s$^{-1}$
for HCN/HCO$^{+}$/HNC J=3--2. 
}
\end{figure}

%--- Figure 6 ---%
\begin{figure}
\begin{center}
\includegraphics[angle=0,scale=.8]{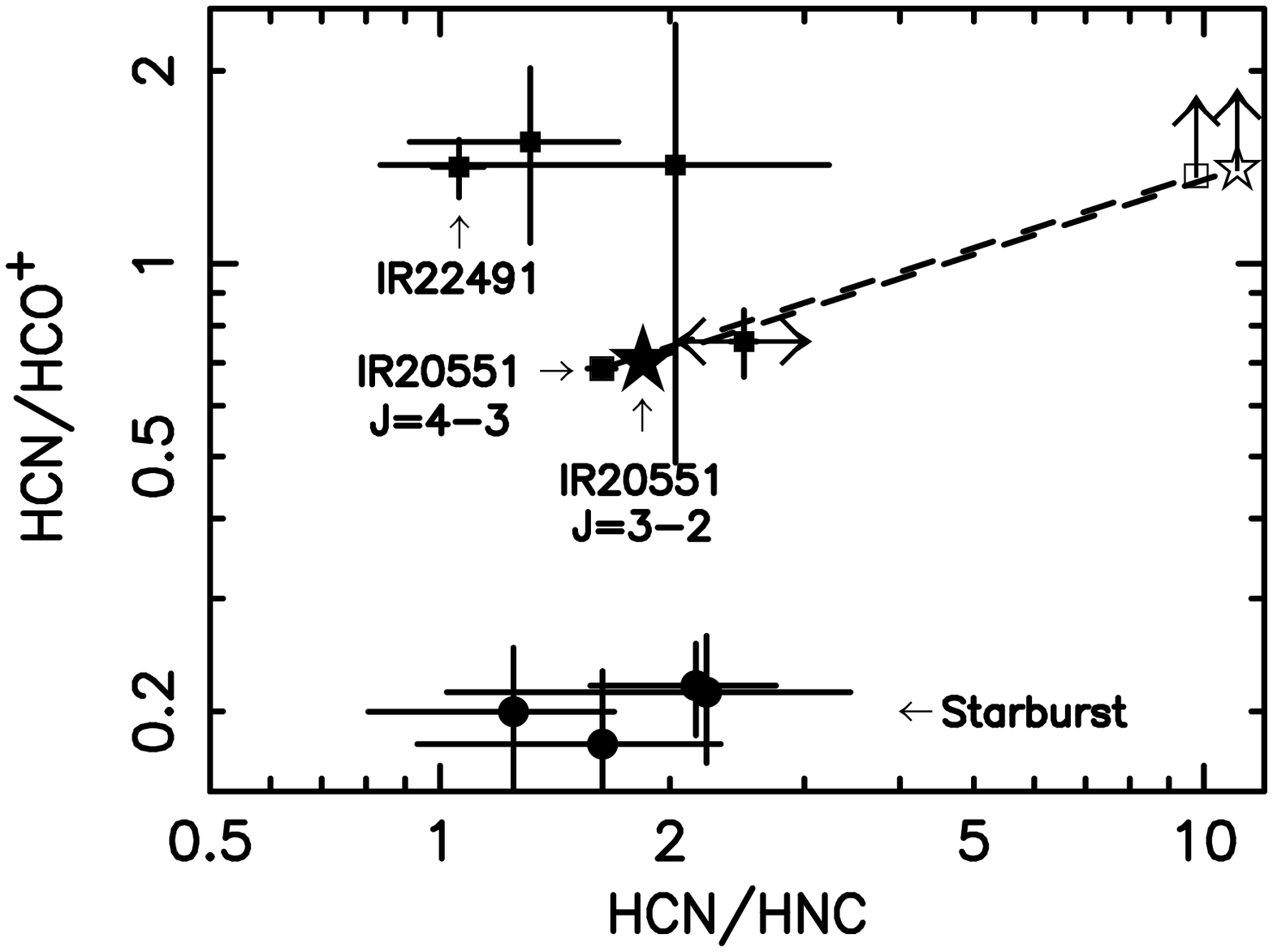} 
\end{center}
\caption{
Observed HCN-to-HNC (abscissa) and HCN-to-HCO$^{+}$ (ordinate) flux 
ratios, based on ALMA observations.
The filled circles are J=4--3 data at multiple positions
of the starburst-dominated LIRG, NGC 1614, taken by ALMA Cycle 0
observations \citep{ima13a}. 
The filled squares are J=4--3 data of ULIRGs, taken by ALMA Cycle 0 
observations \citep{ima13b,ima14}. 
Although the ratios are unchanged, uncertainties are
recalculated.
All ULIRGs but one (IRAS 22491$-$1808) show infrared-identified 
energetically important buried AGNs \citep{ima14}. 
For IRAS 22491$-$1808, the HCN v$_{2}$=1f J=3--2 emission line has been
detected in our ALMA Cycle 2 data \citep{ima16}, suggesting that this
ULIRG may contain an extremely buried AGN which is infrared-elusive but
(sub)millimeter-detectable. 
The filled star is our new ALMA Cycle 2 J=3--2 data of IRAS 20551$-$4250.
For IRAS 20551$-$4250, ratios after the correction for the derived line
opacities ($\S$4.3.2) are shown as the open square (J=4--3) and open
star (J=3--2), which are connected to the observed ratios with the
dashed lines. 
}
\end{figure}

%--- Figure 7 ---%
\begin{figure}
\begin{center}
\includegraphics[angle=0,scale=.8]{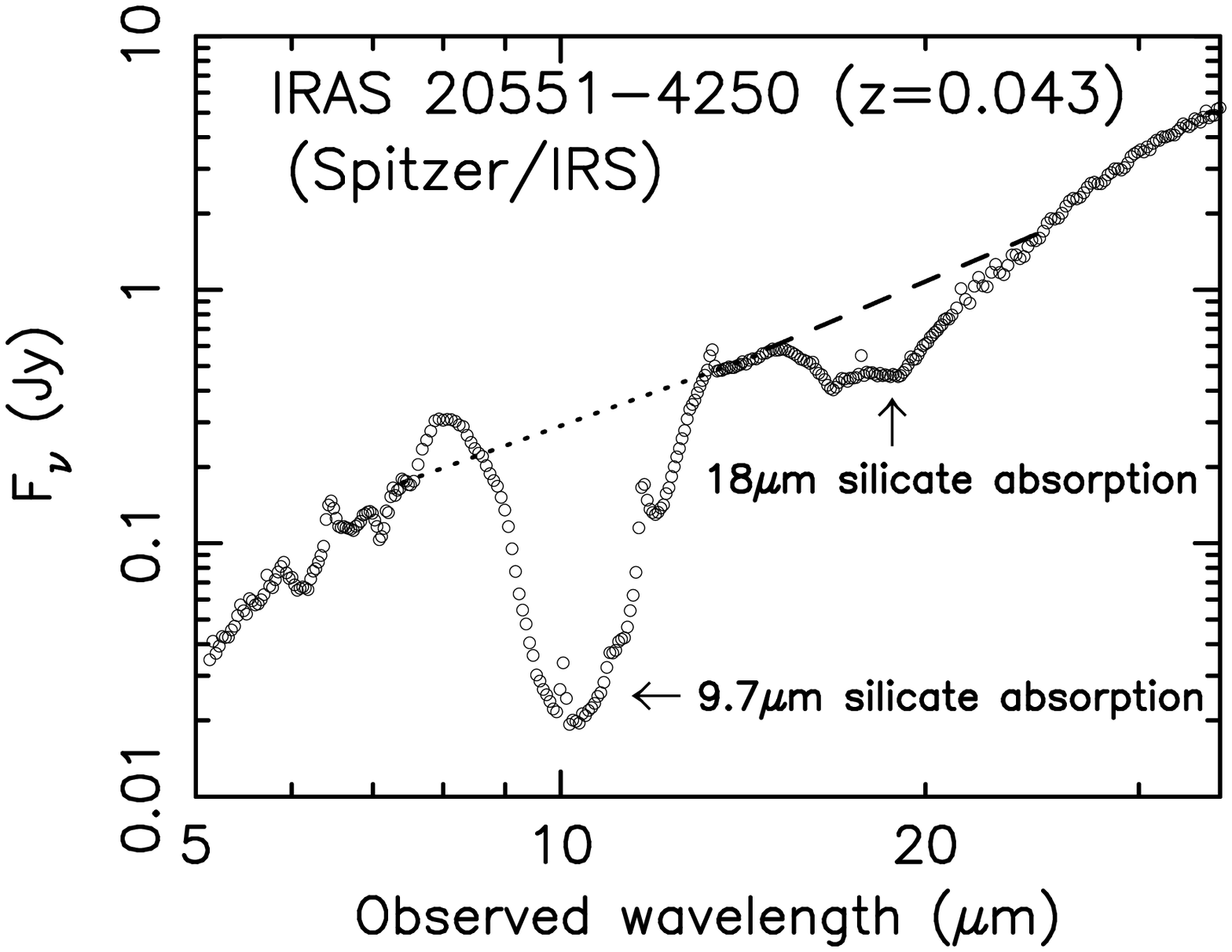} 
\end{center}
\caption{
Spitzer IRS low-resolution (R $\sim$ 100) infrared 5--35 $\mu$m spectrum 
of IRAS 20551$-$4250, originally shown by \citet{ima11}, analyzed in the
same manner as described in our previously published Spitzer IRS papers
\citep{ima07b,ima09b,ima10b}. 
It is virtually identical to the spectrum shown by \citet{sar11}, except
for small gaps among different spectral modules in the latter.
Adopted power law continuum level for the 9.7 $\mu$m (18 $\mu$m)
silicate dust absorption feature is determined, based on data points 
not strongly affected by these absorption features and polycyclic
aromatic hydrocarbon (PAH) emission features \citep{ima07b}, and 
is overplotted as the dotted (dashed) line.
}
\end{figure}

%--- Figure 8 ---%
\begin{figure}
\begin{center}
\includegraphics[angle=0,scale=.8]{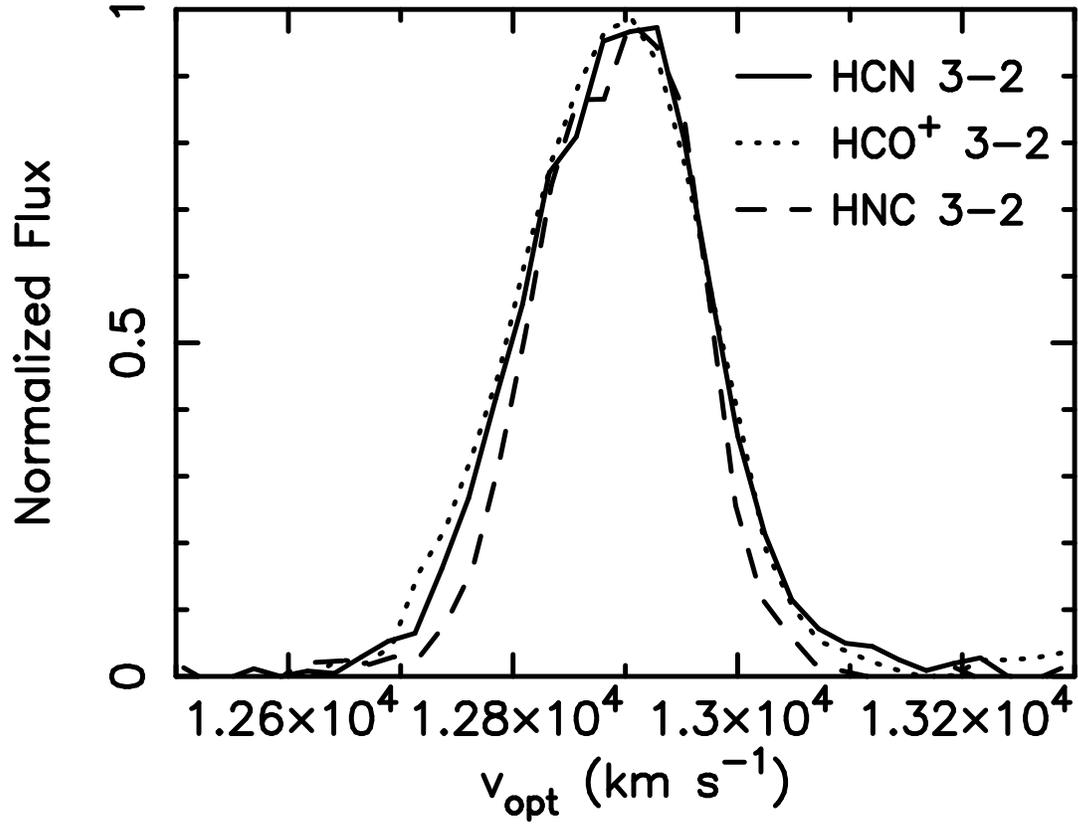} 
\end{center}
\caption{
Comparison of the line profile of HCN J=3--2 (solid line), HCO$^{+}$
J=3--2 (dotted line), and HNC J=3--2 (dashed line) (v=0).
The abscissa is optical LSR velocity (v$_{\rm opt}$ $\equiv$ 
c ($\lambda$$-$$\lambda_{\rm 0}$)/$\lambda_{\rm 0}$), and the
ordinate is flux normalized by the peak value of the Gaussian fits
(Table 3, column 8). 
}
\end{figure}

%--- Figure 9 ---%
\begin{figure}
\begin{center}
\includegraphics[angle=0,scale=.67]{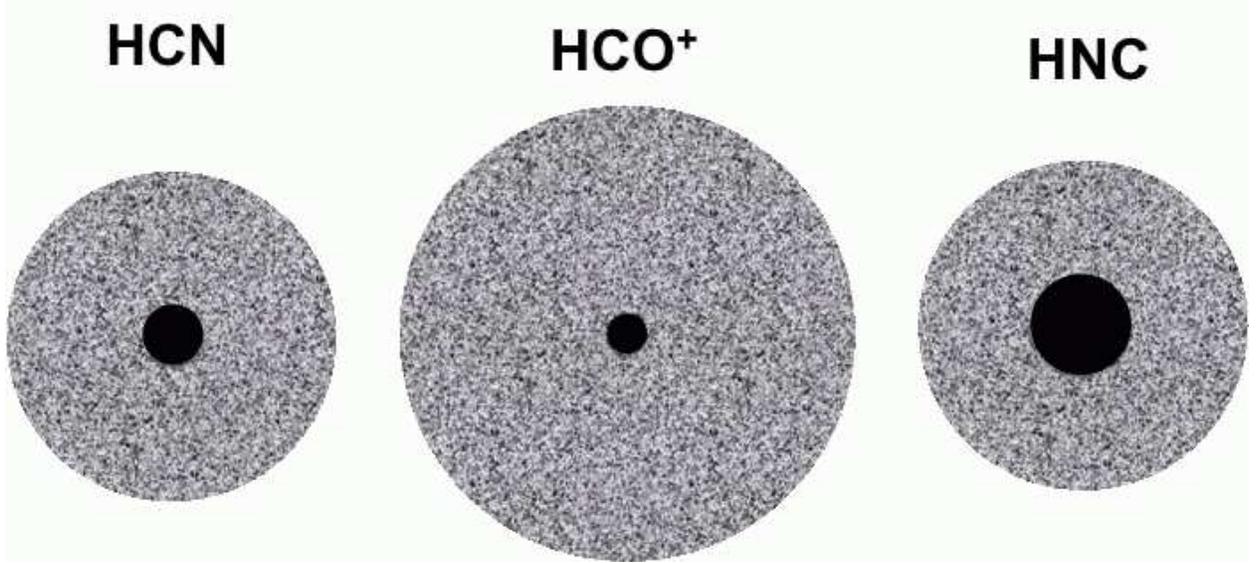} 
\end{center}
\caption{
Schematic diagram of the spatial distribution of nuclear molecular gas  
and continuum emitting regions for infrared radiative pumping for HCN,
HCO$^{+}$, and HNC.  
The large gray circles are the expected sizes of HCN J=3--2, HCO$^{+}$
J=3--2, and HNC J=3--2 (v=0) line emitting regions, assuming gradually 
decreasing radial density distribution of molecular gas from the very
center to the outer region in a galaxy \citep{big12}.    
The size is expected to be larger for HCO$^{+}$ than HCN and HNC, due to 
a factor of $\sim$5 lower critical density for HCO$^{+}$ than HCN and
HNC (see $\S$4.4). 
It is likely that these regions consist of individual molecular gas
clumps (see $\S$4.3.2), and that HCO$^{+}$ is less clumpy
(occupies a larger volume fraction) than HCN and HNC, due to the lower 
HCO$^{+}$ critical density (see $\S$4.4). 
The central filled black circles represent infrared 14 $\mu$m (HCN), 
12 $\mu$m (HCO$^{+}$), 21.5 $\mu$m (HNC) continuum emitting regions. 
The size of the infrared continuum emitting region is expected to be the
smallest for HCO$^{+}$ and the largest for HNC. 
The bulk of the HCN/HCO$^{+}$/HNC v$_{2}$=1f J=3--2 emission is likely
to come from the inner region of the gray circle, where infrared
radiation density from the central energy source is high ($\S$4.4).
}
\end{figure}

%--- Figure 10 ---%
\begin{figure}
\begin{center}
\includegraphics[angle=0,scale=.4]{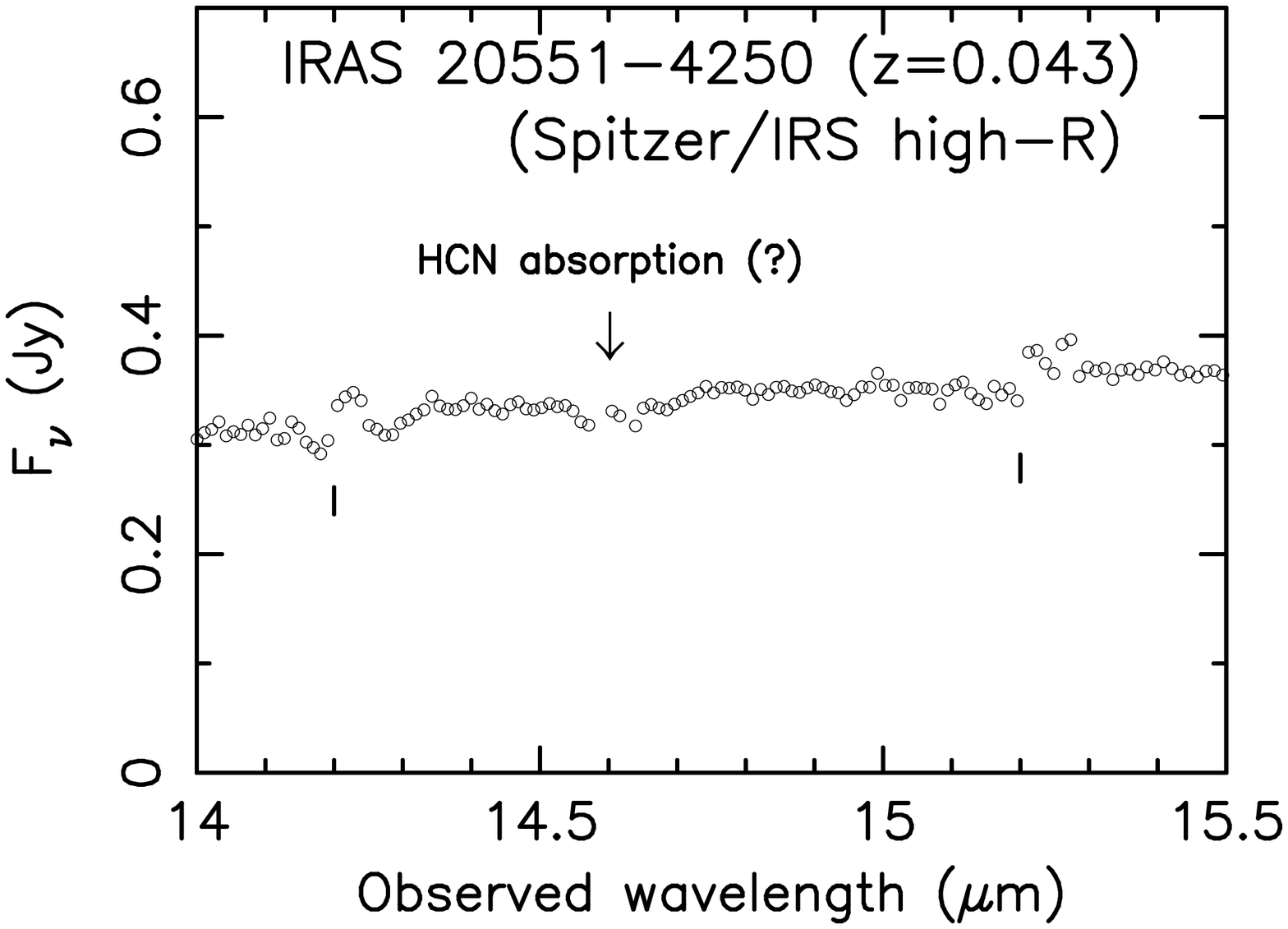} 
\includegraphics[angle=0,scale=.4]{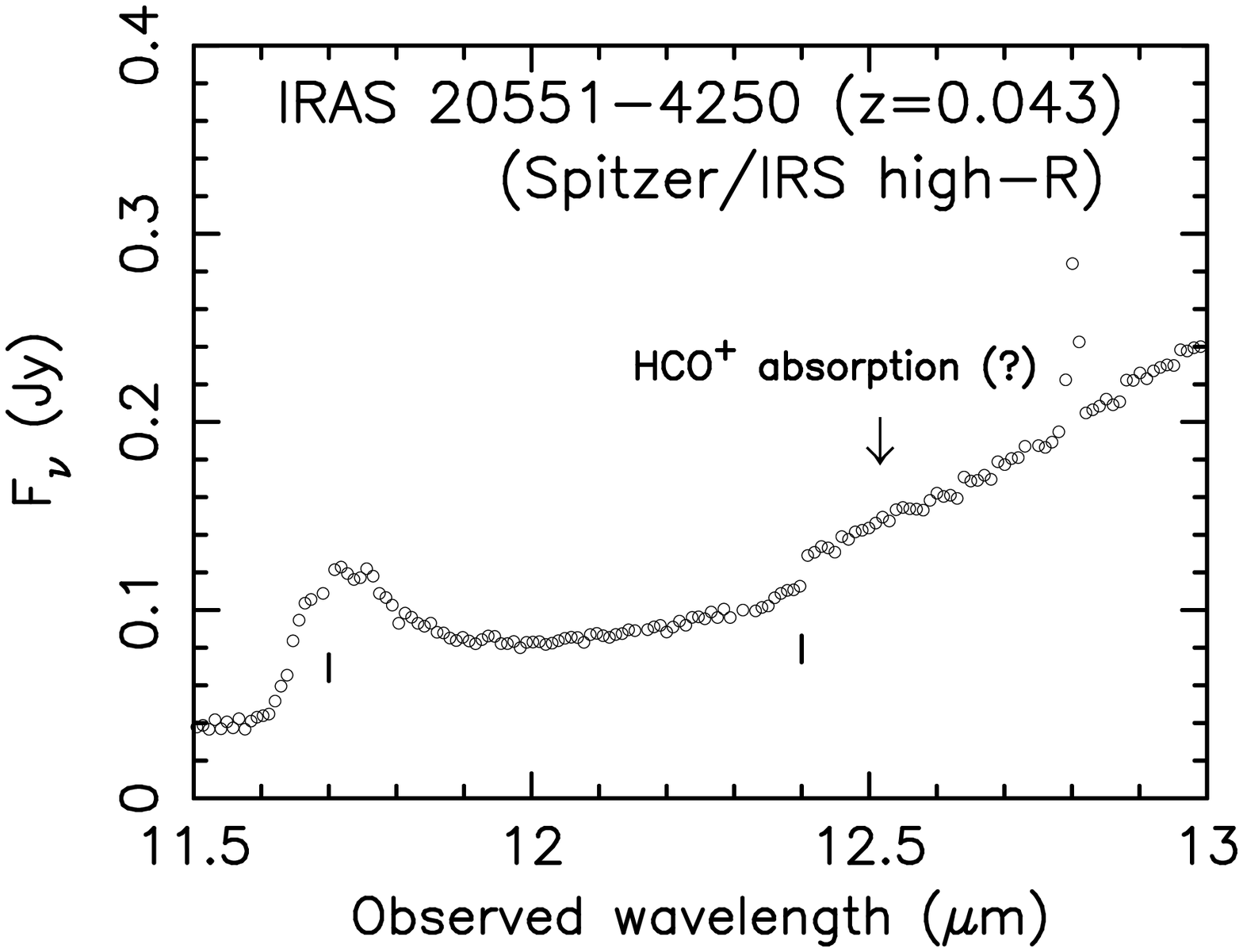} 
\includegraphics[angle=0,scale=.4]{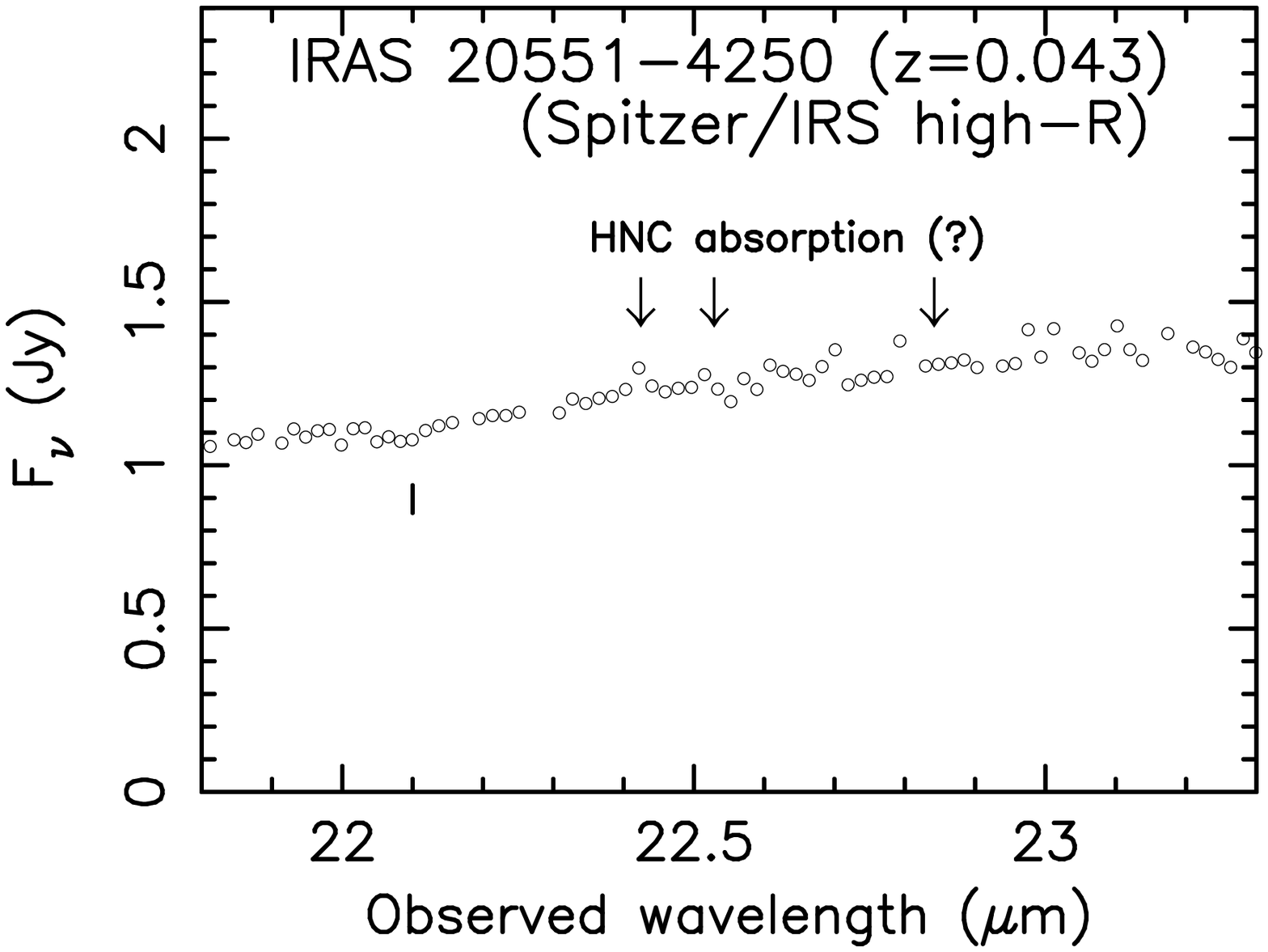} 
\end{center}
\caption{
Spitzer IRS high-resolution (R $\sim$ 600) infrared spectra of IRAS
20551$-$4250 taken from the Cornell Atlas of Spitzer/IRS Sources
(CASSIS) \citep{leb15}.
Zoom-in spectra  around HCN, HCO$^{+}$, and HNC absorption features at  
$\lambda_{\rm rest}$ $\sim$ 14 $\mu$m, 12 $\mu$m, and 21.5 $\mu$m, 
respectively, are shown.
For HCN and HCO$^{+}$, down arrows are shown at $\lambda_{\rm rest}$ =
14.0 $\mu$m and 12.0 $\mu$m, respectively.
For HNC, we plot down arrows at $\lambda_{\rm rest}$ = 21.5 $\mu$m, 21.6
$\mu$m, and 21.9 $\mu$m, which correspond to v$_{2}$=1 transition at 
low-J levels for R, Q, and P branches, respectively \citep{bur87}.
Several small gaps seen in the spectra are due to observations with
different spectral settings.    
These wavelengths at the edge of each spectral setting are indicated as
the vertical solid straight lines. 
}
\end{figure}

\end{document}